\documentclass[english,pra,notitlepage,twocolumn,superscriptaddress]{revtex4-1}
\usepackage[T1]{fontenc}
\usepackage[latin9]{inputenc}
\usepackage{color}
\usepackage{babel}
\usepackage{soul}
\usepackage[noend]{algpseudocode}
\usepackage{subcaption}
\usepackage{amsmath}
\usepackage{amssymb}
\usepackage{graphicx}
\usepackage[ruled,vlined]{algorithm2e}
\usepackage[unicode=true,pdfusetitle,
 bookmarks=true,bookmarksnumbered=false,bookmarksopen=false,
 breaklinks=true,pdfborder={0 0 0},pdfborderstyle={},backref=false,colorlinks=true]
 {hyperref}

\begin{document}

\title{
Solving optimization problems with Rydberg analog quantum computers: 
Realistic requirements for quantum advantage using noisy simulation and classical benchmarks
}

\author{Michel Fabrice Serret}
\affiliation{Atos Quantum Laboratory, Les Clayes-sous-Bois, France}

\affiliation{Ecole Polytechnique, Palaiseau, France}

\author{Bertrand Marchand}
\affiliation{Atos Quantum Laboratory, Les Clayes-sous-Bois, France}

\author{Thomas Ayral}
\affiliation{Atos Quantum Laboratory, Les Clayes-sous-Bois, France}

\begin{abstract}
Platforms of Rydberg atoms have been proposed as promising candidates to solve some combinatorial optimization problems.
Here, we compute quantitative requirements on the system sizes and noise levels that these platforms must fulfill to reach quantum advantage in approximately solving the Unit-Disk Maximum Independent Set problem. 
Using noisy simulations of Rydberg platforms of up to 26  atoms interacting through realistic van der Waals interactions,
we compute the average 
approximation ratio that can be attained with a simple quantum annealing-based heuristic within a fixed temporal computational budget.  
Based on estimates of the correlation lengths measured in the engineered quantum state, we extrapolate the results to large atom numbers and compare them to a simple classical approximation heuristic. 
We find that approximation ratios of at least $\approx 0.84$ are within reach for near-future noise levels.
Not taking into account further classical and quantum algorithmic improvements, we estimate that quantum advantage could be reached by attaining a number of controlled atoms of $\sim8,000$ for a time budget of 2 seconds,
and $\sim 1,000-1,200$ for a time budget of 0.2 seconds, provided the coherence levels of the system can be improved by a factor 10 while maintaining a constant repetition rate. 
\end{abstract}

\maketitle



In recent years, quantum hardware has witnessed very rapid technological progress \cite{monz201114, arute2019quantum},
while still being characterized by a substantial level of imperfections \cite{Preskill2018, national2019quantum}.
It is thus a very timely task to realistically assess the requirements for reaching a break-even 
point with classical algorithms, and hence guide future necessary developments.
This task first requires the identification of problems that are difficult to solve classically, and for which some
quantum platforms are expected to provide some form of speedup.
Among the many classes of computational problems put forth as promising candidates, 
NP-hard combinatorial optimization problems stand out as prototypical hard problems. Many of them have industrial
relevance, and exhibit a natural encoding onto quantum machines, through Ising
Hamiltonians and quantum annealing (see, e.g, \cite{Lucas2014}). Recently, for instance, the MaxCut and the Maximum Independent Set problems have received attention as candidates for quantum advantage \cite{Otterbach2017, Guerreschi2019, pichler2018quantum}. These optimization problems are particularly well suited to noisy intermediate scale quantum (NISQ) computers, as they can be solved using various hybrid quantum-classical approaches \cite{hauke2019perspectives, albash2018adiabatic, Kokail2018, Farhi2014}.

The predictive assessment of a break-even point calls for a realistic modeling of imperfections and a precise definition of the algorithmic success metric.
While
the above-mentioned quantum approaches
(including quantum annealing) provide a general framework for tackling combinatorial optimization problems with quantum machines \cite{hauke2019perspectives, albash2018adiabatic}, very little is known about their performance under realistic hardware models, including a precise modeling of the
microscopic Hamiltonian, decoherence effects, and readout errors.
These imperfections
give these quantum algorithms
a stochastic and heuristic nature.
In particular, approximate (i.e valid, but sub-optimal) solutions to a given instance of a problem might be returned.
These approaches
should therefore be compared with classical algorithms designed for the similar task of \emph{approximately} solving NP-hard problems, namely approximation algorithms \cite{vazirani2013approximation}.
For such a comparison, the most widespread success metric is known as the approximation ratio. This quality factor must be put in perspective with the execution time of both kinds of algorithms.

Due to the spatial structure of their interatomic interactions, platforms of Rydberg atoms \cite{Browaeys2020} have been recently proposed \cite{pichler2018quantum} as candidate quantum processors to solve a subclass of the MIS problem called the Unit-Disk Maximum Independent Set problem (UD-MIS). For instance, Ref.~\cite{pichler2018quantum} compared the probability for finding the UD-MIS solution using
the---analog---quantum annealing algorithm and the---digital (i.e gate-based)---Quantum Approximate Optimization Algorithm (QAOA \cite{Farhi2014}) algorithm in the absence of quantum noise, showing the potential of the QAOA approach. Ref.~\cite{Henriet2020} investigated the generic effect of dephasing and relaxation noise on the QAOA algorithm with a simplified hard-sphere model of the Rydberg interaction.

Despite the promises of the digital (QAOA) approach
and the steady progress to make Rydberg platforms suitable digital platforms \cite{Levine2018, Madjarov2020}, they are still lagging behind the two major digital quantum computing platforms, namely superconducting \cite{Kjaergaard2020} and trapped-ion \cite{Bruzewicz2019} processors, in terms of gate fidelities.
Conversely, the number of controllable atoms in today's most advanced \emph{analog} Rydberg setups is close to a hundred \cite{barredo2018synthetic,OhlDeMello2019} and far exceeds the number of qubits of digital platforms, thus severely challenging classical simulation capabilities.
While recent experimental work \cite{lienhard2018observing} demonstrated the implementation of adiabatic processes that could allow to tackle UD-MIS on grid graphs (square and triangular), other recent works
 \cite{barredo2016atom,labuhn2016tunable,barredo2018synthetic} demonstrated the possibility for arbitrarily positioning tens of neutral atoms, making it possible for the Rydberg platform to tackle almost any of the graph instances coming up in UD-MIS problems.

In this work, we therefore quantitatively compare the potential of Rydberg platforms, described in a realistic way and operated as analog quantum processors, to solve the UD-MIS optimization problem using a quantum annealing-based approximation method, with respect to 
a simple classical approximation approach for this problem.

We start by setting up a generic methodology for comparing quantum and classical randomized approximation algorithms
via the comparison of the maximum approximation ratio achievable within a predefined computational time budget. 
In particular, we study how the system size and number of random samples (or repetitions) impact the approximation ratio. 

On the classical side, we introduce a new (to our best knowledge),
simple locality-based approximation heuristic, inspired from state-of-the-art approximation algorithms \cite{matsui1998approximation,nieberg2004robust,das2018efficient}, and benchmark it on a class of experimentally-implementable random graphs. 
We qualify it with respect to our time-budget metric, and use it to estimate a classical boundary. 

On the quantum side, we construct a hardware model with a realistic account of the van der Waals interactions, of decoherence and readout errors, and we validate it against published experimental results.

We conduct, under this model and improved noise models mimicking near-term and future hardware improvements, numerical noisy simulations of Rydberg systems executing an annealing schedule (similar to the schedules used in \cite{lienhard2018observing} and \cite{pichler2018quantum}) aimed at heuristically tackling UD-MIS. We say this process constitutes a ``quantum-annealing-based heuristic'', as it differs from standard quantum annealing by the fact that the final Hamiltonian (the experimental Rydberg ``resource'' Hamiltonian) is not the ideal ``target'' UD-MIS Hamiltonian.
We reach the unprecedented number of 26 atoms (to our best knowledge, previous noisy simulation work reached 16 \cite{lienhard2018observing} and 18 \cite{Henriet2020} atoms).
In so doing, we identify the existence of a finite and noise-dependent optimal annealing time, a property that is specific to systems with decoherence.

We use these numerical results to
estimate correlation lengths within Rydberg systems for various
noise levels.
Based on this correlation length and on statistical arguments, we extrapolate the numerical results to larger numbers of atoms.
By comparing the so-obtained best quantum approximation ratio to that obtained by our classical approximation approach, we find an estimation of the conditions under which Rydberg platforms should outperform the classical approach.

The paper is organized as follows: 
Section \ref{sec:udmis_theory} introduces the UD-MIS problem and the classical algorithmic approaches to solving it;
in Section \ref{sec:methods}, we describe the methods, quantum (\ref{sec:qa}) and classical (\ref{sec:locality_based_method}), we selected for our comparison;
finally, in Section \ref{sec:overview}, we state our results relative to the break-even point, in terms of noise levels and number of variables, between classical and quantum approaches.

\section{The Unit-Disk Maximum Independent Set problem}\label{sec:udmis_theory}

In this section, we describe the UD-MIS problem, and introduce the concept of approximation algorithm. It allows us
to describe our methodology for comparing quantum and classical approaches to NP-hard problems, which consists
in studying the ability to approximately solve NP-hard problems within a pre-defined time-budget.

\subsection{Definition}

Let $G\equiv(V,E)$ denote a graph with vertex set $V$ and edge set $E$.
Let $N\equiv|V|$ denote the number of vertices of $G$, and $S\equiv(n_{1}^{(S)}\dots n_{N}^{(S)})$
denote a length-$N$ bitstring ($n_{i}^{(S)}\in\{0,1\}$) with Hamming
weight (number of nonzero bits) $|S|=\sum_{i=1}^{N}n_{i}^{S}$.

The
Maximum Independent Set (MIS) problem consists in solving the following maximization problem:
\begin{align}
\max_{S\in\mathcal{B}} & \;|S|\label{eq:UDMIS_problem}\\
\mathrm{s.t} & \;S\in\mathrm{I.S},\nonumber 
\end{align}
where $\mathrm{I.S}$ (for ``Independent Sets'') is the set of bitstrings $(n_{1}...n_{N})$ corresponding to independent sets of $G$,
i.e.
sets of mutually non-connected vertices. Namely, a bitstring $S=(n_{1}...n_{N})$ corresponds to an independent set if $\forall (i,j)\quad n_{i}=n_{j}=1 \Rightarrow (i,j)\notin E$. $\mathcal{B}$
denotes the set of all possible bitstrings. The size of $\mathcal{B}$
is exponential in the graph size, $|\mathcal{B}|=2^{N}$.

In other words, the MIS problem consists in, given a graph $G$, determining the size of the largest possible independent set and returning an example of such a set \cite[p. 108,  on MaxClique, the dual problem of MIS]{hromkovivc2013algorithmics}.

\begin{figure}
\begin{centering}
\includegraphics[width=.8\columnwidth]{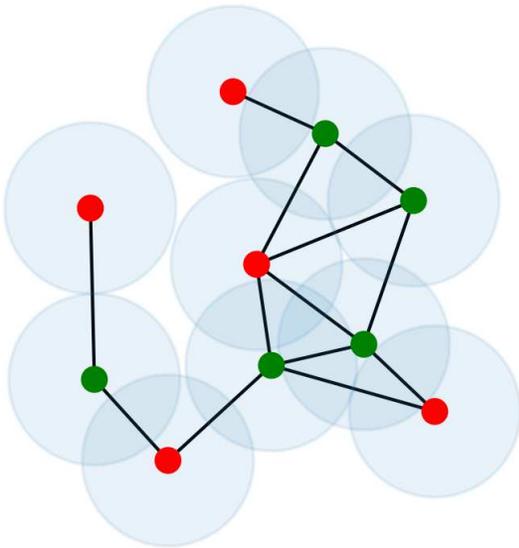}
\par\end{centering}
\caption{
An example of a unit-disk graph with 10 vertices. The red dots correspond to a maximum independent set for this graph, i.e 
a set of mutually non-connected vertices of maximum cardinality.
\label{fig:ud_graph}}
\end{figure}

The \emph{Unit-Disk}
MIS (UD-MIS) problem is the MIS problem restricted to \emph{unit-disk graphs}.
A graph is a unit-disk graph if one can associate a position in the 2D plane to every vertex such that two vertices share an edge if and only if their distance is smaller than unity.
Figure~\ref{fig:ud_graph} displays an example of a unit-disk graph, with an example maximum independent set for that graph, in red.

\subsection{Classical algorithmic approaches}\label{subsec:algo_approaches}

UD-MIS is NP-hard,
which means that any NP optimization problem can be reduced (reformulated) as UD-MIS with polynomial overhead \cite{hromkovivc2013algorithmics}.
Under the widely believed assumption that $P\neq NP$, this implies that there is no generic polynomial-time algorithm for UD-MIS, i.e no algorithm that would take as input any unit-disk graph, and return an independent set of maximum size, with certainty or good probability, in polynomial time. 

However, while restricting the input of the MIS problem to unit-disk graphs preserves NP-hardness, the difficulty of \emph{approximately solving} it drastically changes.
UD-MIS indeed allows for efficient \emph{approximation algorithms}, as we will now discuss.

An approximation algorithm to an NP-hard problem
is a polynomial-time algorithm returning solutions that are guaranteed to be
within a certain percentage of the optimum in terms of a cost function. Here, the cost function is the cardinality $|S|$ of an independent set.

Formally, a polynomial-time algorithm $\mathcal{A}$ may be called \emph{approximation algorithm} \cite{cormen2009introduction,vazirani2013approximation} to a given NP-hard maximization problem if, given an instance $I$
(in our case, a graph $G$), 
it returns a solution $S$ such that:
\begin{equation}
\alpha\equiv\frac{\mathcal{C}(S,I)}{\mathcal{C}(S^{*},I)}\geq \phi(|I|),
\end{equation}

where $\phi$ is a scalar function, sometimes called \emph{approximation factor}, specifying a \emph{guarantee} on the approximation ratio achieved achieved by the algorithm on inputs of size $|I|$.
NP-hard optimization problems may then be classified with respect to the approximation factor achieved by their best approximation algorithm, on what could be called an ``approximability scale'' from least approximable to most approximable.

For instance, a maximization problem only allowing for factors $\phi(|I|)$ decreasing with $|I|$ is usually informally considered as hard to approximate. This is in fact the case for the general MIS problem, which cannot allow for anything better than a polynomially-decreasing factor under specific complexity-theoretic assumptions \cite{vazirani2013approximation}.

Some other NP-hard problems, such as MaxCut, allow for a \emph{constant-factor} approximation algorithm, i.e with $\phi(|I|)=\phi$ independent of the input
size. A value close to $1$, with a simple enough algorithm, may render the problem easy to approximately solve in practice.

UD-MIS falls in yet another category of problems, usually informally considered as the ``most easily approximable problems'', namely those allowing for a PTAS, or \emph{Polynomial-Time Approximation Scheme}.

Formally, a PTAS is an approximation algorithm that, given \emph{any} $\epsilon > 0$ achieves a factor $(1-\epsilon)$ in $O(poly_{\epsilon}(|I|))$. I.e, given \emph{any} fixed $\epsilon$, the complexity of returning a solution within
a factor $(1-\epsilon)$ of the optimal is polynomial in the size of the input.
A PTAS may therefore return solutions arbitrarily close to the optimal, at the expense of a longer computational time (the degree and coefficients of the polynomial may increase when $\epsilon$ decreases). In other words, a PTAS makes a problem ``constant-factor approximable for any factor''.

Table \ref{fig:apx_hardness_table} gives examples of problems with varying degrees of approximabilities. Note how the MIS problem goes from inapproximable (polynomial factor) to constant-factor and PTAS-approximable when one restricts its input to bounded-degree graphs and unit-disk graphs, respectively.

\begin{table}
    \centering
\begin{tabular}{|c|c|}
\hline
\textbf{Problem} & \textbf{Best apx. factor.} ~\cite{vazirani2013approximation}   \\
\hline
general MIS & polynomial \\
\hline
bounded-degree MIS & constant \\
\hline
MaxCut & constant ($\sim 0.878$) \\
\hline
VertexCover & PTAS \\
\hline 
UD-MIS & PTAS \\
\hline 
Knapsack & f-PTAS \\
\hline
\end{tabular}
    \caption{Examples of NP-hard optimization problems with varying degrees of approximation hardness, sorted from least approximable (general MIS) to most approximable (Knapsack, which allows for a strong version of PTAS called f-PTAS \cite{vazirani2013approximation})}
    \label{fig:apx_hardness_table}
\end{table}

All PTAS's for UD-MIS rely on the common idea of partitioning the graph into
small sub-instances that are solved separately. The returned solution
is then the union over the local solutions. Their practical computational complexity as $\epsilon$ increases is notably quite sharp \cite{van2005approximation,das2018efficient}, which raises the question of how quantum approaches may fare compared to them.

Two main strategies can be distinguished for the UD-MIS approximation: 2-level shifting schemes \cite{das2018efficient, matsui1998approximation, da2017shifting,van2005approximation} and Breadth-First-Search-sphere-based schemes \cite{nieberg2004robust}.
We implemented a \emph{locality-based heuristic}
inspired from the latter, as described in section \ref{sec:locality_based_method}.
For this approach, as we shall explain in the next section, we look at the best approximation ratio it can achieve within a certain time-limit.

\subsection{Methodology: comparing quantum and classical approximation algorithms}\label{subsec:methodology}

The basic idea behind our work is to compare the best classical and quantum approaches, given the same time budget, to the UD-MIS problem, with realistic assumptions as to the imperfections of the quantum hardware.

We view our work as a comparison of classical and quantum \emph{approximation algorithms} \cite{vazirani2013approximation} to this
problem. We therefore focus on a standard metric to evaluate the quality of the approximate solutions, namely the approximation
ratio:
\begin{equation}
    \alpha \equiv \frac{\mathcal{C}(S,G)}{\mathcal{C}(S^*,G)},
    \label{eq:apx_ratio_def}
\end{equation}
where $S$ denotes an approximate solution to a particular graph instance $G$, $S^*$ the optimal solution, and $\mathcal{C}$ the cost function that one seeks to maximize. 
In this study, we typically look at \emph{expected approximation ratios}, averaged over both random instances of graphs and independent runs of a given algorithm, whether classical or quantum.

Moreover, in order to take into account the fact that one may have the time to run an algorithm many times within a
predefined time budget, either serially or in parallel, and is then likely to select the \emph{best outcome} as
output, we also look at what we call the \emph{``maximum approximation ratio''} or
\emph{``best-of-$n_\mathrm{shots}$ approximation ratio''}.

Formally, we define it as:
\begin{equation}
    \alpha(n_\mathrm{shots}) \equiv \max_{i\in\{1\dots n_\mathrm{shots}\}}\frac{\mathcal{C}(S_{i},G)}{\mathcal{C}(S^*,G)},
\end{equation}

where $S_{1},\dots,S_{n_\mathrm{shots}}$ are approximate solutions resulting from $n_\mathrm{shots}$ independent runs of a given algorithm $\mathcal{A}$ on the same instance $G$. 
It may be regarded as the plain approximation ratio defined in Eq.~\eqref{eq:apx_ratio_def}, but achieved by a meta-algorithm $\mathcal{A}'$
consisting of $n_\mathrm{shots}$ independent runs of $\mathcal{A}$ and the selection of the best result as output.

As with the basic approximation ratio, we 
then look at that quantity averaged over random instances and/or several independent runs of $\mathcal{A}'$ (each of them in turn consisting in $n_\mathrm{shots}$ independent runs of the algorithm $\mathcal{A}$ under study).

The average \emph{maximum} expectation ratio reflects the typical procedure one would follow with a quantum algorithm
or randomized classical approach (perform many repetitions of the state-preparation/algorithm, either serially or in
parallel, and keep the best solution among all the repetitions).
However, we also keep track of the basic average expectation ratio, which we also find instructive as an assessment of the typical achievable \emph{intrinsic} approximation ratio. Besides, it provides a lower bound for the maximum approximation ratio.

\section{Methods}\label{sec:methods}
\subsection{Quantum approach: Solving UD-MIS using Rydberg atoms\label{sec:qa}}

In this section, we describe how the solution of the UD-MIS problem can be approximated using
a hybrid quantum-classical algorithm whose quantum part makes use of a platform of Rydberg atoms.

\subsubsection{Variational Quantum Simulation approach to the UD-MIS optimization problem}

The UD-MIS problem can be reformulated as a minimization problem that consists in finding the ground-state energy of the following Hamiltonian:

\begin{align}
H_{\mathrm{target}} \equiv-\sum_{i\in V}\hat{n}_{i}+u\sum_{i,j\in E}\hat{n}_{i}\hat{n}_{j}\label{eq:H_target_def}
\end{align}
with $\hat{n}_i = (I - \hat{\sigma}_i^{z})/2$.
The $u$ parameter is a fixed Lagrange multiplier, whose value can be optimized in an outer optimization loop. In this work, we will fix $u$ to a value $u>1$ (namely $u=1.35$), which guarantees that the ground state of $H_\mathrm{target}$ will necessarily be an independent set (IS).
The full derivation of the cost function $H_\mathrm{target}$ is given in Appendix~\ref{sec:udmis_cost_function_def}.

To solve (at least approximately) the problem of finding the ground state of $H_\mathrm{target}$, we follow the Variational Quantum Simulation (VQS) framework, as introduced by \cite{Kokail2018}.
An analog version of the Variational Quantum Eigensolver (VQE, \cite{Peruzzo2014}), VQS consists in minimizing the expectation value of $H_\mathrm{target}$ over a family of variational states $|\Psi(\vec{\theta})\rangle$ constructed using an analog quantum computer:
\begin{equation}
    E_\mathrm{target}(\vec{\theta}) \equiv \langle \Psi(\vec{\theta}) | H_\mathrm{target} | \Psi(\vec{\theta}) \rangle \label{eq:E_target_def}
\end{equation}
The ansatz states are prepared by a (generically time-dependent) "resource Hamiltonian" $H_\mathrm{resource}(\vec{\theta})$ whose parameters $\vec{\theta}$ are optimized with a classical computer to minimize $E_\mathrm{target}(\vec{\theta})$.
Typically, possible variational parameters are the coefficients of the various terms of the resource Hamiltonian (if they are tunable), the duration of application of each term, etc.
In a nutshell, VQS consists in designing a Hamiltonian schedule $H_\mathrm{resource}(\vec{\theta})$ to prepare a final state $|\Psi(\vec{\theta})\rangle$ that is as close as possible to the ground state of $H_\mathrm{target}$.

\subsubsection{Resource Hamiltonian for a Rydberg platform}
We now turn to the available resource terms for $H_\mathrm{resource}(\vec{\theta})$ when working with a Rydberg platform.
One of the most common Rydberg setups implements an Ising Hamiltonian \cite{Browaeys2020}:
\begin{equation}
H_{\mathrm{resource}}^{\mathrm{Rydberg}}(t)=\frac{\omega(t)}{2}\sum_{i\in V}\hat{\sigma}_{i}^{x}-\delta(t)\sum_{i\in V}\hat{n}_{i}+\sum_{i,j\in V^{2}}\frac{V}{r_{ij}^{6}}\hat{n}_{i}\hat{n}_{j},\label{eq:H_rydberg_def}
\end{equation}
with three "resource terms": a Rabi term, a dephasing term and a van der Waals interaction term.
While the third term is time-independent (in the following, we will take $V/h = 2.7\,\mathrm{MHz}$, following \cite{lienhard2018observing}), the first two terms are time-dependent and tunable, and as such $\omega(t)$ and $\delta(t)$ can be regarded as variational parameters $\vec{\theta}$ (after a suitable discrete parametrization). 
Several strategies are then possible to design schedules over $H_{\mathrm{resource}}$ that prepare a (possibly approximate) ground state of $H_\mathrm{target}$.
In this work, we will use the quantum annealing schedules described in \cite{lienhard2018observing} and \cite{pichler2018quantum} to build $H_{\mathrm{resource}}$ from Eq.~(\ref{eq:H_rydberg_def}).

\subsubsection{Quantum annealing-inspired schedule construction}
The goal of the quantum annealing (QA) algorithm is to prepare 
the ground state of a given $H_{\mathrm{target}}$. 
Starting from a $H_{\mathrm{target}}$, QA prescribes a generic form
for the time-dependent Hamiltonian that is meant to prepare the target
ground state.
This Hamiltonian, which, in reference to the VQS notation, we will also refer to as a resource Hamiltonian, usually assumes the following form
\begin{equation}
H_{\mathrm{resource}}^{\mathrm{annealing}}(t)\equiv\Omega(t)\sum_{i}\hat{\sigma}_{i}^{x}+\tilde{\Omega}(t)H_{\mathrm{target}}\label{eq:annealing_Hamiltonian}
\end{equation}
with the initial and final conditions $\Omega(t=0)=1,\tilde{\Omega}(t=0)=0$,  $\Omega(t=t_{\mathrm{f}})=0,\tilde{\Omega}(t=t_{\mathrm{f}})=1$,
and the assumption that the initial
state is $|\Psi(t=0)\rangle=|+\rangle^{\otimes n}$ 
(i.e, the ground state of $H_{\mathrm{resource}}^{\mathrm{annealing}}(t=0)$). We note that $|+\rangle$ is defined as $|+\rangle = (|0\rangle + |1\rangle)/\sqrt{2}$, where $|0\rangle$ (resp. $|1\rangle$) denotes the ground (resp. excited, i.e Rydberg) state of the atom.

A main advantage of QA is that the so-called adiabatic theorem (see, e.g., \cite{albash2018adiabatic}) guarantees that this temporal evolution
prepares the (exact) ground state $|\Psi^{0}\rangle$
of $H_\mathrm{target}$ (provided the annealing time is long enough).

A drawback of QA is that the prescribed form for $H_\mathrm{resource}^\mathrm{annealing}$ may not match the resource Hamiltonian that is available on a given platform.
In our case, indeed, the three resource terms available on the Rydberg platform cannot straightforwardly implement $H_\mathrm{resource}^\mathrm{annealing}$. 
The most consequential difference is the interaction term: while QA would dictate a hard-sphere interaction between neighboring atoms ($u\sum_{i,j\in E}\hat{n}_{i}\hat{n}_{j}$ term in \eqref{eq:H_target_def}), Rydberg atoms interact with a $1/r^6$ potential. Moreover, the value of this potential cannot be turned on and off at will, as would be required by a time-dependent $\tilde{\Omega}(t)$ drive fulfilling the initial and final conditions.

Several advanced strategies could be investigated to make the Rydberg resource Hamiltonian better match the resource Hamiltonian prescribed by QA.
For instance, a stroboscopic scheme was proposed in \cite{pichler2018quantum}. One could also optimize, under the constraint of keeping the graph identical, the \emph{locations} of the atoms themselves, to minimize the effect of long-range tails.
While those strategies could likely lead to a better resource Hamiltonian (insofar as it would be closer to the QA form, and thus be endowed with mathematical guarantees via the adiabatic theorem), they have not yet been implemented experimentally and therefore warrant a detailed separate analysis that is outside the scope of this paper. 

Here, instead, we deliberately stick with already experimentally implemented terms \cite{barredo2016atom} in an effort to quantify what can be achieved \emph{with today's technology}. We use the full experimental Rydberg Hamiltonian as our resource to produce states, and evaluate results through the ideal target Hamiltonian from Eq.~(\ref{eq:H_target_def}).
We merely take inspiration from the QA resource Hamiltonian to design a resource Hamiltonian that prepares a reasonable approximation to the target ground state while complying with the constraints of the Rydberg architecture:

\paragraph{Initialization:}

One cannot start from the superposition state $|\Psi(t=0)\rangle=|+\rangle^{\otimes n}$
, but only from the state $|\Psi(t=0)\rangle=|0\rangle^{\otimes n}$
where no atoms are in the excited (Rydberg) state.

\paragraph{Schedule:}

\begin{figure}
\begin{centering}
\includegraphics[width=1\columnwidth]{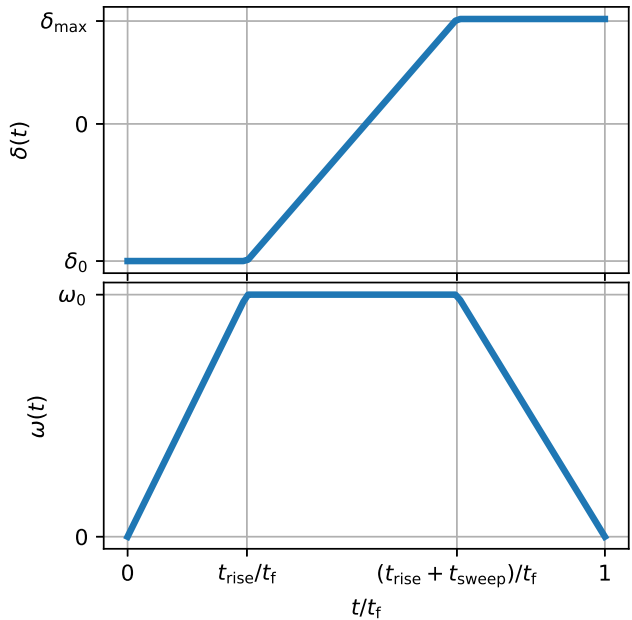}
\par\end{centering}
\caption{Time dependence of the coefficients of the dephasing (top) and Rabi (bottom) terms, see Eq.~\eqref{eq:H_rydberg_def}.\label{fig:schedules}}
\end{figure}

As already mentioned, the van der Waals interaction of Eq.~(\ref{eq:H_rydberg_def}) differs
from the hard-sphere interaction $\sum_{i,j\in E}\hat{n}_{i}\hat{n}_{j}$ required by our problem,
Eq.~(\ref{eq:annealing_Hamiltonian}). 
Moreover, we cannot turn it
off at will as the initial conditions on $\tilde{\Omega}(t)$ require.
As a partial accommodation for these differences,
we design the
time evolution of $H_{\mathrm{resource}}^{\mathrm{Rydberg}}$ as follows,
with three stages (illustrated in Fig.~\ref{fig:schedules}):

(i) for $0<t<t_{\mathrm{rise}}$, we are going to increase $\omega$
from 0 to a maximum value $\omega_{0}$. The goal of this stage is
to bring the system from state $|0\rangle^{\otimes n}$ to state $|+\rangle^{\otimes n}$ (indeed, the ground state at $\omega=0$ (resp. $\omega=\omega_0$) is $|0\rangle^{\otimes n}$ (resp. $|+\rangle^{\otimes n}$));

(ii) for $t_{\mathrm{rise}}<t<t_{\mathrm{rise}}+t_{\mathrm{sweep}}$,
we are going to increase the value of $\delta$ from $\delta_{0}$
to $\delta_{\mathrm{max}}$ in order to favor sets with a large number of atoms, while keeping
the tunneling (Rabi) term on;

(iii) for $t_{\mathrm{rise}}+t_{\mathrm{sweep}}<t<t_{\mathrm{f}}$,
we are going to decrease $\omega$ from $\omega_{0}$ to 0. The goal
is for the state of the system at the end of this stage to be the
ground-state $|\Psi_{\mathrm{resource}}^{0,\mathrm{Rydberg}}\rangle$
of
\[
H_{\mathrm{resource}}^{\mathrm{Rydberg}}(t=t_{\mathrm{f}})=-\delta_{\mathrm{max}}\sum_{i\in V}\hat{n}_{i}+\sum_{i,j\in V^{2}}\frac{V}{r_{ij}^{6}}\hat{n}_{i}\hat{n}_{j}
\]

We note that the very same adiabatic process as presented above has already been implemented experimentally, albeit for grids and not general graphs \cite{lienhard2018observing}.
Importantly, even if this schedule were applied on a noiseless system, and slowly enough to guarantee adiabaticity, the differences between the experimental Rydberg Hamiltonian and the ideal UD-MIS Hamiltonian may imply non-optimal output results. Ref.~\cite{pichler2018computational} gives examples of graph instances where the $1/r^{6}$-tail of the Rydberg Hamiltonian does change the nature of the ground state compared to the ideal UD-MIS Hamiltonian. The two Hamiltonians are however arguably close enough to justify a study of the ability of the former to generate \emph{approximately optimal} states for the latter, within a VQS picture, as we do here.

With the above schedule, the number of variational parameters $\vec{\theta}$ is still daunting. We are thus
going to consider the parameters $\omega_{0},\delta_{0},\delta_{\mathrm{max}},t_{\mathrm{rise}}$
and $t_{\mathrm{sweep}}$ as fixed experimental parameters:
\begin{align}
    \omega_0/(2\pi) & = 1.89\,\mathrm{MHz},\\
    \delta_0/(2\pi) & = - 6.0\,\mathrm{MHz},\\
    \delta_\mathrm{max}/(2\pi) & = 4.59\,\mathrm{MHz},\\
    t_\mathrm{rise} & = 0.25\cdot t_\mathrm{f},\\
    t_\mathrm{sweep} & = 0.44 \cdot t_\mathrm{f}.
\end{align}

The only remaining variational parameter
is thus the total
annealing time $t_{\mathrm{f}}$.
In the absence of noise, the adiabatic theorem would make this parameter useless, as the success probability of
QA gets larger as the annealing time $t_{\mathrm{f}}$ gets larger.
In the presence of noise, however, long annealing times lead to an
increased sensitivity to decoherence, and it thus becomes important
to choose a $t_{\mathcal{\mathrm{f}}}$ large enough that one can
find a good approximation to the ground state, but short enough that
the system has not undergone too much decoherence.

We note that this form
of optimization is nothing but a very crude optimal control protocol,
and that much more sophisticated schedules $\Omega(t),\tilde{\Omega}(t)$
could be designed using quantum optimal control (see, e.g, \cite{Omran2019}).

The approximation ratio for our algorithm is defined, in its "one-shot version" (see discussion in Sec~\ref{subsec:methodology}), as
\begin{align}
\alpha & \equiv\frac{E_{\mathrm{target}}(t_\mathrm{f}^*)}{E_{\mathrm{MIS}}(S^{*})}\label{eq:approx_ratio_def}
\end{align}

with $E_{\mathrm{target}}$ defined in Eq.~(\ref{eq:E_target_def}), and  
$E_{\mathrm{MIS}}(S^{*}) =-|S^{*}|$.
We note that because of the definition of $H_\mathrm{target}$, $E_{\mathrm{target}}$ is usually negative.

\subsubsection{Hardware model\label{subsec:noise_model}}

The Hamiltonian given in Eq.~\eqref{eq:H_rydberg_def} describes the system of neutral atoms in the absence of errors. In this work, based on a validation with respect with previously published experimental data, we take into account two sources of errors: dephasing noise and readout errors.

\paragraph{Decoherence by dephasing}
Decoherence is taken into account via a simple dephasing model that gives a reasonably accurate fit of the published experimental data (as checked in Appendix~\ref{subsubsec:noise_model_determination}).
Following Ref.~\cite{lienhard2018observing}, we describe this dephasing with a Lindblad master equation

\begin{equation}
    \frac{\mathrm{d}\rho}{\mathrm{d}t}=-i\left[H,\rho\right]-\frac{1}{2}\sum_{i\in V} \gamma_i \left[\left\{ L_{i}^{\dagger}L_{i},\rho\right\} -2L_{i}\rho L_{i}^{\dagger}\right]
\end{equation}
where $\rho$ is the density matrix describing the mixed state of the system, and $H \equiv H_\mathrm{resource}^\mathrm{Rydberg}(t)$. The jump operators $L_i$ corresponding to dephasing take the following form:
\begin{equation}
    L_i = \hat{n}_i
\end{equation}
We choose a uniform dephasing described by a single dephasing parameter $\gamma_i = \gamma$.

\paragraph{Readout assignment errors}
Following Ref.~\cite{DeLeseleuc2018a}, we model the readout errors by a simple assignment error model that accounts for the probability $\epsilon$ of erroneously detecting an excited atom $|1\rangle$ while it was in fact in its ground state $|0\rangle$ (false positives), as well as the probability $\epsilon'$ of not detecting an excited atom (false negatives). 
This simple error model is characterized by a so-called assignment probability matrix:

\begin{equation}
    A_{1}=\left[\begin{array}{cc}
p(0|0) & p(0|1)\\
p(1|0) & p(1|1)
\end{array}\right],
\end{equation}
with $P(1|0)=\epsilon$ and $P(0|1)=\epsilon'$, 
that modifies single-qubit probabilities as follows
\begin{equation}
    \left[\begin{array}{c}
\tilde{p}(0)\\
\tilde{p}(1)
\end{array}\right]=A_{1}\left[\begin{array}{c}
p(0)\\
p(1)
\end{array}\right]
\end{equation}

We will neglect any leakage out of the computational subspace, and thus assume $P(0|0) = 1 - \epsilon$ and $P(1|1)= 1 -\epsilon'$. 

The probability distribution for all atoms is thus modified as
$\tilde{P} = \lbrace A_1 \otimes A_1 \dots \otimes A_1 \rbrace \cdot P.$

\paragraph{Repetition rate}
The repetition rate in state-of-the-art experiments is in the 3-5 Hz range \cite{ThierryPrivate, LienhardThesis}.
It is dominated by the time it takes to load atoms into the trap and by the final measurement stage.
In particular, the duration of the annealing ($t_\mathrm{f}$) is negligible compared to the duration of the other stages.


\subsection{Classical approach: A locality-based heuristic}\label{sec:locality_based_method}

\begin{figure}
\includegraphics[width=1\columnwidth]{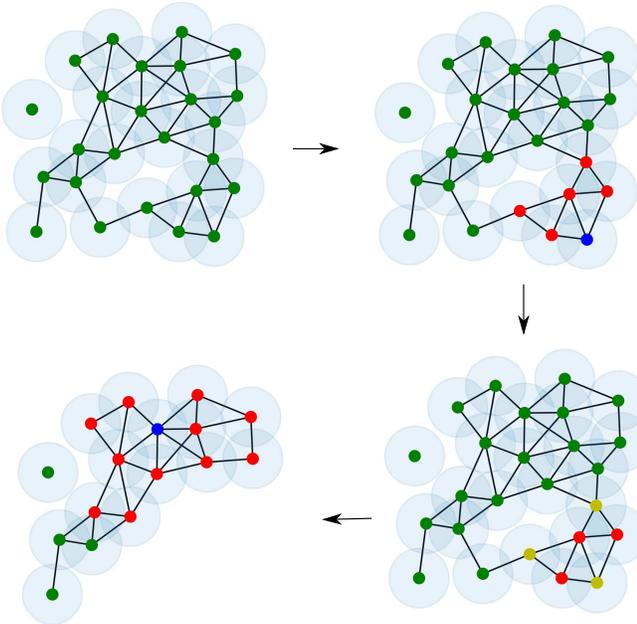}
\caption{Illustration of the execution of lines 4 to 10 of Algorithm~\ref{alg:ptas}, for $d$=2, on a graph with 20 vertices (top left).
\emph{Top right:} a random vertex $u$(in blue) is selected, and the sphere $S_{d}(u)$ of vertices within distance $d$  (red) is computed.
\emph{Bottom right:} An optimal MIS is computed for $S_{d}(u)$ (yellow vertices). All vertices from $S_{d}(u)$, along with all vertices not in $S_{d}(u)$ but connected to a vertex in the MIS of $S_{d}(u)$ are then removed.
\emph{Bottom left:} a new random vertex has been selected for the next iteration of the algorithm.
\label{fig:apx_explain}}
\end{figure}

We now switch, in this section, to a description of a simple locality-based UD-MIS approximation heuristic for UD-MIS, inspired from \cite{nieberg2004robust}. We use it to estimate the classical boundary, given
a fixed time budget, of the UD-MIS approximation.
Algorithm \ref{alg:ptas} and Fig.~\ref{fig:apx_explain} describe its operation.

The design of this heuristic was motivated by the relative complexity of existing Polynomial-Time Approximation Schemes \cite{da2017shifting,das2018efficient,nandy2017faster,nieberg2004robust}. While they all consist in splitting a graph into sub-instances that are solved independently, several decompositions generally have to be considered, before making a choice guaranteeing a lower bound on the approximation ratio. We chose to implement a simpler, intuitive approach that only requires, during one run, to process each vertex exactly \emph{once}. 

We obtain this simplification at the price of renouncing to a provable guarantee on the approximation ratio. However, since we are able to compute numerically these approximation ratios (by solving instances optimally to obtain the optimal maximum independent set size) up to relatively large sizes (500-550 vertices), this is a relatively minor drawback. 

Another motivation is that the local nature of solving subproblems
in our method is reminiscent of the finite \emph{quantum correlation distances} that have been estimated experimentally in Rydberg systems \cite{lienhard2018observing}, and numerically in our simulations (see Appendix~\ref{sec:corr_length_analysis}). It makes it particularly relevant to a comparison with Rydberg adiabatic quantum computing systems.

Our algorithm (Algorithm~\ref{alg:ptas}) consists in an overall \emph{while} loop that continues until the graph is empty. In an iteration of the while loop, UD-MIS is \emph{locally} solved around a randomly selected vertex. We refer to this set of randomly selected vertices as "seeds". 
For each seed $u$, we compute (using Breadth-First Search (BFS, \cite{cormen2009introduction})) the set of 
vertices within a distance $d$ of $u$, that we call $S_{d}(u)$. $d$ is an integer that is given as an execution parameter to the algorithm (by distance we mean \emph{shortest-path distance} within the graph and not \emph{geometric distance} between points). UD-MIS is then
solved optimally for the subgraph induced by $S_{d}(u)$, which is removed from the graph.
This process is repeated until the graph is empty.
Figure~\ref{fig:apx_explain} illustrates an iteration of the algorithm on a specific seed.
The final solution returned by the algorithm is then the union of all the local maximum independent sets obtained
throughout the execution by locally solving UD-MIS around the seeds.
To ensure that the returned solution is indeed an independent set, in addition to removing $S_{d}(u)$ from the set of vertices after
its processing, we further remove any vertex \emph{connected} to the computed Maximum Independent Set of $S_{d}(u)$. Such vertices are indeed constrained to not be part of the final solution. In Figure~\ref{fig:apx_explain},
one can see an instance where vertices not belonging to $S_{d}(u)$ are nonetheless removed from the graph, because they are connected
to the Maximum Independent Set of $S_{d}(u)$.

Note that the iterations of the algorithm cannot be easily parallelized as
the results from solving $S_{d}(u)$ condition which vertices are removed, and therefore the rest of the execution of the algorithm. 

Contrary to \cite{nieberg2004robust} or \cite{das2018efficient}, we do not formally call this simple locality-based randomized approach an ``approximation algorithm'', as we do not provide a \emph{provable} guarantee, regardless of the graph, on the achieved approximation ratio given a certain value of $d$. However, in practice, it qualitatively behaves like a PTAS, as one can see on Fig.~\ref{fig:Approximation-ratio-vs-natoms-vs-runtime} (right), in the sense that the average approximation ratio achieved by this locality-based heuristic gets closer to 1 as $d$ increases. 

Much like the parameters used in \cite{nieberg2004robust} and \cite{das2018efficient}, $d$ therefore acts as a tunable parameter that can be used to refine the quality of an approximation at the expense of a potentially longer computation time.

Convergence to an approximation ratio of $1$ for any graph as $d$ increases is guaranteed by the fact that if $d$ exceeds the \emph{diameter} (the length of the longest shortest path between any pair of vertices) of the graph, then no matter the choice of seed
$u$, $S_{d}(u)$ will cover the graph entirely, and a global optimal Maximum Independent Set will be computed.

\begin{algorithm}
\caption{Simple randomized approximation heuristic for UD-MIS}\label{alg:ptas}
\begin{flushleft}
\textbf{Input:} a unit-disk graph $G=(V,E)$. 

\textbf{Output:} a bit assignment for every vertex ($\{b_{v}\}_{v\in V}$), coding for an independent set.

\textbf{Execution parameter:} An integer $d\geq0$. 
\end{flushleft}
\begin{algorithmic}[1]
\State $B \leftarrow \emptyset$ \Comment{$B$ will contain the ``border'' between exactly-solved patches}
\State $W = V$
\State 
\While{$W\neq \emptyset$}{
\State Pick a vertex $u\in W$ randomly.
\State Compute $S_{d}(u)$, the set of vertices within distance $d$ of $u$ in $G\left[W\right]$ using e.g. Breadth-First Search.
\State Compute bit assignment = solve MIS exactly for $G[S_{d}(u)]$.
\State let $C_{u,d}= \left\{v\in \left[W\setminus S_{d}(u)\right]\text{ s.t } b_{v}\text{ is \emph{constrained} to be 0}  \right\}$
\State $B \leftarrow B\cup C_{u,d} $
\State $W \leftarrow W\setminus S_{d}(u)$
\State $W \leftarrow W\setminus C_{u,d}$
}
\end{algorithmic}
\end{algorithm}

A particular but valid parametrization for Algorithm~\ref{alg:ptas} consists in choosing $d=0$. It appears for instance on Figure~\ref{fig:Approximation-ratio-vs-natoms-vs-runtime} or Figure~\ref{fig:allowed_number_shots_loc}. 
By definition, a BFS sphere $S_{d}(u)$ with $d=0$ simply corresponds to $\{u\}$.
The ``MIS'' of a graph with one vertex $u$ is simply $\{u\}$ and the set $C_{u,d}$ (see line 7 of Algorithm~\ref{alg:ptas}) is simply the neighborhood of $u$ in the graph.
Running the heuristic with $d=0$ therefore corresponds to randomly selecting available vertices and putting them into the returned solution, without really \emph{computing} anything.

\section{Results\label{sec:overview}}

In this section, we summarize our main findings, namely quantum and classical approximability results, and
then explain the practical implications of this comparison.

\begin{figure*}
\begin{centering}
\includegraphics[width=1\textwidth]{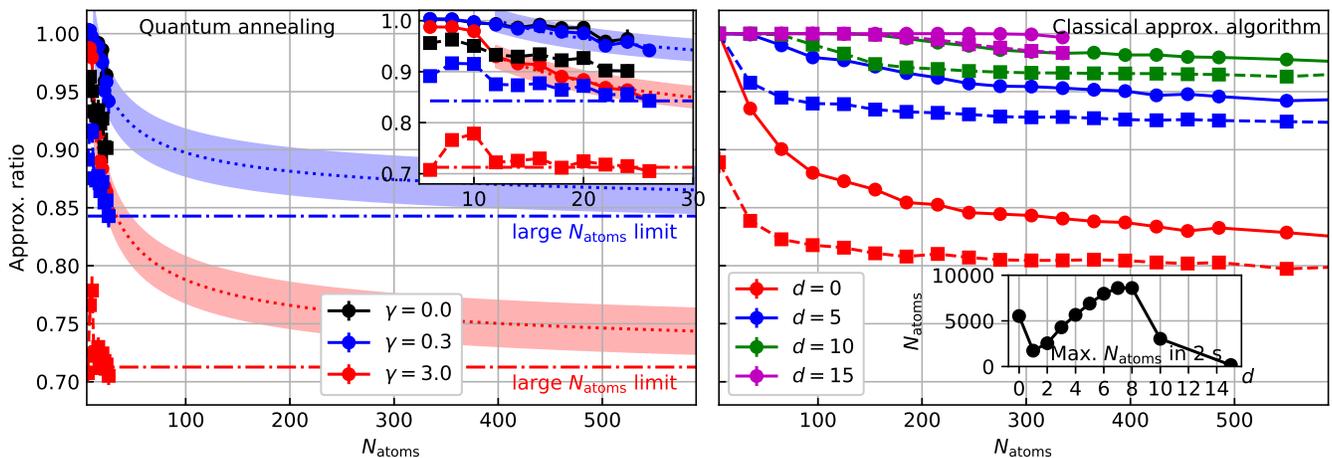}
\par\end{centering}
\caption{
Approximation ratio as a function of the problem size.
\emph{Solid lines and circles}: average maximum approximation ratio over a fixed 2-second run time ($=$10 shots in the quantum case). 
\emph{Dashed lines and squares}: average (single-shot) approximation ratio.
\emph{Left:} Quantum annealing with Rydberg atoms for different noise levels ($\gamma = 3.0$ [red]: state-of-the-art noise level, $\gamma = 0.3$ [blue]: near-term noise level, $\gamma=0.0$ [black]: noiseless case). 
\emph{Dotted lines}: extrapolation using data for 12-26 atoms (see inset), see Eq.~(\ref{eq:fit_nshots_dep}), with 95 \% prediction interval.
\emph{Dash-dotted lines}: constant extrapolation for one-shot case, based on the last ($\gamma = 0.3$) and last three ($\gamma = 3.0$) points. 
Optimized annealing time $t_\mathrm{f}$ (see text). 
\emph{Right:} average maximum approximation ratios and average approximation ratios achieved by classical locality-based approach for maximum resolution distance $d$ of 0 [red], 5 [blue], 10 [green], and 15 [magenta]. Results were obtained by averaging over $100$ random uni-disk graphs.
\emph{Inset:} 
Maximum graph size (atom number) reachable by classical algorithm within 2 seconds as a function of $d$.
\label{fig:Approximation-ratio-vs-natoms-vs-runtime}
}
\end{figure*}

The main results are shown in Fig.~\ref{fig:Approximation-ratio-vs-natoms-vs-runtime}, where we compare the approximation ratios as defined above as a function of the problem size for various external parameters, both in the quantum and the classical approach.

\subsection{Results of the quantum approach}

The quantum results are shown in the left panel of Fig.~\ref{fig:Approximation-ratio-vs-natoms-vs-runtime}.
We show the average approximation ratios obtained in three hardware scenarios for Rydberg platforms: (i) \emph{state-of-the-art hardware}, corresponding to already published experimental results~\cite{lienhard2018observing}, with noise levels captured by one parameter, $\gamma = 3.0$; (ii) \emph{near-term hardware}, corresponding to current noise levels on the most advanced platforms, with noise levels divided by a factor of 10 with respect to the first scenario \cite{ThierryPrivate}, $\gamma=0.3$; (iii) \emph{noiseless hardware}, with a vanishing noise parameter $\gamma=0.0$. The definition of, and justification for these models is given in section~\ref{subsec:noise_model} above, and in Appendix~\ref{sec:supporting_data_quantum} below.

We show the average results obtained with a time budget of 0.2 seconds and 2 seconds.
With a realistic experimental repetition rate of about 5 Hz \cite{ThierryPrivate,LienhardThesis} (comprising initialization, evolution and readout), these two cases correspond to a single repetition (measurement) of the algorithm, and ten repetitions, respectively.
They thus correspond to the two possible definitions for the approximation ratio that we discussed in section \ref{subsec:methodology}.

As expected, decreasing noise levels yield improved approximation ratios, as do longer computational time budgets.
For small system sizes (here number of neutral atoms), the approximation ratio  decreases as the number of atoms increases.
The behavior then differs for different run times:

\emph{For one repetition} (0.2 s) the approximation ratio (corresponding to the statistical expectation value of the approximation over the quantum state distribution) saturates for the largest values of noise ($\gamma=3.0$, and to a lesser extent $\gamma=0.3$), while it slowly decreases for smaller noise parameters ($\gamma=0.0$).

\emph{For several (10) repetitions}, the approximation ratio does not seem to saturate but instead to slowly decrease with the number of atoms.

Both behaviors---saturation for one shot and slow decrease for several shots---can be explained, and we will use this explanation to extrapolate the curves to much larger number of atoms.

The saturation one observes for the average one-shot approximation ratio and the larger noise levels can be traced back to the finite correlation lengths induced by an imperfect preparation of the sought-after MIS solution.

By analyzing the spin-spin correlation function, we compute rough estimates of the spin correlation length $\xi$ for various system sizes and noise levels (see details in Appendix~\ref{subsubsec:corrlength}), and conclude that the correlation length is set primarily by the noise (decoherence) level, and is roughly independent of the system size (see Fig.~\ref{fig:xi_vs_natoms}). (The value we find for $\gamma = 3.0$, $\xi \approx 1.4$, is very similar to the correlation length extracted in previous experiments, albeit on a regular lattice \cite{lienhard2018observing}.)
This allows us to compute the approximate system size $N^*$ above which finite size effects should become irrelevant in the numerical simulation.
Such a size can be estimated by counting
the number of atoms comprised in a $\xi \times \xi$ square. For the atom density we consider, $\nu = 2$, we find $N^* \approx 4$ for $\gamma=3.0$ and $N^* \approx 30$ for $\gamma = 0.3$ (for $\gamma=0$, we find much larger, although finite estimates, of > 1000: our resource Hamiltonian prepares only an approximate ground state of $H_\mathrm{target}$, as expected).
We argue that the average approximation ratio, which is a function of the average (local) spin value $\langle z_i \rangle $ (see Eq.~\eqref{eq:E_target_def}), should not depend on $N$ beyond $N^*$. 
(It also depends, a priori, on $\langle z_i z_j \rangle$, but this correlation can be regarded as weak because the dynamics are dominated, by design, by independent-set configurations, see Appendix~\ref{subsubsec:IS_restriction}.)

We therefore extrapolate the value computed numerically for $N=N^*$ to larger values, provided we can reach a size of $N^*$ atoms with our numerical simulations. 
By reaching 26 atoms, we achieve this for $N^*(\gamma=3.0)$($\approx 4$) and $N^*(\gamma = 0.3)$($\approx 30$). 
Since we can even go beyond $N^*$ (especially so for $\gamma = 3.0$), 
we can not only perform the extrapolation, but also check that the approximation ratio has already reached saturation
for number of atoms close to $N^*$.
The extrapolated values are shown as the dash-dotted lines in Fig.~\ref{fig:Approximation-ratio-vs-natoms-vs-runtime}.
We can regard these values as a lower bound for the quantum approximation ratio.

The shape of the slow decrease in the maximum approximation ratio after several runs 
(Fig.~\ref{fig:Approximation-ratio-vs-natoms-vs-runtime}, left side, solid lines and dotted extrapolation)
can be rationalized by statistical arguments.
In the simpler case of a uniform final state distribution and mild assumptions detailed in
Appendix~\ref{sec:exp_of_max}, one can show that the maximum approximation ratio after $n_\mathrm{shots}$ repetitions
on a system of $N_\mathrm{atoms}$ atoms is upper bounded by
$\alpha + \sqrt{\frac{\log{n_\mathrm{shots}}}{2 N_\mathrm{atoms}}}$, with $\alpha$ the one-shot expectation value ($\alpha=1/2$ for a uniform state distribution).
It turns out that the actual approximation ratio for a uniform state distribution obeys the same $\propto \sqrt{\frac{\log{n_\mathrm{shots}}}{2 N_\mathrm{atoms}}}$ dependence as its upper bound. 

Interestingly, at fixed number of shots, for the uniform distribution, the maximum approximation ratio decreases quite slowly ($\propto 1/\sqrt{N_\mathrm{atoms}}$) with the number of atoms. Conversely, at a fixed number of atoms, the increase in approximation ratio is very slow ($\propto\sqrt{\log{n_\mathrm{shots}}}$), and even more so with larger number of atoms. This has the practical implication that a very large number of repetitions (shots) may not be the most efficient way to improve the quality of the algorithm.
In particular, the $\sqrt{\frac{\log{n_\mathrm{shots}}}{2 N_\mathrm{atoms}}}$ scaling of the approximation ratio
gain limits the potential added value of running several quantum machines in parallel within the same time budget, and selecting the best result.
For $\sim 1000$ atoms, using our simplified model, running a hundred machines in parallel would
only amount to a gain of about $\sqrt{\frac{\log 100}{2\cdot 1000}}\simeq 0.047$ in terms of approximation ratio.

A mathematically rigorous extension of this observation to the generically non uniform state distribution produced by the quantum algorithm is beyond the scope of this work.
Nevertheless, not only is the approximation ratio obtained for a uniform distribution a lower bound to the approximation ratio we obtain via quantum annealing, but the $\sqrt{\frac{\log{n_\mathrm{shots}}}{2 N_\mathrm{atoms}}}$ dependence is arguably a purely statistical property; it is therefore not a feature of the sole uniform distribution. 

We thus assume that the $(n_\mathrm{shots}, N_\mathrm{atoms})$ dependence of the maximum approximation ratio obtained numerically can be fitted and extrapolated by the following function:
\begin{equation}
    \alpha(N_\mathrm{atoms}, n_\mathrm{shots}, \gamma) = \alpha(\gamma) + \beta(\gamma)  \sqrt{\frac{\log{n_\mathrm{shots}}}{2 N_\mathrm{atoms}}},
    \label{eq:fit_nshots_dep}
\end{equation}
with $\alpha(\gamma)$ being given by the saturation value computed for the one-shot case using the $N^*$ extrapolation explained in the previous paragraphs.

We then use the numerical data we obtained for $N_\mathrm{atoms} = 6$ to 26 to find the value of the remaining parameter, $\beta(\gamma)$, using a least-squares minimization.
With this parameter, we can extrapolate the approximation ratio after several shots to larger numbers of atoms (see the dotted lines in Fig.~\ref{fig:Approximation-ratio-vs-natoms-vs-runtime}, left panel), and compute the corresponding 95\% prediction interval.

\subsection{Results of the classical approach}

\subsubsection{Identification of an efficient locality-based heuristic}

In order to provide a meaningful classical reference benchmark,
we reviewed several approximation algorithms \cite{das2018efficient,nieberg2004robust,da2017shifting,das2015approximation,nandy2017faster,van2005approximation} for UD-MIS, and more specifically
Polynomial-Time Approximation Schemes (PTAS, see Section~\ref{subsec:algo_approaches} above).

We experimented in particular with a simplified version of \cite{nieberg2004robust}, namely
a locality-based approach that solves the MIS problems \emph{exactly} within a certain distance 
$d$ around randomly chosen points.
Whereas the algorithm presented in \cite{nieberg2004robust} chooses this distance adaptively to ensure a specified approximation ratio,
our
simplified 
method takes a specific value
as input and uses it everywhere (see Section~\ref{sec:locality_based_method} for more details)

We see this classical locality-based heuristic as loosely
``quantum-similar'', the finite solving distance $d$
playing a similar role to that of the quantum correlation length $\xi$, as characterized experimentally in Rydberg systems in, e.g, \cite{lienhard2018observing}, and estimated numerically from our simulations (see Appendix~\ref{sec:corr_length_analysis}).

\subsubsection{Classically attainable approximation ratios}

The approximation ratio achieved by our locality-based approach, as a function of the number of vertices, and for several values
of the fixed ``correlation distance'' $d$, is plotted on the right panel of Fig.~\ref{fig:Approximation-ratio-vs-natoms-vs-runtime}. 
Curves stop when even one run of the algorithm
exceeds the time limit.
Note that, for the range of atoms considered in Fig.~\ref{fig:Approximation-ratio-vs-natoms-vs-runtime}, this only happens with $d=15$.

For other values of $d$, the approach may run within the specified time budget well up to a several thousand vertices, as one can see on the lower right inset of Fig.~\ref{fig:Approximation-ratio-vs-natoms-vs-runtime}, 
showing the maximum attainable graph size as a function of $d$. In all cases, the approximation ratio appears to stabilize to an asymptotic value for large number of vertices. These asymptotic values are therefore accessible up to very large graph sizes ($\gtrsim 5000$). 

The evolution of the approximation ratio's asymptotic value with respect to $d$ is rather intuitive: as the solving distance $d$ increases, the quality of approximation to the optimal improves. This makes sense within the analogy to the quantum correlation distance that we drew above: as vertices are allowed to ``interact'' across larger distances, global behavior improves. 

On all curves (whether corresponding to the average approximation ratio or average maximum approximation ratio, see Section~\ref{sec:overview} for details on these definitions), the approximation seems to start close to $\sim 1$ for small sizes.
This makes sense, as a small graph is quite likely to be covered entirely by a BFS sphere $S_{d}(u)$, no matter the random choice
of the initial vertex $u$.

Note that, in order to compute the approximation ratio achieved by an approach on a given graph, one needs to determine the size of a maximum independent set for this graph, i.e. to solve the problem optimally.
This was carried out using the algorithm presented in \cite{li2017minimization} for graphs containing up to $550$ vertices to compute the results of the right panel of Fig.~\ref{fig:Approximation-ratio-vs-natoms-vs-runtime}. 
For our class of random graphs (see Section \ref{sec:random_graph_generation} for details), instances with $\leq 200$ vertices are optimally solved in $\leq 0.05s$, while the ``exponential explosion'' of the solver happens around $\sim 300-400$ vertices (See Figure~\ref{fig:runtime_opt} for more precise benchmark results).

\subsubsection{Execution time}

The largest attainable graph size given a value of $d$ within the specified time-budget displays a much more counter-intuitive behavior, namely the bell-shaped curve we see on the lower-right inset of Figure~\ref{fig:Approximation-ratio-vs-natoms-vs-runtime}, and in Figure~\ref{fig:breakeven}. This behavior comes from the fact that the execution time of the locality-based heuristic is determined by a subtle interplay between the number of sub-instances to solve and their sizes, and does not necessarily increase with $d$ (See
Appendix~\ref{sec:supp_classical}, in particular Figs.~\ref{fig:rt_benchmark} and 
\ref{fig:supp_rt_loc} for more details).

This explains why, for instance, the heuristic may be run with $d=5$ up to larger graph sizes
than $d=2$ within the prescribed time-budget ($7000$ compared to $3000$), even though $d=5$ 
involves solving larger sub-instances and achieves better approximation ratios.

Asymptotic values of the approximation ratio, plotted against the highest number of vertices reachable within two seconds, are reported on Figure~\ref{fig:breakeven}. These values can be found on Figure~\ref{fig:Approximation-ratio-vs-natoms-vs-runtime} for $d=0,5,10,15$ (and Figure~\ref{fig:supp_apx} for other values of $d$).

A striking fact concerning our locality-based approximation heuristic is therefore that one can run it with $d=8$ for up to $\sim 8000$ atoms (see Figure~\ref{fig:allowed_number_shots_loc}) while staying within our time-budget of 2 seconds, and achieving an asymptotic approximation ratio value of $\sim 0.95$.

Note that our benchmarks were carried out with a single-core implementation of our locality-based
heuristic. If the heuristic itself is not parallelizable (the state of the graph at some point
of the execution depends on the result of previous iterations, because of line 7 of Algorithm~\ref{alg:ptas}), one could of course launch several runs of the algorithms in parallel.

With a serial implementation, the value of the ``maximum average approximation ratio'' coincides
with the basic average approximation ratio at the time limit (as only one run can be carried out, see Fig. \ref{fig:Approximation-ratio-vs-natoms-vs-runtime}). Launching several independent runs in parallel could allow keep on achieving approximation ratios higher than the average baseline at the time limit.

However, based on the statistical arguments developed in the previous subsection, and detailed
in Appendix \ref{sec:exp_of_max}, it is reasonable to believe that the gain on approximation ratio
will scale as $\sqrt{\frac{\log{n_\mathrm{shots}}}{2\cdot N_\mathrm{atoms}}}$, and therefore be of limited magnitude.

In addition, importantly, parallelism does not affect the maximum attainable graph sizes within a given time budget. As the heuristic itself is not parallelizable, several independent runs will still
need the incompressible time of one run (i.e the values reported on Figs \ref{fig:allowed_number_shots_loc}
and \ref{fig:supp_rt_loc}) to execute.

As the run time of the heuristic is dominated by the optimal solving of subinstances (see above, Fig. \ref{fig:rt_benchmark} and Appendix \ref{sec:supp_classical}), and as we neglect the ``input-output'' overhead of the solver we use (\cite{li2017minimization}), it is reasonable to say that improvements
upon the classical benchmark results we report here may only either come from pure algorithmic improvements on \cite{li2017minimization}, or from running the algorithm on more modern CPU chips.

\subsection{Discussion of the break-even point}

\begin{figure}
\begin{centering}
\includegraphics[width=1\columnwidth]{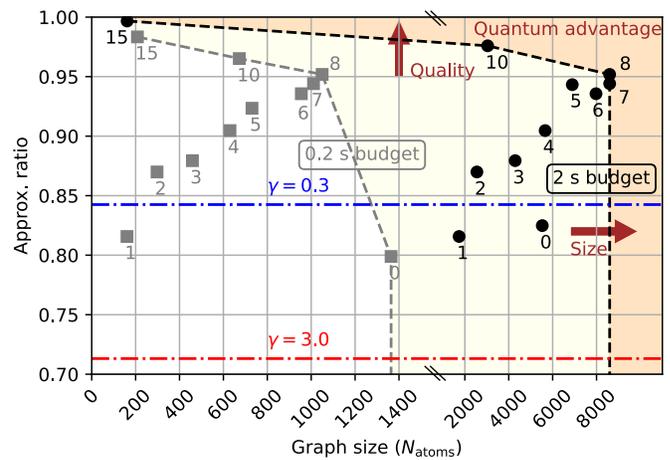}
\par\end{centering}
\caption{Break-even diagram.
Black dots (resp. grey squares): average classical approximation ratio reached for the maximum graph size reachable in 2 seconds (resp. 0.2 seconds) (asymptotic approximation ratio extrapolation from Figure~\ref{fig:Approximation-ratio-vs-natoms-vs-runtime}~and~\ref{fig:supp_apx}, and maximum reachable graph size estimation from Figure~\ref{fig:Approximation-ratio-vs-natoms-vs-runtime} and~\ref{fig:supp_rt_loc}).
Dash-dotted lines: lower bound to quantum approximation ratio reached for decoherence levels of $\gamma = 3.0$ (red) and $\gamma=0.3$ (blue), assuming constant repetition rate with system size. 
Black (resp. grey) dashed line: break-even line for a time budget of 2 seconds (resp. 0.2 seconds). 
\label{fig:breakeven}}
\end{figure}

The comparison of the quantum and classical approaches we just described allows us to discuss the conditions that must be satisfied for the quantum approach to become more competitive than the classical approach.
The approximation ratio as a raw metric must be put in perspective with the time-to-solution and the system size, as we stressed in the beginning of this section.

The classical frontier can readily be mapped for the classical approximation heuristic we have investigated. It is shown as the black (resp. grey) dashed line of Fig.~\ref{fig:breakeven}, that we constructed by computing the maximum size and approximation ratio that could be reached within a 2-second (resp. 0.2-second) time budget.
The orange (resp. yellow) region above and right of this black (resp. grey) dashed line is the parameter area that the quantum algorithm should reach to become advantageous compared to the classical one.

We superimposed, in Fig.~\ref{fig:breakeven}, the extrapolated average approximation ratios obtained through the quantum approach (i.e, corresponding to one repetition, or a budget of 0.2 seconds; we did not show the average maximum approximation ratio that could be reached within 2 seconds, which would be slightly, yet not substantially, higher).

In extrapolating these two lines to very large atom numbers, we made the important assumption that the experimental repetition rate will remain constant up to these large numbers of atoms. 
While a naive individual rearrangement of atoms would lead to a repetition rate roughly scaling with the inverse atom count, more advanced techniques could lead to a mitigation of this dependence (see, e.g, \cite[Supp. Mat. 3.3]{Endres2016}, for an example of a more collective rearrangement, with a $O(1/\sqrt{N_\mathrm{atoms}})$ scaling of the repetition rate). Further improvements at the level of image acquisition are also possible, so that increases of the repetition rate by one order of magnitude are claimed to be realistic \cite{Henriet2020a}.

Working at a fixed time budget, there are two non-exclusive ways a quantum algorithm can outperform a classical algorithm: either it reaches larger system sizes (with possibly the same approximation ratio), or it reaches higher approximation ratios.
There are thus two possible \emph{hardware-related} conditions for quantum advantage:
(i) an increase in the quantum system size to several thousands of atoms ($>8000$ for a 2-second time budget, according to our estimates); (ii) an improvement of the coherence properties of the quantum system (to be able to reach approximation ratios above $\gtrsim 0.95$). 
We note that meeting both conditions would not be necessary to demonstrate practical quantum advantage, although it definitely appears as a desirable goal.

We emphasize that higher quantum approximation ratios could also be reached by \emph{purely algorithmic improvements}, i.e by going beyond the simple quantum annealing-based approach we adopted in this study. In particular, adapting digital approaches like QAOA to an analog setting (via, e.g, the VQS approach \cite{Kokail2018}) seems a promising route, given the enhanced success probabilities of QAOA compared to quantum annealing in a noiseless setting \cite{pichler2018quantum}.

These quantitative estimates are, by nature, tied to the predefined time budget.
For instance, a budget of $0.2$s would allow to perform 
exactly $1$ shot of the quantum approach, making the maximum average approximation ratio equal to the average (single-shot) approximation ratio. Importantly, this would not change the approximation ratio limit for large number of atoms, since
(see Figure~\ref{fig:Approximation-ratio-vs-natoms-vs-runtime}) the relative influence of the selection of the best outcome out of $n_\mathrm{shots}$ candidate solutions decreases for large system sizes (as $\sqrt{\frac{\log n_\mathrm{shots}}{2N_{atoms}}}$).
The asymptotic limit of the single-shot average approximation ratio is indeed already, in the case of a 2s-time budget, the value onto which we base our extrapolation of the approximation ratio to large numbers of atoms.  
On the classical side, however, for our locality-based heuristic, a
time budget of 0.2s would only allow to reach graph sizes of
$1,000-1,200$ (see Figure~\ref{fig:allowed_number_shots_loc}), bringing
the quantum advantage boundary far closer than a budget of $2$s
(which corresponds to Figure~\ref{fig:breakeven}). Regarding the quality of approximation, the asymptotic values that we used on Figure~\ref{fig:breakeven} are already reached for 1000-1100 atoms (see Figure~\ref{fig:Approximation-ratio-vs-natoms-vs-runtime}, right panel), and therefore would not change when switching to a time budget of $0.2$s.

\section{Conclusions}

In this work, we compared the approximation ratios reachable in a finite computational run time using state-of-the-art, yet realistically implementable, quantum and classical approximation algorithms to the UD-MIS combinatorial optimization problem.

We set up a precise comparison methodology based on the computation of the average maximum approximation ratio reachable within a fixed time budget, as opposed to the more usual average approximation ratio used for characterizing randomized algorithms.
We studied the dependence of both metrics with respect to system size and run time, under a realistic hardware model of the Rydberg quantum platform where this quantum algorithm could be executed thanks to the specific form of its interparticle interactions.

Based on simulations with up to 26 atoms, and on estimations of the spin correlation lengths, we predicted the large-size limit of the quantum average approximation ratio.
We also inferred, from statistical arguments, the asymptotics of the quantum maximal approximation ratio.
We found average approximation ratios of $\approx 0.72$ for the noise levels corresponding to recently published data, and of $\approx 0.84$ for near-future noise levels.
We found that two key aspects for reaching quantum advantage with respect to classical algorithms are the coherence level and the repetition rate.
Keeping a fixed time budget of 2 seconds, if the coherence level is substantially improved, or if the repetition rate is maintained while scaling to larger atom counts, Rydberg platforms could reach quantum advantage if approximation ratios above $\approx 95\%$, or atom counts of about 8000, can be attained.

These quite drastic size requirements point to the importance of developing and implementing quantum algorithms with better success probabilities and better suited for actual hardware. 

Such algorithms would effectively lower the quantum advantage bar.
In this study, we deliberately simulated a rather simple quantum annealing-based algorithm that can be implemented on today's Rydberg platforms. Including more sophisticated existing and forthcoming algorithmic improvements will likely bring the quantum advantage frontier closer, provided the hardware specificities are duly taken into account: 
we did not seek, as it is done in optimal control (see, e.g, \cite{Omran2019}), to optimize the classical control parameters of the quantum algorithm; nor did we try to use improved optimization costs such as the Conditional Value at Risk (CVaR, \cite{Barkoutsos2017,Henriet2020}).
Other possible routes towards improved approximation ratios could include algorithmic refinements of the annealing procedure (see, e.g, \cite{Yu2020}), attempts to better tailor the resource Hamiltonian to the target Hamiltonian (like optimizing the placement of atoms to mitigate unwanted tail interactions), or even a parallel implementation of the algorithm on multiple Rydberg platforms.

\acknowledgments
We acknowledge numerous fruitful discussions with A. Browaeys and T. Lahaye, as well as V. Lienhard (IOGS) and L. Henriet (Pasqal).
The computations were performed on the Atos Quantum Learning Machine (QLM). 
This work received funding from the European Union's Horizon 2020 research and innovation programme under grant agreement No 817482 (PASQuanS).

\appendix


\section{Supporting data for the quantum approach}\label{sec:supporting_data_quantum}

In this section, we address the question of the maximum approximation ratio one can reach using a Rydberg platform as a function of the
run time and the number of atoms. In the context of the hybrid quantum-classical
algorithm we described in the Methods section, the run time of the
algorithm is dominated by the physical set-up of the Rydberg atom platform.

The results we show are averaged over a certain number of random unit-disk graphs
(the detailed graph generation procedure is described in Appendix~\ref{sec:random_graph_generation}).
The main metric for describing such graphs is the vertex density $\nu$. Small ($\nu\ll1$) and large ($\nu\gg1$) densities result in easy optimization tasks (see e.g \cite{Henriet2020}). We thus choose an intermediate density $\nu=2$ to tackle a hard optimization regime.

The numerical methods used to obtain the results below are described in Appendix~\ref{subsec:numerical_methods}.

\subsection{Determination of the noise model\label{subsubsec:noise_model_determination}}

\begin{figure}
\begin{centering}
\includegraphics[width=1\columnwidth]{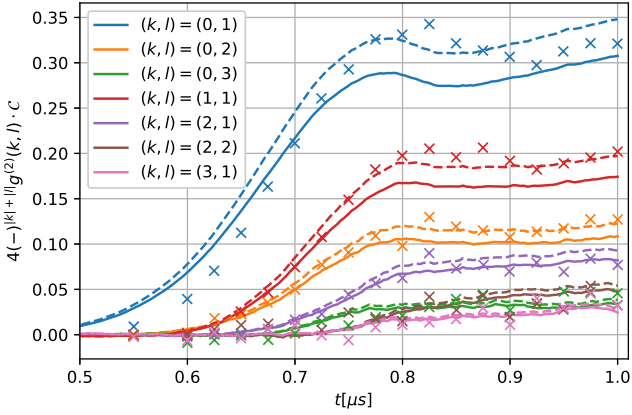}
\par\end{centering}
\caption{N\'eel structure factor as a function of time. Same parameters as Fig.
4.a of \cite{lienhard2018observing}: $4\times4$ square lattice of atoms,
$\gamma=3.0$. Solid lines: with readout error $\epsilon=\epsilon' = 0.03$. Dashed lines: without
readout error. Crosses: experimental data, reproduced from \cite{lienhard2018observing} .\label{fig:Neel-structure-factor-Lienhard}}
\end{figure}

The experimental results of Ref.~\cite{lienhard2018observing}
can be reproduced by adjusting $\gamma=3.0$ in the noise model described above (section \ref{subsec:noise_model})). (This corresponds to $\gamma/2\pi = 0.48 \mathrm{MHz}$ when reinstating dimensions).

In Figure~\ref{fig:Neel-structure-factor-Lienhard}, we show the temporal evolution of the N\'eel structure factor measured experimentally and compared to noisy simulations with and without readout errors, for a square lattice geometry (as opposed to the rest of the text, where we consider unit-disk graphs).

A value of $\gamma=3.0$ gives an accurate agreement between the noisy simulation and the experimental data.
As for the readout error, we take $\epsilon = \epsilon' = 3\%$ to reflect typical readout errors: Ref.~\cite{DeLeseleuc2018a} estimates $\epsilon\approx 1-2\%$, $\epsilon'<5\%$.

\subsection{Annealing time optimization}\label{subsubsec:anneal_time_opt}

\begin{figure}
\begin{centering}
\includegraphics[width=1\columnwidth]{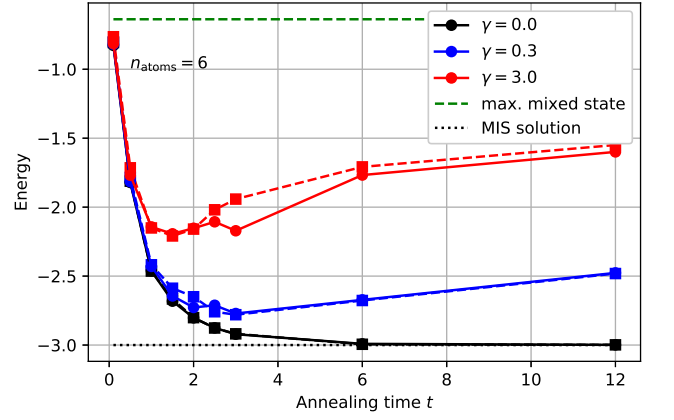}
\par\end{centering}
\caption{Target energy as a function of the annealing time (in $\mu$s) for various noise levels.
Solid lines: full Hilbert space.
Dashed lines: keeping only IS states in Hilbert space (see Appendix~\ref{subsubsec:IS_restriction}).
Dashed green line: energy of the maximally mixed state.
Dotted black line: energy of the optimal (MIS) solution. 
\label{fig:Target-energy-vs-anneal-time}}
\end{figure}

\begin{figure}
\begin{centering}
\includegraphics[width=1\columnwidth]{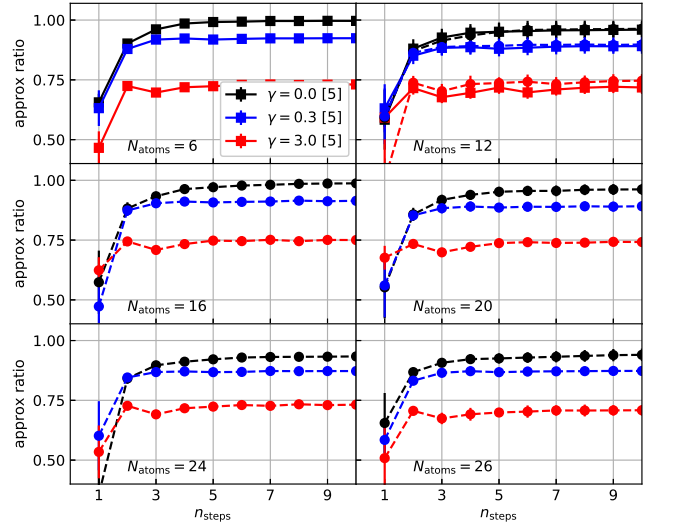}
\par\end{centering}
\caption{Approximation ratio as a function of the  number of
annealing time optimization steps. Solid lines and square
symbols: full Hilbert space. Dashed lines and circles: IS-Hilbert
space.\label{fig:Approximation-ratio-vs-optim-steps}
}
\end{figure}

\begin{figure} 
\begin{centering}
\includegraphics[width=1\columnwidth]{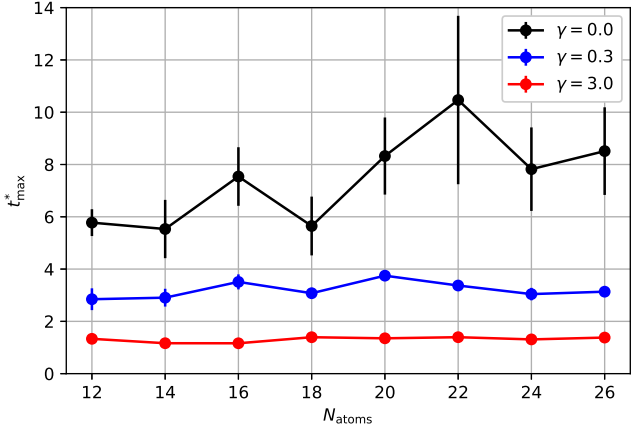}
\par\end{centering}
\caption{Optimal annealing time $t_\mathrm{f}^*$ (in $\mu$s) as a function of the number of atoms for various noise levels. \label{fig:tf-opt-vs-natoms} }
\end{figure}

Here, we focus on the optimization of the annealing time for reaching the lowest target energy (see Appendix \ref{sec:udmis_cost_function_def} for more details about the construction of $H_\mathrm{target}$).

We first show numerical evidence that in the presence of noise, there exists an optimal annealing time $t_{\mathrm{f}}^{*}$. 

Fig.~\ref{fig:Target-energy-vs-anneal-time} shows the dependence of the target energy on the annealing time.
We observe that in the absence of noise, the larger the annealing time, the lower the target energy, as expected from the theory of adiabatic computing.
By contrast, in the presence of noise, there is an optimal annealing time.
Indeed, an increased annealing time entails a larger susceptibility to decoherence, and thus an increase in the minimal energy reached by the system
(we note that for large noise values and large annealing times one obtains a target energy that gets closer to the energy of the maximally mixed state, as expected). 
Contrary to the noiseless case, there is thus a balance to be found between increasing the success probability of annealing by slowing down the annealing process, and fighting against decoherence by completing the computation as quickly as possible.

In the following, we thus use a classical optimizer (COBYLA) to find the annealing time that accommodates both constraints.
In Fig.~\ref{fig:Approximation-ratio-vs-optim-steps}, we show the dependence of the approximation ratio on the number of optimization steps for different numbers of atoms and various noise levels.

For small numbers of atoms $N_{\mathrm{atoms}}\leq12$,
we use the full Hilbert space to carry out the simulation. For larger
atom numbers ($N_{\mathrm{atoms}}\geq12$), we use a restriction of
the Hilbert space to independent sets, as described in Appendix \ref{subsubsec:IS_restriction}.
We observe that convergence is reached after a couple of iterations.

In Fig.~\ref{fig:tf-opt-vs-natoms}, we show the obtained optimized annealing times as a function of the number of atoms.
We observe that in the presence of noise, the optimal time is roughly independent of the atom number.
In the noiseless case, it shows a slight increase with respect to system size, possibly indicating a decreasing value of the minimum gap.

Because of this weak, or absence of, dependence of the annealing time on the system size, the computation of the optimal annealing time can essentially be neglected in the computation of the total run time.
It can indeed be regarded as a heuristic parameter that can be determined once and for all by studying small (and therefore negligibly quick to simulate numerically) instances of the problem.

We thus do not take this optimization step into account in the total run time estimations shown in this work.

\subsection{Impact of the number of repetitions}

\begin{figure}
\begin{centering}
\includegraphics[width=1\columnwidth]{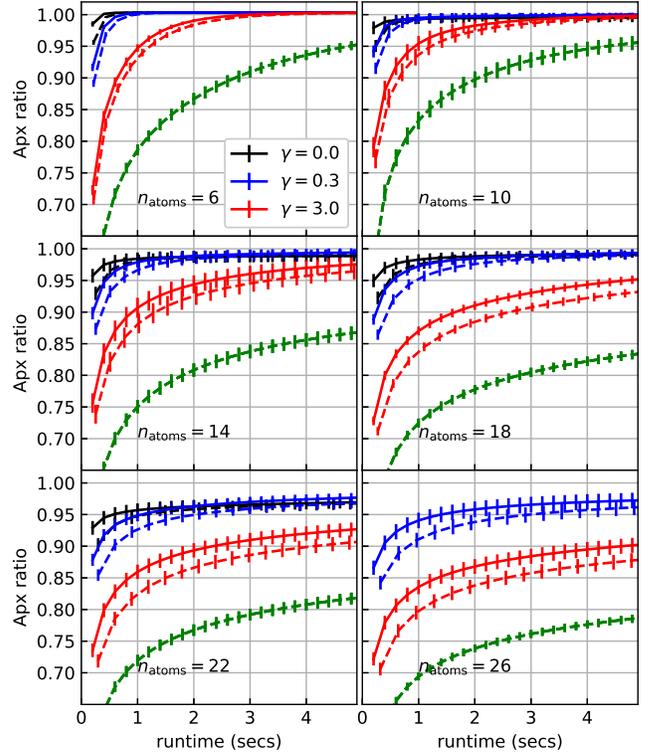}
\par\end{centering}
\caption{Approximation ratio as a function of the run time (computed as the number of repetitions $n_\mathrm{shots}$ times the repetition rate, $f=5\mathrm{Hz}$ here), after optimization of the annealing time $t_\mathrm{f}$.
Solid lines: without readout noise. Dashed lines: with readout noise $\epsilon = \epsilon' = 3\%$.
Black, blue and red curves: approximation ratio for the final state distribution obtained with noise parameters $\gamma=0.0$, $0.3$, and $3.0$, respectively.
Green curves: for a uniform distribution over the IS states.  State space restricted to IS-Hilbert
space.
\label{fig:Approximation-ratio-vs-nshots}}
\end{figure}

\begin{figure}
\begin{centering}
\includegraphics[width=1\columnwidth]{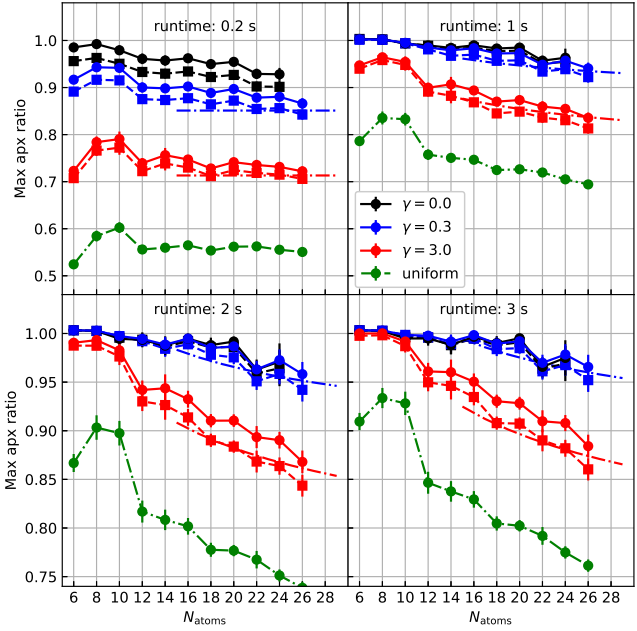}
\par\end{centering}
\caption{Approximation ratio as a function of the number of atoms in quantum
annealing for various fixed run times (same data as Fig.~\ref{fig:Approximation-ratio-vs-nshots}).
Solid lines: without readout noise. Dashed lines: with readout noise $\epsilon = \epsilon' = 3\%$. Dash-dotted lines: fit function Eq.~\eqref{eq:fit_nshots_dep}.
Black, blue and red curves: approximation ratio for the final state distribution obtained with noise parameters $\gamma=0.0$, $0.3$, and $3.0$, respectively, averaged over 20 random graphs (except for $N_\mathrm{atoms} = 24, 26$: 10 graphs).
Green dash-dotted lines: average approximation ratio for a uniform probability over IS states.
\label{fig:Approximation-ratio-vs-natoms-vs-runtimes} }
\end{figure}

Once the annealing time is fixed, the system can be evolved in time until its final state $\Psi^0(t_\mathrm{f})$ (or $\rho(t_\mathrm{f})$) is projected by a fluorescence measurement.
The outcome of the measurement is a string of bits $i \equiv (b_1 \dots b_n)$ indicating whether the $k$th atom is in its Rydberg state ($b_k = 1$) or its ground state ($b_k= 0$).
The probability for this particular bitstring to be measured is given by Born's rule $p_i = |\langle i|\Psi^0 \rangle|^2 = \left[\rho(t_\mathrm{f})\right]_{ii}$.
This bitstring can be mapped to a target energy $E_\mathrm{target}(i)$ to evaluate its proximity to the optimal solution.

The value for the target energy that is obtained on average by performing one measurement is thus given by
\begin{equation}
    \mathbb{E}\left[ E_\mathrm{target} \right] = \sum_i p_i E_\mathrm{target}(i).
\end{equation}

One can repeat this measurement process $n_\mathrm{shots}$ times and keep the maximum value obtained over these $n_\mathrm{shots}$ repetitions.
The average value of this maximum over $n_\mathrm{shots}$ repetitions can be readily computed using $p_i$. For this,  we define the cumulative energy density function as:
\begin{equation}
F(x) \equiv \sum_{i=0}^{d-1} p_i\,\theta(x - E_\mathrm{target}(i)/E_\mathrm{min}),
\end{equation}
with $d$ the dimension of the space (number of IS states if we restrict the dynamics to IS states), $E_\mathrm{target}(i)$ the target energy of state $i$, and $\theta$ the Heaviside function. We note that due to the restriction to the IS states, the support of $F$ is $x\in[0, 1]$.

As shown in Appendix \ref{sec:exp_of_max}, $F$ can be used to compute the average maximum approximation ratio after $n_\mathrm{shots}$ shots:
\begin{equation}
	\alpha(n_\mathrm{shots}) = 1 - \int_0^{1} \left [F(x)\right]^{n_\mathrm{shots}} \mathrm{d}x.
\end{equation}

The overall run time of the quantum part of the computation can thus be evaluated as the number of repetitions times the repetition rate. This rate is in the range 3-5 Hz \cite{ThierryPrivate, LienhardThesis} in the current experimental setups.

We show the corresponding results in Fig.~\ref{fig:Approximation-ratio-vs-nshots}.
As expected, the average approximation ratio increases with the number of repetitions, and decreases with increasing noise and in the presence of readout errors.
For the latter, we assume that those $\epsilon'$ errors (erroneous measurement of a $|1\rangle$) that lead to a non-IS state can be corrected (by detecting the fact that the solution is non IS) at the price of increasing the number of repetitions used to compute the maximum expectation ratio so as to compensate for the discarded solutions.
The data shown in Fig~ \ref{fig:Approximation-ratio-vs-nshots} takes into account this correction and compensation mechanism.  

We can also fix the maximum run time to a set value and show the evolution of the average approximation ratio obtained for this run time as a function of the number of atoms.
The corresponding results are shown in Fig.~\ref{fig:Approximation-ratio-vs-natoms-vs-runtimes}.

\subsection{Extrapolation to larger number of atoms: the role of the coherence length\label{subsubsec:corrlength}}

\begin{figure}
\includegraphics[width=1\columnwidth]{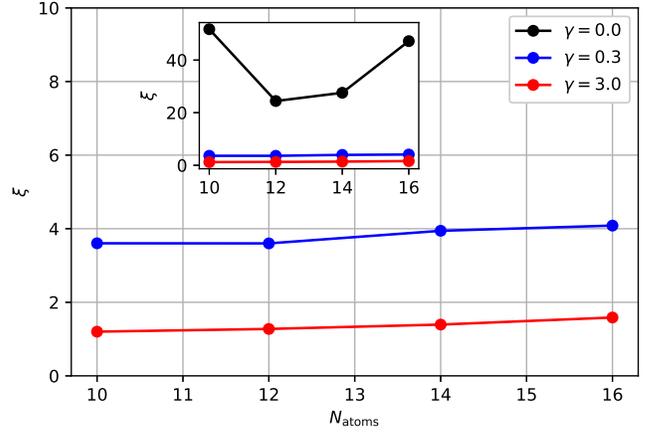}
\caption{Spin-spin correlation length $\xi$ extracted by fitting the spin-spin correlation function for various noise levels (black: $\gamma=0.0$, blue: $\gamma=0.3$, red: $\gamma=3.0$). (See Fig.~\ref{fig:corr_func_vs_distance} for fitting details.)
\label{fig:xi_vs_natoms}}
\end{figure}

The numerical simulations presented in the previous subsections are limited to a number of atoms (26) that is small compared to the graph sizes that can be reached in reasonable compute times by classical algorithms.

Yet, the numerical data for the higher noise level ($\gamma = 3.0$, see Fig.~\ref{fig:Approximation-ratio-vs-natoms-vs-runtime}) display an apparent saturation of the approximation ratio.
This saturation occurs already for quite small atom counts.
Such an observation can be accounted for by considering the correlation length associated with the MIS ordered state.
By conducting an analysis of the spin-spin correlation function (see Appendix~\ref{sec:corr_length_analysis} for details), we extract a graph-averaged correlation length and study its dependence on the graph size and noise level. As shown in Fig.~\ref{fig:xi_vs_natoms}, this correlation length is roughly independent of the graph size, and gets shorter as the intensity of decoherence ($\gamma$) increases.
This physically-expected behavior can explain the saturation observed for $\gamma = 3.0$ and the onset of a saturation for $\gamma = 0.3$.
Indeed, the value of the approximation ratio can be intuitively related to the correlation distance observed in the system: on average, the first "defects" in the MIS state will occur when perfect correlation is lost between two atoms. This happens at a length scale $\xi$.
Increasing the system size beyond the number $N^*(\xi)$ of atoms contained in a volume of size $\xi$ will therefore not increase the approximation ratio. Consequently, the numerically computed approximation ratio will saturate beyond system sizes of the order of $N^*(\xi)$.

We can estimate $N^*(\xi)$ based on our numerical estimates of $\xi$: by counting the number of atoms in a square of length $\xi$, we find $N^*(\gamma=3.0) = 4$ and $N^*(\gamma=0.3) = 30$. This explains the observed saturation for $\gamma = 3.0$. For $\gamma = 0.3$, it is likely that the saturation is not entirely reached, although the maximum atom number of 26 is in the same range as 30, allowing us to use, as a rough estimate of the approximation ratio at large atom numbers, the value that we obtained numerically for $N_\mathrm{atoms}= 26$.

Interestingly, the ratio of correlation length to the optimal annealing time (see subsection~\ref{subsubsec:anneal_time_opt} above) seems to be independent of $\gamma$ (its value is close to 1).
This behavior deserves further investigation.

\section{Supporting data for the classical approach \label{sec:supp_classical}}

\subsection{Approximation ratio}

Figure~\ref{fig:Approximation-ratio-vs-natoms-vs-runtime} (right-panel) shows the approximation ratio achieved by our locality-based heuristic (Algorithm~\ref{alg:ptas}) on a class of random unit-disk graphs with constant density (see Appendix \ref{sec:random_graph_generation} for more details on our graph generation procedure), for several value of the resolution distance $d$.

As the time to perform
a single run of Algorithm~\ref{alg:ptas} increases with graph size (see Figure~\ref{fig:allowed_number_shots_loc}, the number $k$ of allowed runs within the time budget of 2 seconds (Figure~\ref{fig:allowed_number_shots_loc}, bottom) decreases. Therefore, because the average maximum approximation ratio describes the
average ratio obtained when running $k$ runs and selecting the best outcome, its value is bound to converge to the ``single-run'' approximation ratio, equalling it when the time to carry out a single run is $\simeq 2$ seconds. 

\subsection{Run time and classical limit}

Figure~\ref{fig:rt_benchmark} shows the execution time of our locality-based approach along with a numerical lower bound solely taking into account the time to exactly solve MIS on sub-instances. Their agreement show that Breadth-First Search and other graph manipulations in Algorithm~\ref{alg:ptas} are negligible compared to MIS-solving, and that the run time of the heuristic can be expressed as:
\begin{equation}
\mathrm{run\_time}\equiv \sum_{u\in seeds} \mathrm{exact\_solving\_time}(S_{d}(u))
\end{equation}

The exact-solving time
of a unit-disk graph is, as one can expect from the NP-hard nature of UD-MIS, exponential in the graph size. Figure~\ref{fig:runtime_opt} 
shows the average exact-solving time of our random graph class (see Appendix~\ref{sec:random_graph_generation} for details) using a state-of-the-art solver, freely available on the Internet \cite{li2017minimization}. One can see the exponential ``taking off'' around 300-400 vertices with graphs of up to $\simeq 200$ vertices routinely solved in $\leq 0.05$ seconds.

\begin{figure}
\includegraphics[width=1\columnwidth]{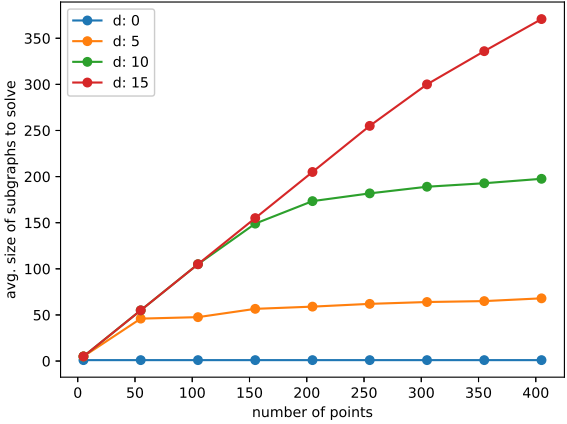}
\caption{0.9-quantile of sizes (i.e 90\% of the sizes fall below this threshold) of sub-instances to solve exactly when running the locality-based heuristic Algorithm \ref{alg:ptas}.
\label{fig:subgraph_size}}
\end{figure}

This can explain why, although providing a better approximation ratio and involving the solving of larger instances, the heuristic with $d=5$ is not slower than $d=2$, as one can see on Figure~\ref{fig:rt_benchmark}. Indeed, as one can see on Figure~\ref{fig:subgraph_size}, reporting the 0.9-quantile of the sizes of subgraphs to solve, $d=5$ involves solving graphs largely falling under $200$ vertices, which is below the ``exponential explosion'' of the exact solving run time at $300-400$. Therefore, it involves fewer instances than, say, $d=2$ that are bigger but not significantly longer to solve.

\begin{figure}
\includegraphics[width=1\columnwidth]{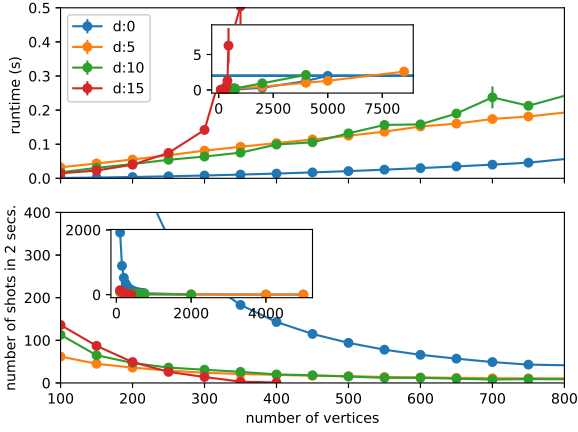}
\caption{
\emph{Top:} Run time of locality-based approximation heuristic as a function of the number of atoms. 
\emph{Bottom:} corresponding number of allowed shots within a 2-second time window. 
\label{fig:allowed_number_shots_loc}}
\end{figure}

Figure~\ref{fig:allowed_number_shots_loc} shows the run time (top)
and corresponding number of allowed shots (bottom) within the time-budget we chose (2 seconds). The $d=15$ curve fairly resembles the optimal solving run-time curve of Figure~\ref{fig:runtime_opt}, going over time-budget at around $300-400$ vertices. This can be explained by the size of sub-instances to solve being roughly equal to the entire system size up to $300-400$ vertices, as exemplified by Figure~\ref{fig:subgraph_size}. 

Other curves involve instances whose sizes largely fall below the ``exponential explosion threshold'' of optimal solving. Therefore, they are able to stay within time-budget for much longer, timing out at $\sim 4000$, $\sim5000$ and $\sim 8500$ for $d=10$, $0$ and $5$ respectively.

\subsection{Benchmark and implementation}

\begin{figure}
\includegraphics[width=1\columnwidth]{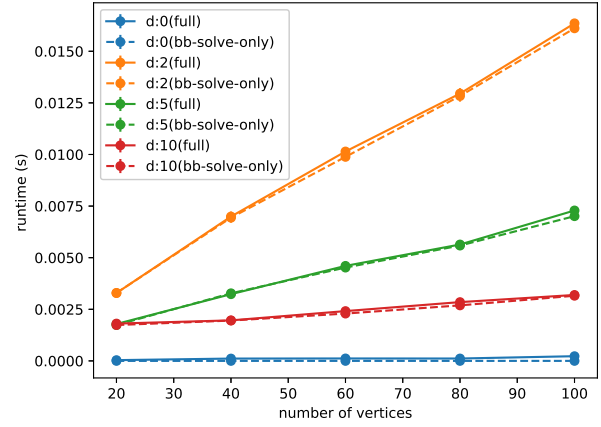}
\caption{Run time of locality-based heuristic from Algorithm \ref{alg:ptas}. \emph{Solid lines:} run time including graph manipulations (such as BFS) and solving, not taking file I/O overhead inherent to the execution of \cite{li2017minimization}. \emph{Dashed lines:} solving time only, disregarding BFS and other graph manipulation.
Overall, run time is governed by the number and sizes of graph instances to solve.
\label{fig:rt_benchmark}}
\end{figure}

Algorithm~\ref{alg:ptas} involves, given a graph, producing \emph{sub-instances} that will be solved and removed from the graph, until it is empty. 

These sub-instances are solved with \cite{li2017minimization} (see next subsection). It comes as an executable, which is called onto a graph-describing
text file. In our implementation, such text files are created, filled and erased automatically when executing Algorithm~\ref{alg:ptas}. 

In our reported execution times, on Figure~\ref{fig:allowed_number_shots_loc} or \ref{fig:rt_benchmark}, for instance, we neglect the time taken by these file manipulations. We consider that it does not represent actual ``computation time'', see subsection below and Figure~\ref{fig:runtime_opt} for further discussion.

The parametrization integer $d$ governs the degree of approximation to the optimal solution. A larger value of $d$ involves solving larger instances, as one can see on Figure~\ref{fig:subgraph_size}.

\subsection{Run times of exact branch-and-bound solver}

\begin{figure}
\begin{centering}
\includegraphics[width=1\columnwidth]{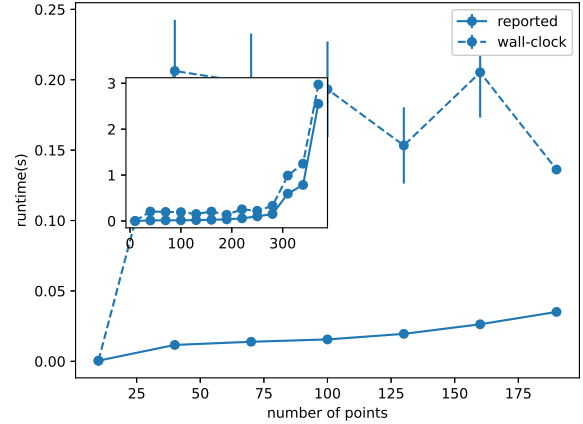}
\par\end{centering}
\caption{Run time of exact branch-and-bound solver \cite{li2017minimization} as a function of the number of vertices,
for the class of random graphs we considered (described in Appendix~\ref{sec:random_graph_generation}). 
\label{fig:runtime_opt}
}
\end{figure}

In Fig.~\ref{fig:runtime_opt}, we show the average run times of the optimal branch-and-bound solver, as a function of the number of vertices,
for our class of random graphs (see Appendix~\ref{sec:random_graph_generation}).

We used a state-of-the-art MIS solver \cite{li2017minimization} freely available on the web.
It is a generic solver, not restricted nor specifically optimized for unit-disk graphs.

\begin{figure}
    \centering
    \includegraphics[width=\columnwidth]{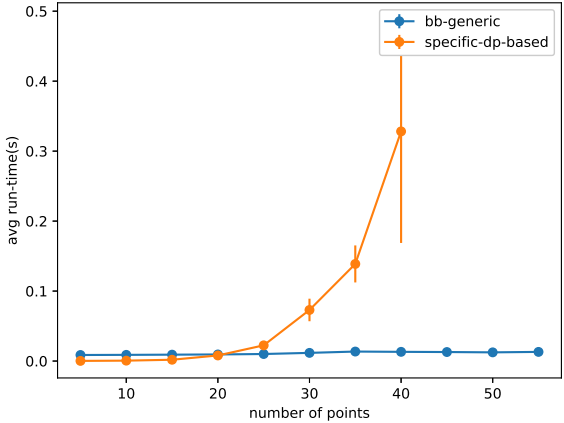}
    \caption{Comparison of execution times of generic state-of-the-art branch-and-bound MIS solver (bb-generic) \cite{li2017minimization} and an implementation of a specialized UD-MIS solver, based on dynamic programming \cite{cormen2009introduction}, as can be readily derived from \cite{matsui1998approximation} (specific-dp-based) . 
    }
    \label{fig:comp_bb_own}
\end{figure}

It is a legitimate question to ask whether, given the specific geometric structure of unit-disk graphs, a specialized solver would not fare better in practice. We tested this hypothesis by implementing a dynamic programming technique
directly derivable from \cite{matsui1998approximation}. In algorithmic terms, it exploits the fixed-parameter tractability \cite{downey2013fundamentals} of UD-MIS, with the \emph{thickness} of input graphs taken as a parameter.
Figure~\ref{fig:comp_bb_own} compares the execution times of the specialized dynamic-programming and the generic branch-and-bound approaches, largely in favour of the latter, namely \cite{li2017minimization}, whose exponential explosion happens around 300-400 vertices (See Figure~\ref{fig:runtime_opt}) and not $\sim30$ as in Figure~\ref{fig:comp_bb_own}.

In estimating the run time ("reported" run time in Fig.~\ref{fig:runtime_opt}), we remove the input-output overhead specific to the solver of \cite{li2017minimization} so as to measure only the actual computational time (\cite{li2017minimization} works as an executable that reads and writes graphs from files). We report both run times (with and without I/O) as "wall-clock" time and "reported" run time, respectively. We observe that the I/O does represents a large part of the run time for small graphs.

\subsection{Supplementary data}
\begin{figure}
    \centering
    \includegraphics[width=\columnwidth]{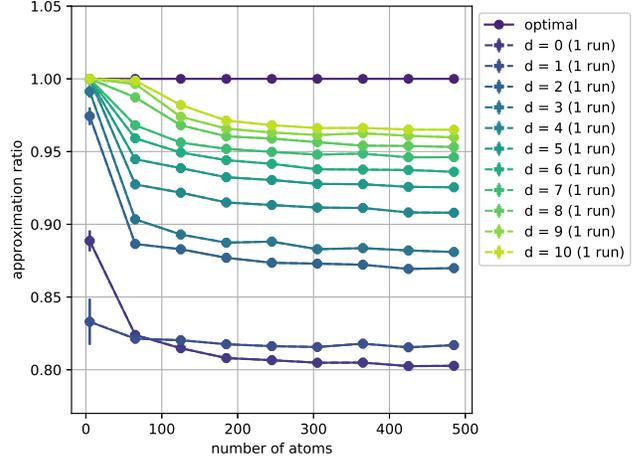}
    \caption{Single-shot average (over $100$ random graphs) approximation ratio achieved by classical locality-based heuristic (Algorithm~\ref{alg:ptas}) on random unit-disk graphs (see Section~\ref{sec:random_graph_generation})}
    \label{fig:supp_apx}
\end{figure}

\begin{figure}
    \centering
    \includegraphics[width=\columnwidth]{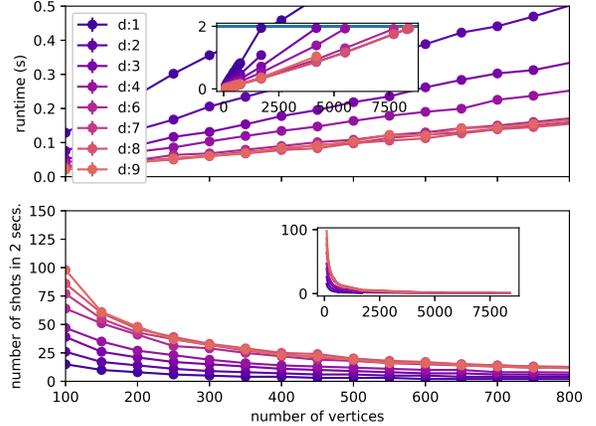}
    \caption{
    \emph{Top:} Execution time of classical locality-based approximation heuristic, as a function
    of the number of atoms.
    \emph{Bottom:} Corresponding
    allowed number of shots in $2$s}
    \label{fig:supp_rt_loc}
\end{figure}

Figure~\ref{fig:supp_apx} shows the average approximation
ratio achieved by locality-based approximation heuristic for different values of the parameter $d$. For all values, the approximation ratio is observed to stabilize to an asymptotic value, which increases with $d$.

Figure~\ref{fig:supp_rt_loc} documents the execution time of our locality-based approximation heuristic for values
of $d$ not shown on
Figure~\ref{fig:allowed_number_shots_loc}, and the corresponding number of allowed shots. This data is used in Figure~\ref{fig:breakeven} to estimate the largest graph size attainable within a time-budget of $2s$.

\section{Reformulation of the UDMIS problem as a minimization problem}\label{sec:udmis_cost_function_def}

To be amenable to a resolution using a quantum processor, the problem has to be reformulated as an unconstrained minimization problem, ultimately in the form of an Ising Hamiltonian.

\subsection{Boolean reformulation}\label{par:udmis_def}

Let us reformulate the UD-MIS problem defined in Section~\ref{sec:udmis_theory} entirely in terms of Boolean functions
by defining
\begin{align}
f(S) & \equiv\sum_{i\in V}n_{i}^{(S)}=|S|\label{eq:def_f}\\
h(S) & \equiv\sum_{i,j\in E}n_{i}^{(S)}n_{j}^{(S)}\label{eq:h_def}
\end{align}
with $S=(n_1, \dots n_{N_\mathrm{atoms}})$ a given bitstring.

Let us note that $S\in\mathrm{I.S}$ iff $h(S)=0$, so that the problem
described in Eq.~(\ref{eq:UDMIS_problem}) can be reformulated as:
\begin{align}
\max_{S\in\mathcal{B}} & f(S)\label{eq:UDMIS_problem_boolean}\\
\mathrm{s.t} & \;h(S)=0\nonumber 
\end{align}

Let us then denote $S^{*}$ the solution of this optimization problem,
called (UD)MIS, and $p^{*}$ its optimum:
\begin{equation}
p^{*}\equiv f(S^{*})\label{eq:prime_def}
\end{equation}

\subsection{Lagrangian relaxation of UD-MIS}

We want to transform the above constrained optimization problem, Eq.
(\ref{eq:UDMIS_problem_boolean}), into an unconstrained optimization
problem, while requiring that the optimal solution of this new optimization
problem is still a Maximum Independent Set.

One way to achieve this is to construct a Lagrangian relaxation (see, e.g, Ref.~\cite{Lemarechal2001}) of
the problem by introducing the Lagrangian
\begin{equation}
\Lambda(S,u)\equiv f(S)-u \cdot h(S)\label{eq:Lagrangian_def}
\end{equation}

with the Lagrange multiplier $u$. Since $h(S)\geq0$ for any $S$,
we see that whenever $u>0$, the second term penalizes non feasible
solutions (i.e non-IS states). Let us also define the Lagrangian dual
function:
\begin{equation}
g(u)\equiv\max_{S\in\mathcal{B}}\Lambda(S,u)\label{eq:g_def}
\end{equation}

At this point, we cannot ensure yet that the maximum $S_{u}^{*}$
of the maximization problem Eq.~(\ref{eq:g_def}) is an IS. To impose
this condition, one can prove (see subsection \ref{sec:proof_U_bound} below) that it is enough to require $u>1.$ Thus, we are going to solve
the following maximization problem:
\begin{align}
\max_{u\in\mathbb{R}} & \max_{S\in\mathcal{B}}\Lambda(S,u)\label{eq:UDMIS_relaxed}\\
\mathrm{s.t} & \;u>1\nonumber 
\end{align}

Since we have proven that the solution $S_{u}^{*}$ is IS,
we have $\Lambda(S_{u}^{*},u)=f(S_{u}^{*})$. As in addition, among all IS solutions, the ones that maximize
$f$ are the Maximum Independent Sets, we have that $S_{u}^{*}$ is indeed a Maximum Independent Set of the graph, 
and that the optimal solutions to both optimization problems are the same.

\subsection{Reformulation as a minimization problem}

Our goal is to use a quantum algorithm to solve the inner optimization
problem, defined by Eq.~(\ref{eq:g_def}). As described in the main text, our
quantum algorithm is going to perform a \emph{minimization} task.
Therefore, we first reformulate the above maximization problem as
a minimization problem. This is achieved by redefining the objective
function as $-f$ instead of $f$. Thus, we want to solve
\begin{align*}
p^{*}=\min_{S\in\mathcal{B}} & \left\{ -f(S)\right\} \\
\mathrm{s.t} & \;h(S)=0
\end{align*}

with the following Lagrangian relaxation:
\begin{align}
q^{*}=\min_{u\in\mathbb{R}} & \min_{S\in\mathcal{B}}\left\{ -f(S)+u \cdot h(S)\right\} \label{eq:lagrangian_relaxation_minimization}\\
\mathrm{s.t} & \;u>1\nonumber 
\end{align}

\begin{figure}
\begin{centering}
\includegraphics[width=1\columnwidth]{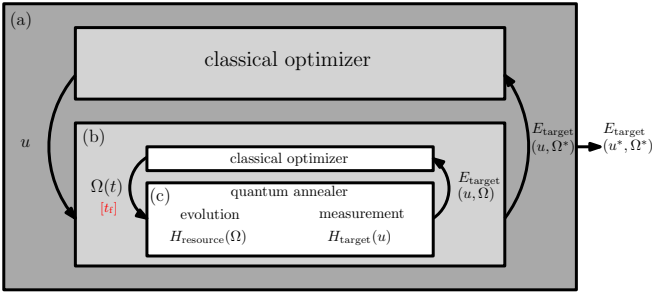}
\par\end{centering}
\caption{Hybrid quantum-classical algorithm for the approximate solution of
the UD-MIS problem.\label{fig:Hybrid-quantum-classical-algorit}
\emph{Box (a)}: outer minimization.
\emph{Box (b)}: inner minimization.
\emph{Box (c)}: quantum part of the algorithm, performed with a Rydberg platform.
Red labels in brackets: simplified variational parameters
used in this work. All variational parameters are lumped
into a single parameter $\vec{\theta}$.
$u$ is set to $u=1.35$ for all shown results.}
\end{figure}

Following the reasoning of the previous subsection, we have the guarantee
that the optimal solution is indeed an optimal Maximum Independent Set.
One can
now solve the double minimization problem of Eq.~(\ref{eq:lagrangian_relaxation_minimization})
in two steps:

(A) \emph{the inner minimization}
\begin{equation}
g(u)=\min_{S\in\mathcal{B}}\left\{ -f(S)+u \cdot h(S)\right\} ,u>1\label{eq:inner_minimization}
\end{equation}
can be performed using a quantum algorithm; since this quantum algorithm
itself comes with parameters that can be optimized (like the annealing time), the inner minimization
will itself be a quantum-classical algorithm; this inner minimization
is represented as the light gray box (box (b)) in Fig.~\ref{fig:Hybrid-quantum-classical-algorit} 

(B) \emph{the outer minimization}
\[
\min_{u>1}g(u)
\]
can be performed using a classical minimization algorithm. This outer
minimization is represented by the outermost box (box (a)) in Fig.
\ref{fig:Hybrid-quantum-classical-algorit}. 
In this work, we do not study the influence of this outer minimization, and instead consider the Lagrange multiplier to be set to a fixed value $u=1.35$.

\subsection{Lagrangian relaxation: Proof of bound on $u$}\label{sec:proof_U_bound}
Here, we prove the condition $u>1$ that  ensures that the cost function is maximal for a feasible solution (independent set).

Let us suppose that $u > 1$, and that we have a solution (bitstring) $\tilde{S}_{u}^{*}$ of the
maximization problem Eq.~(\ref{eq:g_def}) that is not an IS.

Since it is a solution, we have:
\[
f(\tilde{S}_{u}^{*})-u \cdot h(\tilde{S}_{u}^{*})\geq f(S)-u \cdot h(S),\forall S\in\Omega
\]

In particular, for IS states, 
\begin{equation}
f(\tilde{S}_{u}^{*})-u \cdot h(\tilde{S}_{u}^{*})\geq f(S),\forall S\in\mathrm{I.S}\label{eq:inequality_IS_states}
\end{equation}
Let us call $S_{u}^{*}$ the IS one finds when removing (some) edge-sharing vertices from $\tilde{S}_{u}^{*}$ 
(one can always construct such
a state: a possible (but non optimal) procedure is, for instance, to consider every conflict, i.e every edge $(i,j)$ such that both ends are in the set $\tilde{S}_{u}^{*}$, and remove one of the two vertices from $\tilde{S}_{u}^{*}$).
Let us call $k>0$ this number of vertices.

Then:
\begin{align*}
f(S_{u}^{*}) & =f(\tilde{S}_{u}^{*})-k\\
h(S_{u}^{*}) & =0\\
h(\tilde{S}_{u}^{*}) & \geq k
\end{align*}

The last inequality comes from the fact that a new occupied vertex
can create more than one edge. Thus:

\begin{align*}
f(\tilde{S}_{u}^{*})-u \cdot h(\tilde{S}_{u}^{*}) & \leq\left(f(S_{u}^{*})+k\right)-uk\\
 & =f(S_{u}^{*})-\left(u-1\right)k
\end{align*}

Since $u>1$, 
\[
f(\tilde{S}_{u}^{*})-u \cdot  h(\tilde{S}_{u}^{*})<f(S_{u}^{*}),
\]

which contradicts Eq.~(\ref{eq:inequality_IS_states}).

\section{Random graph generation procedure}\label{sec:random_graph_generation}
We use the following procedure to generate random graphs:

\begin{algorithm}[H]
Input parameters: density $\nu$, number of points $N$, exclusion radius $r$.

Initialize an empty list to be filled by the points: $V$.

 \While{$|V|<N$:}{
 pick two Cartesian coordinates $x$, $y$ uniformly at random in $[0, \sqrt{\frac{N}{\nu}}]$;
 
  \eIf{if the point $(x,y)$ comes within less than $r$ of any point in $V$}{
   reject point
   }{
 add it to $V$
  }
 }
 \KwResult{$V$ }
 \caption{Random point generation algorithm}
\end{algorithm}

The parameter $\nu$ is the vertex density, i.e. the average number of atoms per unit square, while $r$ plays the role of an exclusion radius.

We choose a density $\nu = 2$ that corresponds to a hard computational regime, i.e it is above the percolation threshold of $\nu_p \approx 1.4$, so that the generated graphs are connected on average, and not too high to be compatible with our exclusion radius, $r = 0.3$.
This exclusion radius is chosen to be above the minimum distance between two atoms, and below the blockade radius of 0.5 (our condition for two vertices to be connected is that their distance should be $\leq 1$, which means that disks of radius $0.5$ around them should intersect).

\section{Numerical methods\label{subsec:numerical_methods}}

\subsection{Restriction to independent sets subspace} \label{subsubsec:IS_restriction}
The Rydberg interaction (Eq.~\eqref{eq:H_rydberg_def}) favors configurations where neighboring atoms have a different internal state. These configurations correspond to the independent sets (IS) of the underlying graph.

In our simulations, we therefore assume that the temporal dynamics are limited to the vector space spanned by such configurations. 

We check that this approximation, which becomes exact only in the hard-sphere limit with $V\rightarrow \infty$, has little quantitative impact on the final results: see, e.g., Fig.~\ref{fig:Target-energy-vs-anneal-time} (the solid lines show a computation within the full Hilbert space, the dashed lines with our restriction to the IS subspace), or Fig.~\ref{fig:Approximation-ratio-vs-optim-steps} (see the panel for $N_\mathrm{atoms}=12$). 
As can be expected, the difference is largest for high noise levels and unoptimized anneal times, both of which may favor non-IS states.

\subsection{Quantum trajectories approach}
The Lindblad master equation is solved using the quantum trajectories or quantum jump approach \cite{Dalibard1992}  as implemented in Qutip \cite{Johansson2013}. We take 100 trajectories per run.

\subsection{Classical optimization}
For the classical optimization of the annealing time, we choose the COBYLA optimizer as implemented in scipy.optimize \cite{2020SciPy-NMeth}.


\section{Dependence of the correlation length on the noise level and the number of atoms}\label{sec:corr_length_analysis}

\begin{figure}
\includegraphics[width=1\columnwidth]{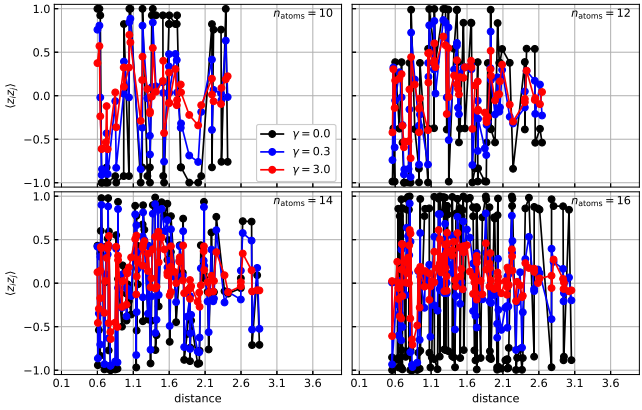}
\caption{Spin-spin correlation function $\langle z_{i}z_{j}\rangle_\Psi$ (see text) as a function of the Euclidean distance for one graph instance and for various noise levels (black: $\gamma=0.0$, blue: $\gamma=0.3$, red: $\gamma=3.0$) and graph sizes (clockwise from top left: $N_\mathrm{atoms}=10, 12, 16, 14$). 
\label{fig:corr_func_vs_distance_raw}}
\end{figure}

\begin{figure}
\includegraphics[width=1\columnwidth]{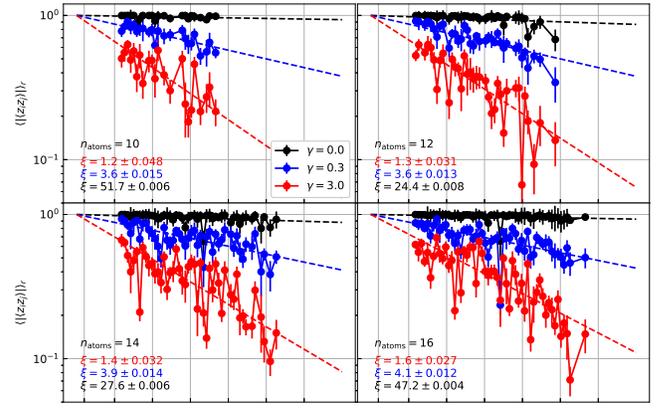}
\caption{Spin-spin correlation function $g^{(2)}(r)$ (see text) as a function of the distance for various noise levels (black: $\gamma=0.0$, blue: $\gamma=0.3$, red: $\gamma=3.0$) and various graph sizes (clockwise from top left: $N_\mathrm{atoms}=10, 12, 16, 14$). The dashed lines show the result of a fit with the exponential decay function $e^{-r/\xi(\gamma, N_\mathrm{atoms})}$.
\label{fig:corr_func_vs_distance}}
\end{figure}

In this section, we give data supporting the claim that the correlation length is roughly independent of the graph size, and depends on the noise level $\gamma$.

At the end of the optimization procedure for $t_\mathrm{f}$, we compute the correlation functions 
\begin{equation}
	\langle z_{i}z_{j}\rangle_\rho \equiv \mathrm{Tr} \left [ \rho  \sigma^z_i \sigma^z_j \right ],
\end{equation}
with $i,j=1\dots N_\mathrm{atoms}$.
We do not choose connected correlation functions because the sought-after solution is a classical state in general, which would entail a vanishing connected correlation function.

Such a correlation function displays oscillations corresponding to the alternation of correlations (1) and anticorrelations (-1) between the occupancy of site $i$ and that of site $j$, as shown in Fig.~\ref{fig:corr_func_vs_distance_raw}. These oscillations appear to be damped as the (Euclidean) intersite distance $d_{ij} = | r_i - r_j|$ increases, with a damping that increases as the decoherence level $\gamma$ increases. Our goal is to extract the envelope of these curves to compute a corresponding spin-spin correlation length $\xi$.

To this aim, we then compute the following binned correlation function
\begin{equation}
	g^{(2)}(r) = \max_{\substack{G\in \mathcal{G}\\i,j\\
|\boldsymbol{r}_{i}-\boldsymbol{r}_{j}|\approx r
}
}|\langle z_{i}z_{j}\rangle_{\rho_G}|
\label{eq:g2_def}
\end{equation}

We thus extract the envelope by finding, for each bin around $r$, the maximum over several (here 5) random graph instances, and over all pairs such that $|\boldsymbol{r}_{i}-\boldsymbol{r}_{j}|$ is in the vicinity $\delta r$ of $r$ (here, we choose $\delta r = 0.04$, but the results do not depend on this value provided each bin $\delta r$ around $r$ contains enough points).
Similar (although quantitatively slightly different) results can be obtained by averaging over the bin, instead of maximizing.

We plot the corresponding curves in Fig.~\ref{fig:corr_func_vs_distance}. An exponential fit $\exp(-r/\xi(\gamma, N_\mathrm{atoms}))$ turns out to be an accurate description of the data. We extract the corresponding correlation length $\xi$, which we plot in Fig.~\ref{fig:xi_vs_natoms}.

We observe that for $\gamma > 0.0$, the correlation length is reasonably constant with respect to the number of atoms, and decreases substantially with increasing noise level $\gamma$. For $\gamma=0.0$, the extracted correlation lengths are larger than those obtained in the noiseless case, but they do not appear to be constant with respect to the graph size.

We henceforth consider the value of $\xi$ to be constant with respect to the number of atoms: taking the average over the largest three atom counts (12, 14, 16), we get the estimates $\xi(\gamma=0.3) = 3.9$ and $\xi(\gamma=3.0) = 1.4$ (and $\xi(\gamma = 0.0) \approx 33$ if we average despite the variations).

We can roughly estimate the size (in terms of number of atoms) of the system beyond which the approximation ratio is going to get stable because the spatial extent of the atoms gets larger than the correlation length.
For this, we estimate the number $N^*$ of atoms contained in a square of length $\xi$. It is given by $N^* = \nu \xi^2$. For the density under consideration ($\nu = 2.0$), and the above estimates for the correlation length, we obtain  $N^*(\gamma=0.3) = 30$, $N^*(\gamma=3.0) = 4$ (and $N^*(\gamma=0.0)\approx 2000$). 
 
In practice, $N^*$ also gives us the atom number beyond which it is legitimate to extrapolate the numerical values obtained for lower atom counts.

\section{Expectation value of the maximum over a series of runs\label{sec:exp_of_max}}

In this section, we rationalize the asymptotic behavior of the maximum approximation ratio with respect to the number of repetitions (shots) and system size.

\subsection{Definitions}

Let us denote by $p_{i}=|\langle i |\Psi \rangle |^2$, $i=0\dots d-1$ ($d$ is dimension of the
space, $d=2^{N_{\mathrm{atoms}}}$ if all states are kept), the probabilities
of the computational basis states in the final state $\Psi$ at the
end of the annealing
and 
\[
E_{i}=-\left[\hat{H}_{\mathrm{target}}\right]_{ii}
\]
the corresponding target energies (we put a minus sign because due
to the definition of $\hat{H}_{\mathrm{target}}$, the minimum target
energy, which measures the size of the maximum independent set, is
negative). Let us further define the expectation value 
\[
\langle E\rangle_{p}\equiv\langle\psi|-\hat{H}_{\mathrm{target}}|\psi\rangle=\sum_{i=0}^{2^{n}-1}p_{i}E_{i}
\]

We want to compute the expectation value $M(N)$ of the maximum of the value of
$-H_{\mathrm{target}}$ over $N$ repetitions of the preparation and
measurement of $\Psi$ ($N=n_\mathrm{shots}$ in the main text). It is given by:
\begin{align}
M(N) & \equiv\mathbb{E}_{P}\left[\max\left(E_{i_{1}},\dots,E_{i_{N}}\right)\right] \\
& = \sum_{i_{1}\dots i_{N}}\max\left(E_{i_{1}},\dots,E_{i_{N}}\right)P\left(i_{1},i_{2}\dots i_{N}\right)\label{eq:M_N_def}
\end{align}

where $P\left(i_{1},i_{2}\dots i_{N}\right)$ is the probability of
observing the states $i_{1},\dots i_{N}$ as the $N$ outcomes of the
readouts. Since the repetitions are independent, 
$P\left(i_{1},i_{2}\dots i_{N}\right)=p_{i_{1}}\dots p_{i_{N}}$
and thus
\begin{equation}
    M(N)=\sum_{i_{1}\dots i_{N}}\max\left(E_{i_{1}},\dots,E_{i_{N}}\right)p_{i_{1}}\dots p_{i_{N}}\label{eq:M_N_expr}
\end{equation}

We note that
\[
M(N=1)=\langle E\rangle_{p}
\]

Furthermore, denoting by $E_{\mathrm{max}}$ the maximum target energy,
we expect that, if the support of $p$ contains the state with maximum
energy (the MIS),

\begin{align*}
M(N=\infty) & =E_{\mathrm{max}}.
\end{align*}

\subsection{Computation using the density of states}

We now define the following "target" density of states:
\begin{equation}
D(\varepsilon)\equiv\sum_{i=0}^{2^{n}-1}p_{i}\delta(\varepsilon-E_{i})\label{eq:dos_def}
\end{equation}

It gives the probability of observing a given target energy over one
repetition.  Starting from Eq.~\eqref{eq:M_N_expr}, we can then rewrite $M(N)$ as:
\begin{align*}
M(N) & =\sum_{i_{1}\dots i_{N}}\int d\varepsilon_{1}\dots\int d\varepsilon_{N}\\
 & \;\;\times\delta(\varepsilon_{1}-E_{i_{1}})\dots\delta(\varepsilon_{N}-E_{i_{N}})\\
 & \;\;\times\max\left(E_{i_{1}},\dots,E_{i_{N}}\right)p_{i_{1}}\dots p_{i_{N}}\\
 & =\int d\varepsilon_{1}\dots\int d\varepsilon_{N}\\
 & \;\;\times\sum_{i_{1}}p_{i_{1}}\delta(\varepsilon_{1}-E_{i_{1}})\dots\sum_{i_{N}}p_{i_{N}}\delta(\varepsilon_{N}-E_{i_{N}})\\
 & \;\;\times\max\left(E_{1},\dots,E_{N}\right)
\end{align*}

i.e
\begin{align}
M(N) & =\int d\varepsilon_{1}\dots\int d\varepsilon_{N}D(\varepsilon_{1})\dots D(\varepsilon_{N})\max\left(\varepsilon_{1},\dots,\varepsilon_{N}\right)\label{eq:M_N_vs_dos}\\
 & \equiv\mathbb{E}_{R_{N}}\left[\max\left(\varepsilon_{1},\dots,\varepsilon_{N}\right)\right]\label{eq:M_N_as_exp_over_max}
\end{align}

with $R_{N}(\varepsilon_{1},\dots\varepsilon_{N})=D(\varepsilon_{1})\dots D(\varepsilon_{N})$.

Let us define the random variable
\[
\mathcal{E}_{N}\equiv\max\left(\varepsilon_{1},\dots,\varepsilon_{N}\right)
\]

Let us call $\mathcal{F}_{N}$ the cumulative distribution function
(CDF) of $\mathcal{E}_{N}$.
Then:

\begin{align}
M(N)  &=\mathbb{E}_{R_{N}}\nonumber\\
&=\int_{E_{\mathrm{\text{min}}}}^{E_{\mathrm{max}}}\varepsilon\mathcal{F}_{N}'(\varepsilon)d\varepsilon\nonumber\\
 & =\left[\varepsilon\mathcal{F}_{N}(\varepsilon)\right]_{E_{\mathrm{min}}}^{E_{\mathrm{max}}}-\int_{E_{\mathrm{min}}}^{E_{\mathrm{max}}}\mathcal{F}_{N}(\varepsilon)d\varepsilon \label{eq:M_N_interm}
\end{align}
where we have used integration by parts to obtain the last line.

We can now simplify the expression for $\mathcal{F}_{N}$. Let us denote by $F$ the CDF corresponding to the probability distribution function $D$,
i.e $D=F'$. Then
\begin{align*}
\mathcal{F}_{N}(\varepsilon) & =P\left(\max\left(\varepsilon_{1},\dots,\varepsilon_{N}\right)\leq\varepsilon\right)\\
 & =P(\varepsilon_{1}\leq\varepsilon,\dots\varepsilon_{N}\leq\varepsilon)\\
 & =P(\varepsilon_{1}\leq\varepsilon)\dots P(\varepsilon_{N}\leq\varepsilon)\\
 & =\left[ F(\varepsilon) \right]^N
\end{align*}

We thus obtain the final expression:
\begin{equation}
 M(N) =E_{\mathrm{max}}-\int_{E_{\mathrm{min}}}^{E_{\mathrm{max}}}\left[ F(\varepsilon) \right]^Nd\varepsilon\label{eq:M_N_final}
\end{equation}

Since $F(\varepsilon)<1 (\forall\varepsilon\in]E_{\mathrm{min}},E_{\mathrm{max}}[$),
we have, as expected:
\[
M(N=\infty)=E_{\mathrm{max}}
\]

\section{Approximate upper bound for the approximation ratio corresponding to a uniform state distribution}\label{sec:approx_upper_bound_exp_max}

\begin{figure}
\includegraphics[width=1\columnwidth]{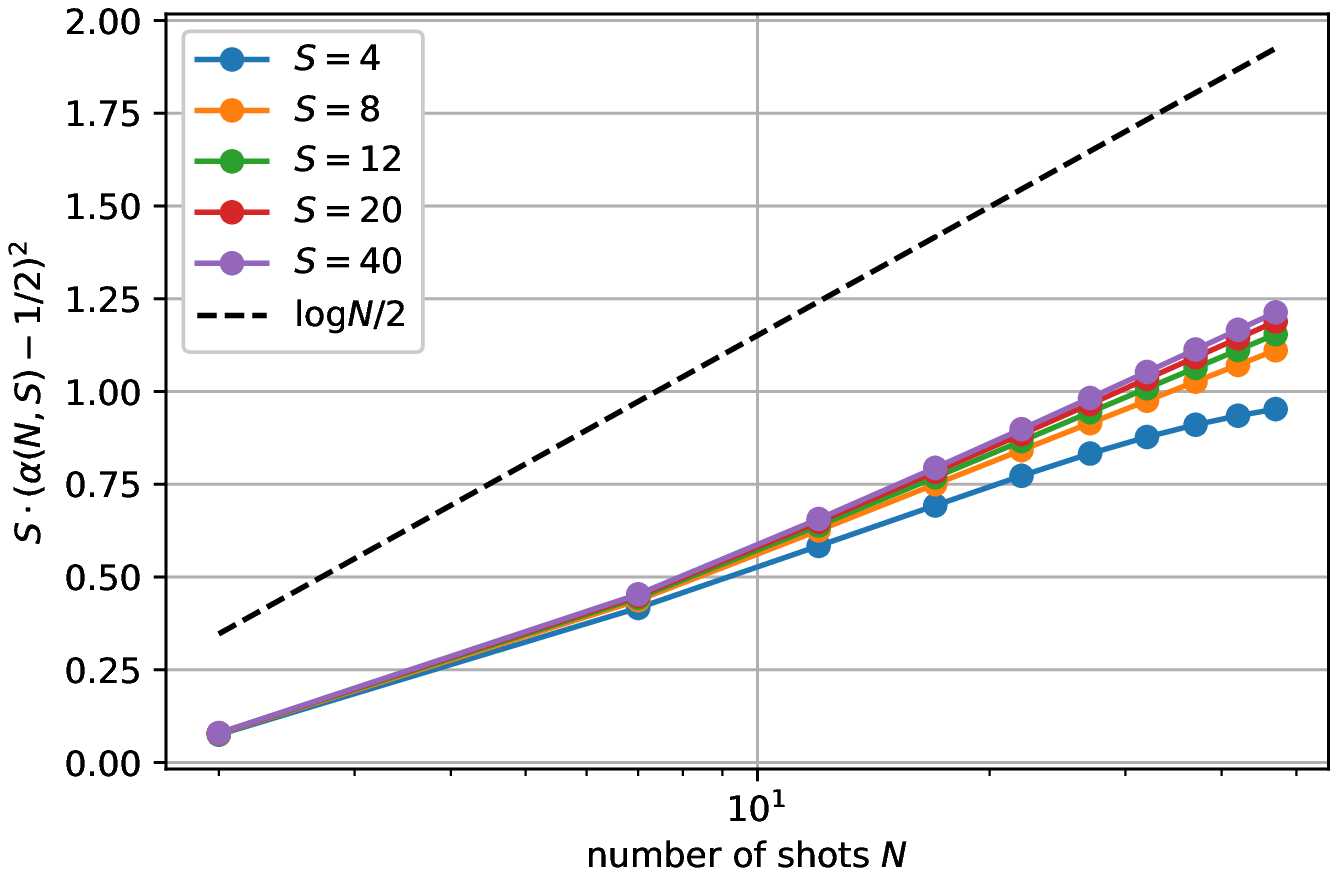}
\caption{Comparison of approximation ratio with its upper bound in the uniform case. Solid lines: $S\cdot (\alpha(N, S)-1/2)^2$, with $\alpha(N, S) = \mathbb{E}\left [ \max_{p=1\dots N}\lbrace E_i(S; p) \rbrace \right ] / S$.
Dashed black line: $\log{N}/2$.
\label{fig:apx_ratio_upper_bound}}
\end{figure}

In this section, we derive an approximate upper bound for the expectation value of the maximum of the approximation ratio computed over $N$ samples of a \emph{uniform} distribution of states,
\begin{equation}
    p_i = \frac{1}{d}, \;\forall i = 0 \dots d-1
\end{equation}
with $d=2^S$, and $S$ denotes the system size ($S = N_\mathrm{atoms}$ and $N=n_\mathrm{shots}$ in the main text).

The cost function we consider for a solution (bitstring) $i=(n^i_1, \dots, n^i_S)$ is its Hamming weight (sum of bits):

\begin{equation}
E_i(S) = \sum_{k = 1}^{S} n^i_k
\end{equation}

It corresponds to a simplification of the target energy of the main text that corresponds to taking $U=0$ (and inverting the sign). 
The approximation ratio is then $\alpha_i = E_i(S)/S$ as the maximum Hamming weight is $S$.

To study the expectation value of the maximum of $\alpha_i$ over $N$ repetitions,
we first notice that $E_i$ is a random variable with binomial distribution law, $E_i \sim \mathcal{B}(p=1/2, S)$.
For large enough $S$, it is well approximated by a Gaussian-distributed random variable, $E_i \approx E^G_i$, with $E^G_i \sim \mathcal{N}(\mu = S / 2, \sigma^2 = S / 4)$. The expectation value of the maximum of $N$ Gaussian samples obeys the inequality:

\begin{equation}
	\mathbb{E}\left [ \max_{p=1\dots N}\lbrace E^G_i(S; p) \rbrace \right ] \leq \mu(S) + \sigma(S) \sqrt{2\log{N}}
 \label{eq:exp_of_max_gaussians}
\end{equation}

(Indeed, using Jensen's inequality to obtain the first line, we have, $\forall \kappa > 0$:
\begin{align}
  e^{\kappa\mathbb{E}\left[M_{N}\right]}&\leq\mathbb{E}\left[e^{\kappa M_{N}}\right]=\mathbb{E}\left[\max_{p=1\dots N}e^{\kappa E_{p}^{G}}\right]\nonumber\\
  &\leq\sum_{p=1}^{N}\mathbb{E}\left[e^{\kappa E_{p}^{G}}\right]=Ne^{\frac{\kappa^{2}\sigma^{2}}{2}}\nonumber
\end{align}
Taking the logarithm of this expression and picking $\kappa=\frac{\sqrt{2\log(N)}}{\sigma}$ yields the $\sigma\sqrt{2\log(N)}$ upper bound)

Approximating $E_i \approx E^G_i$ in inequality \eqref{eq:exp_of_max_gaussians}, and dividing by $S$, we obtain the final approximate upper bound on the approximation ratio:
\begin{equation}
\alpha(N, S) \lesssim \frac{1}{2} + \sqrt{ \frac{\log{N}}{2 S}}.
\label{eq:approx_exp_of_max_bound}
\end{equation}

In Fig.~\ref{fig:apx_ratio_upper_bound}, we check numerically that $\alpha(N, S)$ follows the same $(N, S)$ dependence as its upper bound, i.e that $\alpha(N, S) \approx 1/2 + \beta \sqrt{\frac{\log{N}}{2 S}}$, motivating the fitting function that is used in the main text.

\bibliographystyle{apsrev4-1}
\bibliography{biblio}

\begin{thebibliography}{49}%
\makeatletter
\providecommand \@ifxundefined [1]{%
 \@ifx{#1\undefined}
}%
\providecommand \@ifnum [1]{%
 \ifnum #1\expandafter \@firstoftwo
 \else \expandafter \@secondoftwo
 \fi
}%
\providecommand \@ifx [1]{%
 \ifx #1\expandafter \@firstoftwo
 \else \expandafter \@secondoftwo
 \fi
}%
\providecommand \natexlab [1]{#1}%
\providecommand \enquote  [1]{``#1''}%
\providecommand \bibnamefont  [1]{#1}%
\providecommand \bibfnamefont [1]{#1}%
\providecommand \citenamefont [1]{#1}%
\providecommand \href@noop [0]{\@secondoftwo}%
\providecommand \href [0]{\begingroup \@sanitize@url \@href}%
\providecommand \@href[1]{\@@startlink{#1}\@@href}%
\providecommand \@@href[1]{\endgroup#1\@@endlink}%
\providecommand \@sanitize@url [0]{\catcode `\\12\catcode `\$12\catcode
  `\&12\catcode `\#12\catcode `\^12\catcode `\_12\catcode `\%12\relax}%
\providecommand \@@startlink[1]{}%
\providecommand \@@endlink[0]{}%
\providecommand \url  [0]{\begingroup\@sanitize@url \@url }%
\providecommand \@url [1]{\endgroup\@href {#1}{\urlprefix }}%
\providecommand \urlprefix  [0]{URL }%
\providecommand \Eprint [0]{\href }%
\providecommand \doibase [0]{http://dx.doi.org/}%
\providecommand \selectlanguage [0]{\@gobble}%
\providecommand \bibinfo  [0]{\@secondoftwo}%
\providecommand \bibfield  [0]{\@secondoftwo}%
\providecommand \translation [1]{[#1]}%
\providecommand \BibitemOpen [0]{}%
\providecommand \bibitemStop [0]{}%
\providecommand \bibitemNoStop [0]{.\EOS\space}%
\providecommand \EOS [0]{\spacefactor3000\relax}%
\providecommand \BibitemShut  [1]{\csname bibitem#1\endcsname}%
\let\auto@bib@innerbib\@empty
\bibitem [{\citenamefont {Monz}\ \emph {et~al.}(2011)\citenamefont {Monz},
  \citenamefont {Schindler}, \citenamefont {Barreiro}, \citenamefont {Chwalla},
  \citenamefont {Nigg}, \citenamefont {Coish}, \citenamefont {Harlander},
  \citenamefont {H{\"{a}}nsel}, \citenamefont {Hennrich},\ and\ \citenamefont
  {Blatt}}]{monz201114}%
  \BibitemOpen
  \bibfield  {author} {\bibinfo {author} {\bibfnamefont {T.}~\bibnamefont
  {Monz}}, \bibinfo {author} {\bibfnamefont {P.}~\bibnamefont {Schindler}},
  \bibinfo {author} {\bibfnamefont {J.~T.}\ \bibnamefont {Barreiro}}, \bibinfo
  {author} {\bibfnamefont {M.}~\bibnamefont {Chwalla}}, \bibinfo {author}
  {\bibfnamefont {D.}~\bibnamefont {Nigg}}, \bibinfo {author} {\bibfnamefont
  {W.~A.}\ \bibnamefont {Coish}}, \bibinfo {author} {\bibfnamefont
  {M.}~\bibnamefont {Harlander}}, \bibinfo {author} {\bibfnamefont
  {W.}~\bibnamefont {H{\"{a}}nsel}}, \bibinfo {author} {\bibfnamefont
  {M.}~\bibnamefont {Hennrich}}, \ and\ \bibinfo {author} {\bibfnamefont
  {R.}~\bibnamefont {Blatt}},\ }\href {\doibase 10.1103/PhysRevLett.106.130506}
  {\bibfield  {journal} {\bibinfo  {journal} {Physical Review Letters}\
  }\textbf {\bibinfo {volume} {106}},\ \bibinfo {pages} {130506} (\bibinfo
  {year} {2011})},\ \Eprint {http://arxiv.org/abs/1009.6126} {arXiv:1009.6126}
  \BibitemShut {NoStop}%
\bibitem [{\citenamefont {Arute}\ \emph {et~al.}(2019)\citenamefont {Arute},
  \citenamefont {Arya}, \citenamefont {Babbush}, \citenamefont {Bacon},
  \citenamefont {Bardin}, \citenamefont {Barends}, \citenamefont {Biswas},
  \citenamefont {Boixo}, \citenamefont {Brandao}, \citenamefont {Buell},
  \citenamefont {Burkett}, \citenamefont {Chen}, \citenamefont {Chen},
  \citenamefont {Chiaro}, \citenamefont {Collins}, \citenamefont {Courtney},
  \citenamefont {Dunsworth}, \citenamefont {Farhi}, \citenamefont {Foxen},
  \citenamefont {Fowler}, \citenamefont {Gidney}, \citenamefont {Giustina},
  \citenamefont {Graff}, \citenamefont {Guerin}, \citenamefont {Habegger},
  \citenamefont {Harrigan}, \citenamefont {Hartmann}, \citenamefont {Ho},
  \citenamefont {Hoffmann}, \citenamefont {Huang}, \citenamefont {Humble},
  \citenamefont {Isakov}, \citenamefont {Jeffrey}, \citenamefont {Jiang},
  \citenamefont {Kafri}, \citenamefont {Kechedzhi}, \citenamefont {Kelly},
  \citenamefont {Klimov}, \citenamefont {Knysh}, \citenamefont {Korotkov},
  \citenamefont {Kostritsa}, \citenamefont {Landhuis}, \citenamefont
  {Lindmark}, \citenamefont {Lucero}, \citenamefont {Lyakh}, \citenamefont
  {Mandr{\`{a}}}, \citenamefont {McClean}, \citenamefont {McEwen},
  \citenamefont {Megrant}, \citenamefont {Mi}, \citenamefont {Michielsen},
  \citenamefont {Mohseni}, \citenamefont {Mutus}, \citenamefont {Naaman},
  \citenamefont {Neeley}, \citenamefont {Neill}, \citenamefont {Niu},
  \citenamefont {Ostby}, \citenamefont {Petukhov}, \citenamefont {Platt},
  \citenamefont {Quintana}, \citenamefont {Rieffel}, \citenamefont {Roushan},
  \citenamefont {Rubin}, \citenamefont {Sank}, \citenamefont {Satzinger},
  \citenamefont {Smelyanskiy}, \citenamefont {Sung}, \citenamefont
  {Trevithick}, \citenamefont {Vainsencher}, \citenamefont {Villalonga},
  \citenamefont {White}, \citenamefont {Yao}, \citenamefont {Yeh},
  \citenamefont {Zalcman}, \citenamefont {Neven},\ and\ \citenamefont
  {Martinis}}]{arute2019quantum}%
  \BibitemOpen
  \bibfield  {author} {\bibinfo {author} {\bibfnamefont {F.}~\bibnamefont
  {Arute}}, \bibinfo {author} {\bibfnamefont {K.}~\bibnamefont {Arya}},
  \bibinfo {author} {\bibfnamefont {R.}~\bibnamefont {Babbush}}, \bibinfo
  {author} {\bibfnamefont {D.}~\bibnamefont {Bacon}}, \bibinfo {author}
  {\bibfnamefont {J.~C.}\ \bibnamefont {Bardin}}, \bibinfo {author}
  {\bibfnamefont {R.}~\bibnamefont {Barends}}, \bibinfo {author} {\bibfnamefont
  {R.}~\bibnamefont {Biswas}}, \bibinfo {author} {\bibfnamefont
  {S.}~\bibnamefont {Boixo}}, \bibinfo {author} {\bibfnamefont {F.~G. S.~L.}\
  \bibnamefont {Brandao}}, \bibinfo {author} {\bibfnamefont {D.~A.}\
  \bibnamefont {Buell}}, \bibinfo {author} {\bibfnamefont {B.}~\bibnamefont
  {Burkett}}, \bibinfo {author} {\bibfnamefont {Y.}~\bibnamefont {Chen}},
  \bibinfo {author} {\bibfnamefont {Z.}~\bibnamefont {Chen}}, \bibinfo {author}
  {\bibfnamefont {B.}~\bibnamefont {Chiaro}}, \bibinfo {author} {\bibfnamefont
  {R.}~\bibnamefont {Collins}}, \bibinfo {author} {\bibfnamefont
  {W.}~\bibnamefont {Courtney}}, \bibinfo {author} {\bibfnamefont
  {A.}~\bibnamefont {Dunsworth}}, \bibinfo {author} {\bibfnamefont
  {E.}~\bibnamefont {Farhi}}, \bibinfo {author} {\bibfnamefont
  {B.}~\bibnamefont {Foxen}}, \bibinfo {author} {\bibfnamefont
  {A.}~\bibnamefont {Fowler}}, \bibinfo {author} {\bibfnamefont
  {C.}~\bibnamefont {Gidney}}, \bibinfo {author} {\bibfnamefont
  {M.}~\bibnamefont {Giustina}}, \bibinfo {author} {\bibfnamefont
  {R.}~\bibnamefont {Graff}}, \bibinfo {author} {\bibfnamefont
  {K.}~\bibnamefont {Guerin}}, \bibinfo {author} {\bibfnamefont
  {S.}~\bibnamefont {Habegger}}, \bibinfo {author} {\bibfnamefont {M.~P.}\
  \bibnamefont {Harrigan}}, \bibinfo {author} {\bibfnamefont {M.~J.}\
  \bibnamefont {Hartmann}}, \bibinfo {author} {\bibfnamefont {A.}~\bibnamefont
  {Ho}}, \bibinfo {author} {\bibfnamefont {M.}~\bibnamefont {Hoffmann}},
  \bibinfo {author} {\bibfnamefont {T.}~\bibnamefont {Huang}}, \bibinfo
  {author} {\bibfnamefont {T.~S.}\ \bibnamefont {Humble}}, \bibinfo {author}
  {\bibfnamefont {S.~V.}\ \bibnamefont {Isakov}}, \bibinfo {author}
  {\bibfnamefont {E.}~\bibnamefont {Jeffrey}}, \bibinfo {author} {\bibfnamefont
  {Z.}~\bibnamefont {Jiang}}, \bibinfo {author} {\bibfnamefont
  {D.}~\bibnamefont {Kafri}}, \bibinfo {author} {\bibfnamefont
  {K.}~\bibnamefont {Kechedzhi}}, \bibinfo {author} {\bibfnamefont
  {J.}~\bibnamefont {Kelly}}, \bibinfo {author} {\bibfnamefont {P.~V.}\
  \bibnamefont {Klimov}}, \bibinfo {author} {\bibfnamefont {S.}~\bibnamefont
  {Knysh}}, \bibinfo {author} {\bibfnamefont {A.}~\bibnamefont {Korotkov}},
  \bibinfo {author} {\bibfnamefont {F.}~\bibnamefont {Kostritsa}}, \bibinfo
  {author} {\bibfnamefont {D.}~\bibnamefont {Landhuis}}, \bibinfo {author}
  {\bibfnamefont {M.}~\bibnamefont {Lindmark}}, \bibinfo {author}
  {\bibfnamefont {E.}~\bibnamefont {Lucero}}, \bibinfo {author} {\bibfnamefont
  {D.}~\bibnamefont {Lyakh}}, \bibinfo {author} {\bibfnamefont
  {S.}~\bibnamefont {Mandr{\`{a}}}}, \bibinfo {author} {\bibfnamefont {J.~R.}\
  \bibnamefont {McClean}}, \bibinfo {author} {\bibfnamefont {M.}~\bibnamefont
  {McEwen}}, \bibinfo {author} {\bibfnamefont {A.}~\bibnamefont {Megrant}},
  \bibinfo {author} {\bibfnamefont {X.}~\bibnamefont {Mi}}, \bibinfo {author}
  {\bibfnamefont {K.}~\bibnamefont {Michielsen}}, \bibinfo {author}
  {\bibfnamefont {M.}~\bibnamefont {Mohseni}}, \bibinfo {author} {\bibfnamefont
  {J.}~\bibnamefont {Mutus}}, \bibinfo {author} {\bibfnamefont
  {O.}~\bibnamefont {Naaman}}, \bibinfo {author} {\bibfnamefont
  {M.}~\bibnamefont {Neeley}}, \bibinfo {author} {\bibfnamefont
  {C.}~\bibnamefont {Neill}}, \bibinfo {author} {\bibfnamefont {M.~Y.}\
  \bibnamefont {Niu}}, \bibinfo {author} {\bibfnamefont {E.}~\bibnamefont
  {Ostby}}, \bibinfo {author} {\bibfnamefont {A.}~\bibnamefont {Petukhov}},
  \bibinfo {author} {\bibfnamefont {J.~C.}\ \bibnamefont {Platt}}, \bibinfo
  {author} {\bibfnamefont {C.}~\bibnamefont {Quintana}}, \bibinfo {author}
  {\bibfnamefont {E.~G.}\ \bibnamefont {Rieffel}}, \bibinfo {author}
  {\bibfnamefont {P.}~\bibnamefont {Roushan}}, \bibinfo {author} {\bibfnamefont
  {N.~C.}\ \bibnamefont {Rubin}}, \bibinfo {author} {\bibfnamefont
  {D.}~\bibnamefont {Sank}}, \bibinfo {author} {\bibfnamefont {K.~J.}\
  \bibnamefont {Satzinger}}, \bibinfo {author} {\bibfnamefont {V.}~\bibnamefont
  {Smelyanskiy}}, \bibinfo {author} {\bibfnamefont {K.~J.}\ \bibnamefont
  {Sung}}, \bibinfo {author} {\bibfnamefont {M.~D.}\ \bibnamefont
  {Trevithick}}, \bibinfo {author} {\bibfnamefont {A.}~\bibnamefont
  {Vainsencher}}, \bibinfo {author} {\bibfnamefont {B.}~\bibnamefont
  {Villalonga}}, \bibinfo {author} {\bibfnamefont {T.}~\bibnamefont {White}},
  \bibinfo {author} {\bibfnamefont {Z.~J.}\ \bibnamefont {Yao}}, \bibinfo
  {author} {\bibfnamefont {P.}~\bibnamefont {Yeh}}, \bibinfo {author}
  {\bibfnamefont {A.}~\bibnamefont {Zalcman}}, \bibinfo {author} {\bibfnamefont
  {H.}~\bibnamefont {Neven}}, \ and\ \bibinfo {author} {\bibfnamefont {J.~M.}\
  \bibnamefont {Martinis}},\ }\href {\doibase 10.1038/s41586-019-1666-5}
  {\bibfield  {journal} {\bibinfo  {journal} {Nature}\ }\textbf {\bibinfo
  {volume} {574}},\ \bibinfo {pages} {505} (\bibinfo {year} {2019})},\ \Eprint
  {http://arxiv.org/abs/1911.00577} {arXiv:1911.00577} \BibitemShut {NoStop}%
\bibitem [{\citenamefont {Preskill}(2018)}]{Preskill2018}%
  \BibitemOpen
  \bibfield  {author} {\bibinfo {author} {\bibfnamefont {J.}~\bibnamefont
  {Preskill}},\ }\href {\doibase 10.22331/q-2018-08-06-79} {\bibfield
  {journal} {\bibinfo  {journal} {Quantum}\ }\textbf {\bibinfo {volume} {2}},\
  \bibinfo {pages} {79} (\bibinfo {year} {2018})},\ \Eprint
  {http://arxiv.org/abs/1801.00862} {arXiv:1801.00862} \BibitemShut {NoStop}%
\bibitem [{\citenamefont {Grumbling}\ and\ \citenamefont
  {Horowitz}(2019)}]{national2019quantum}%
  \BibitemOpen
  \bibinfo {editor} {\bibfnamefont {E.}~\bibnamefont {Grumbling}}\ and\
  \bibinfo {editor} {\bibfnamefont {M.}~\bibnamefont {Horowitz}},\ eds.,\ \href
  {\doibase 10.17226/25196} {\emph {\bibinfo {title} {{Quantum Computing:
  Progress and Prospects}}}},\ Vol.\ \bibinfo {volume} {9781461418}\ (\bibinfo
  {publisher} {National Academies Press},\ \bibinfo {address} {Washington,
  D.C.},\ \bibinfo {year} {2019})\ pp.\ \bibinfo {pages}
  {2388--2405}\BibitemShut {NoStop}%
\bibitem [{\citenamefont {Lucas}(2014)}]{Lucas2014}%
  \BibitemOpen
  \bibfield  {author} {\bibinfo {author} {\bibfnamefont {A.}~\bibnamefont
  {Lucas}},\ }\href {\doibase 10.3389/fphy.2014.00005} {\bibfield  {journal}
  {\bibinfo  {journal} {Frontiers in Physics}\ }\textbf {\bibinfo {volume}
  {2}},\ \bibinfo {pages} {1} (\bibinfo {year} {2014})},\ \Eprint
  {http://arxiv.org/abs/1302.5843} {arXiv:1302.5843} \BibitemShut {NoStop}%
\bibitem [{\citenamefont {Otterbach}\ \emph {et~al.}(2017)\citenamefont
  {Otterbach}, \citenamefont {Manenti}, \citenamefont {Alidoust}, \citenamefont
  {Block}, \citenamefont {Bloom}, \citenamefont {Caldwell}, \citenamefont
  {Didier}, \citenamefont {Fried}, \citenamefont {Hong}, \citenamefont
  {Karale~kas}, \citenamefont {Osborn}, \citenamefont {Papageorge},
  \citenamefont {Peterson}, \citenamefont {Prawiroatmodjo}, \citenamefont
  {Rubin}, \citenamefont {Ryan}, \citenamefont {Scheer}, \citenamefont {Sete},
  \citenamefont {Sivarajah}, \citenamefont {Smith}, \citenamefont {Staley},
  \citenamefont {Tezak}, \citenamefont {Ze~ng}, \citenamefont {Hudson},
  \citenamefont {Johnson}, \citenamefont {Reagor}, \citenamefont {da~Silva},\
  and\ \citenamefont {Rigetti}}]{Otterbach2017}%
  \BibitemOpen
  \bibfield  {author} {\bibinfo {author} {\bibfnamefont {J.~S.}\ \bibnamefont
  {Otterbach}}, \bibinfo {author} {\bibfnamefont {R.}~\bibnamefont {Manenti}},
  \bibinfo {author} {\bibfnamefont {A.}~\bibnamefont {Alidoust}, \bibfnamefont
  {N.~a nd~Bestwick}}, \bibinfo {author} {\bibfnamefont {M.}~\bibnamefont
  {Block}}, \bibinfo {author} {\bibfnamefont {B.}~\bibnamefont {Bloom}},
  \bibinfo {author} {\bibfnamefont {S.}~\bibnamefont {Caldwell}}, \bibinfo
  {author} {\bibfnamefont {N.}~\bibnamefont {Didier}}, \bibinfo {author}
  {\bibfnamefont {E.~S.}\ \bibnamefont {Fried}}, \bibinfo {author}
  {\bibfnamefont {S.}~\bibnamefont {Hong}}, \bibinfo {author} {\bibfnamefont
  {P.}~\bibnamefont {Karale~kas}}, \bibinfo {author} {\bibfnamefont {C.~B.}\
  \bibnamefont {Osborn}}, \bibinfo {author} {\bibfnamefont {A.}~\bibnamefont
  {Papageorge}}, \bibinfo {author} {\bibfnamefont {E.~C.}\ \bibnamefont
  {Peterson}}, \bibinfo {author} {\bibfnamefont {G.}~\bibnamefont
  {Prawiroatmodjo}}, \bibinfo {author} {\bibfnamefont {N.}~\bibnamefont
  {Rubin}}, \bibinfo {author} {\bibfnamefont {D.}~\bibnamefont {Ryan},
  \bibfnamefont {Colm A. an d~Scarabelli}}, \bibinfo {author} {\bibfnamefont
  {M.}~\bibnamefont {Scheer}}, \bibinfo {author} {\bibfnamefont {E.~A.}\
  \bibnamefont {Sete}}, \bibinfo {author} {\bibfnamefont {P.}~\bibnamefont
  {Sivarajah}}, \bibinfo {author} {\bibfnamefont {R.~S.}\ \bibnamefont
  {Smith}}, \bibinfo {author} {\bibfnamefont {A.}~\bibnamefont {Staley}},
  \bibinfo {author} {\bibfnamefont {N.}~\bibnamefont {Tezak}}, \bibinfo
  {author} {\bibfnamefont {W.~J.}\ \bibnamefont {Ze~ng}}, \bibinfo {author}
  {\bibfnamefont {A.}~\bibnamefont {Hudson}}, \bibinfo {author} {\bibfnamefont
  {B.~R.}\ \bibnamefont {Johnson}}, \bibinfo {author} {\bibfnamefont
  {M.}~\bibnamefont {Reagor}}, \bibinfo {author} {\bibfnamefont {M.~P.}\
  \bibnamefont {da~Silva}}, \ and\ \bibinfo {author} {\bibfnamefont
  {C.}~\bibnamefont {Rigetti}},\ }\href {http://arxiv.org/abs/1712.05771} {\
  (\bibinfo {year} {2017})},\ \Eprint {http://arxiv.org/abs/1712.05771}
  {arXiv:1712.05771} \BibitemShut {NoStop}%
\bibitem [{\citenamefont {Guerreschi}\ and\ \citenamefont
  {Matsuura}(2019)}]{Guerreschi2019}%
  \BibitemOpen
  \bibfield  {author} {\bibinfo {author} {\bibfnamefont {G.~G.}\ \bibnamefont
  {Guerreschi}}\ and\ \bibinfo {author} {\bibfnamefont {A.~Y.}\ \bibnamefont
  {Matsuura}},\ }\href {\doibase 10.1038/s41598-019-43176-9} {\bibfield
  {journal} {\bibinfo  {journal} {Scientific Reports}\ }\textbf {\bibinfo
  {volume} {9}},\ \bibinfo {pages} {6903} (\bibinfo {year} {2019})},\ \Eprint
  {http://arxiv.org/abs/1812.07589} {arXiv:1812.07589} \BibitemShut {NoStop}%
\bibitem [{\citenamefont {Pichler}\ \emph
  {et~al.}(2018{\natexlab{a}})\citenamefont {Pichler}, \citenamefont {Wang},
  \citenamefont {Zhou}, \citenamefont {Choi},\ and\ \citenamefont
  {Lukin}}]{pichler2018quantum}%
  \BibitemOpen
  \bibfield  {author} {\bibinfo {author} {\bibfnamefont {H.}~\bibnamefont
  {Pichler}}, \bibinfo {author} {\bibfnamefont {S.-t.}\ \bibnamefont {Wang}},
  \bibinfo {author} {\bibfnamefont {L.}~\bibnamefont {Zhou}}, \bibinfo {author}
  {\bibfnamefont {S.}~\bibnamefont {Choi}}, \ and\ \bibinfo {author}
  {\bibfnamefont {M.~D.}\ \bibnamefont {Lukin}},\ }\href
  {http://arxiv.org/abs/1808.10816} {\  (\bibinfo {year}
  {2018}{\natexlab{a}})},\ \Eprint {http://arxiv.org/abs/1808.10816}
  {arXiv:1808.10816} \BibitemShut {NoStop}%
\bibitem [{\citenamefont {Hauke}\ \emph {et~al.}(2019)\citenamefont {Hauke},
  \citenamefont {Katzgraber}, \citenamefont {Lechner}, \citenamefont
  {Nishimori},\ and\ \citenamefont {Oliver}}]{hauke2019perspectives}%
  \BibitemOpen
  \bibfield  {author} {\bibinfo {author} {\bibfnamefont {P.}~\bibnamefont
  {Hauke}}, \bibinfo {author} {\bibfnamefont {H.~G.}\ \bibnamefont
  {Katzgraber}}, \bibinfo {author} {\bibfnamefont {W.}~\bibnamefont {Lechner}},
  \bibinfo {author} {\bibfnamefont {H.}~\bibnamefont {Nishimori}}, \ and\
  \bibinfo {author} {\bibfnamefont {W.~D.}\ \bibnamefont {Oliver}},\
  }\href@noop {} {\bibfield  {journal} {\bibinfo  {journal} {arXiv preprint
  arXiv:1903.06559}\ } (\bibinfo {year} {2019})}\BibitemShut {NoStop}%
\bibitem [{\citenamefont {Albash}\ and\ \citenamefont
  {Lidar}(2018)}]{albash2018adiabatic}%
  \BibitemOpen
  \bibfield  {author} {\bibinfo {author} {\bibfnamefont {T.}~\bibnamefont
  {Albash}}\ and\ \bibinfo {author} {\bibfnamefont {D.~A.}\ \bibnamefont
  {Lidar}},\ }\href {\doibase 10.1103/RevModPhys.90.015002} {\bibfield
  {journal} {\bibinfo  {journal} {Reviews of Modern Physics}\ }\textbf
  {\bibinfo {volume} {90}},\ \bibinfo {pages} {015002} (\bibinfo {year}
  {2018})},\ \Eprint {http://arxiv.org/abs/0502014} {arXiv:0502014 [quant-ph]}
  \BibitemShut {NoStop}%
\bibitem [{\citenamefont {Kokail}\ \emph {et~al.}(2019)\citenamefont {Kokail},
  \citenamefont {Maier}, \citenamefont {van Bijnen}, \citenamefont {Brydges},
  \citenamefont {Joshi}, \citenamefont {Jurcevic}, \citenamefont {Muschik},
  \citenamefont {Silvi}, \citenamefont {Blatt}, \citenamefont {Roos},\ and\
  \citenamefont {Zoller}}]{Kokail2018}%
  \BibitemOpen
  \bibfield  {author} {\bibinfo {author} {\bibfnamefont {C.}~\bibnamefont
  {Kokail}}, \bibinfo {author} {\bibfnamefont {C.}~\bibnamefont {Maier}},
  \bibinfo {author} {\bibfnamefont {R.}~\bibnamefont {van Bijnen}}, \bibinfo
  {author} {\bibfnamefont {T.}~\bibnamefont {Brydges}}, \bibinfo {author}
  {\bibfnamefont {M.~K.}\ \bibnamefont {Joshi}}, \bibinfo {author}
  {\bibfnamefont {P.}~\bibnamefont {Jurcevic}}, \bibinfo {author}
  {\bibfnamefont {C.~A.}\ \bibnamefont {Muschik}}, \bibinfo {author}
  {\bibfnamefont {P.}~\bibnamefont {Silvi}}, \bibinfo {author} {\bibfnamefont
  {R.}~\bibnamefont {Blatt}}, \bibinfo {author} {\bibfnamefont {C.~F.}\
  \bibnamefont {Roos}}, \ and\ \bibinfo {author} {\bibfnamefont
  {P.}~\bibnamefont {Zoller}},\ }\href {\doibase 10.1038/s41586-019-1177-4}
  {\bibfield  {journal} {\bibinfo  {journal} {Nature}\ }\textbf {\bibinfo
  {volume} {569}},\ \bibinfo {pages} {355} (\bibinfo {year} {2019})},\ \Eprint
  {http://arxiv.org/abs/1810.03421} {arXiv:1810.03421} \BibitemShut {NoStop}%
\bibitem [{\citenamefont {Farhi}\ \emph {et~al.}(2014)\citenamefont {Farhi},
  \citenamefont {Goldstone},\ and\ \citenamefont {Gutmann}}]{Farhi2014}%
  \BibitemOpen
  \bibfield  {author} {\bibinfo {author} {\bibfnamefont {E.}~\bibnamefont
  {Farhi}}, \bibinfo {author} {\bibfnamefont {J.}~\bibnamefont {Goldstone}}, \
  and\ \bibinfo {author} {\bibfnamefont {S.}~\bibnamefont {Gutmann}},\ }\href
  {http://arxiv.org/abs/1411.4028} {\  (\bibinfo {year} {2014})},\ \Eprint
  {http://arxiv.org/abs/1411.4028} {arXiv:1411.4028} \BibitemShut {NoStop}%
\bibitem [{\citenamefont {Vazirani}(2013)}]{vazirani2013approximation}%
  \BibitemOpen
  \bibfield  {author} {\bibinfo {author} {\bibfnamefont {V.~V.}\ \bibnamefont
  {Vazirani}},\ }\href@noop {} {\emph {\bibinfo {title} {Approximation
  algorithms}}}\ (\bibinfo  {publisher} {Springer Science \& Business Media},\
  \bibinfo {year} {2013})\BibitemShut {NoStop}%
\bibitem [{\citenamefont {Browaeys}\ and\ \citenamefont
  {Lahaye}(2020)}]{Browaeys2020}%
  \BibitemOpen
  \bibfield  {author} {\bibinfo {author} {\bibfnamefont {A.}~\bibnamefont
  {Browaeys}}\ and\ \bibinfo {author} {\bibfnamefont {T.}~\bibnamefont
  {Lahaye}},\ }\href {\doibase 10.1038/s41567-019-0733-z} {\bibfield  {journal}
  {\bibinfo  {journal} {Nature Physics}\ }\textbf {\bibinfo {volume} {16}},\
  \bibinfo {pages} {132} (\bibinfo {year} {2020})},\ \Eprint
  {http://arxiv.org/abs/2002.07413} {arXiv:2002.07413} \BibitemShut {NoStop}%
\bibitem [{\citenamefont {Henriet}(2020)}]{Henriet2020}%
  \BibitemOpen
  \bibfield  {author} {\bibinfo {author} {\bibfnamefont {L.}~\bibnamefont
  {Henriet}},\ }\href {\doibase 10.1103/PhysRevA.101.012335} {\bibfield
  {journal} {\bibinfo  {journal} {Physical Review A}\ }\textbf {\bibinfo
  {volume} {101}},\ \bibinfo {pages} {012335} (\bibinfo {year} {2020})},\
  \Eprint {http://arxiv.org/abs/1910.10442} {arXiv:1910.10442} \BibitemShut
  {NoStop}%
\bibitem [{\citenamefont {Levine}\ \emph {et~al.}(2018)\citenamefont {Levine},
  \citenamefont {Keesling}, \citenamefont {Omran}, \citenamefont {Bernien},
  \citenamefont {Schwartz}, \citenamefont {Zibrov}, \citenamefont {Endres},
  \citenamefont {Greiner}, \citenamefont {Vuleti{\'{c}}},\ and\ \citenamefont
  {Lukin}}]{Levine2018}%
  \BibitemOpen
  \bibfield  {author} {\bibinfo {author} {\bibfnamefont {H.}~\bibnamefont
  {Levine}}, \bibinfo {author} {\bibfnamefont {A.}~\bibnamefont {Keesling}},
  \bibinfo {author} {\bibfnamefont {A.}~\bibnamefont {Omran}}, \bibinfo
  {author} {\bibfnamefont {H.}~\bibnamefont {Bernien}}, \bibinfo {author}
  {\bibfnamefont {S.}~\bibnamefont {Schwartz}}, \bibinfo {author}
  {\bibfnamefont {A.~S.}\ \bibnamefont {Zibrov}}, \bibinfo {author}
  {\bibfnamefont {M.}~\bibnamefont {Endres}}, \bibinfo {author} {\bibfnamefont
  {M.}~\bibnamefont {Greiner}}, \bibinfo {author} {\bibfnamefont
  {V.}~\bibnamefont {Vuleti{\'{c}}}}, \ and\ \bibinfo {author} {\bibfnamefont
  {M.~D.}\ \bibnamefont {Lukin}},\ }\href {\doibase
  10.1103/PhysRevLett.121.123603} {\bibfield  {journal} {\bibinfo  {journal}
  {Physical Review Letters}\ }\textbf {\bibinfo {volume} {121}},\ \bibinfo
  {pages} {123603} (\bibinfo {year} {2018})},\ \Eprint
  {http://arxiv.org/abs/1806.04682} {arXiv:1806.04682} \BibitemShut {NoStop}%
\bibitem [{\citenamefont {Madjarov}\ \emph {et~al.}(2020)\citenamefont
  {Madjarov}, \citenamefont {Covey}, \citenamefont {Shaw}, \citenamefont
  {Choi}, \citenamefont {Kale}, \citenamefont {Cooper}, \citenamefont
  {Pichler}, \citenamefont {Schkolnik}, \citenamefont {Williams},\ and\
  \citenamefont {Endres}}]{Madjarov2020}%
  \BibitemOpen
  \bibfield  {author} {\bibinfo {author} {\bibfnamefont {I.~S.}\ \bibnamefont
  {Madjarov}}, \bibinfo {author} {\bibfnamefont {J.~P.}\ \bibnamefont {Covey}},
  \bibinfo {author} {\bibfnamefont {A.~L.}\ \bibnamefont {Shaw}}, \bibinfo
  {author} {\bibfnamefont {J.}~\bibnamefont {Choi}}, \bibinfo {author}
  {\bibfnamefont {A.}~\bibnamefont {Kale}}, \bibinfo {author} {\bibfnamefont
  {A.}~\bibnamefont {Cooper}}, \bibinfo {author} {\bibfnamefont
  {H.}~\bibnamefont {Pichler}}, \bibinfo {author} {\bibfnamefont
  {V.}~\bibnamefont {Schkolnik}}, \bibinfo {author} {\bibfnamefont {J.~R.}\
  \bibnamefont {Williams}}, \ and\ \bibinfo {author} {\bibfnamefont
  {M.}~\bibnamefont {Endres}},\ }\href {http://arxiv.org/abs/2001.04455} {\
  \textbf {\bibinfo {volume} {1}},\ \bibinfo {pages} {1} (\bibinfo {year}
  {2020})},\ \Eprint {http://arxiv.org/abs/2001.04455} {arXiv:2001.04455}
  \BibitemShut {NoStop}%
\bibitem [{\citenamefont {Kjaergaard}\ \emph {et~al.}(2019)\citenamefont
  {Kjaergaard}, \citenamefont {Schwartz}, \citenamefont {Braum{\"{u}}ller},
  \citenamefont {Krantz}, \citenamefont {Wang}, \citenamefont {Gustavsson},\
  and\ \citenamefont {Oliver}}]{Kjaergaard2020}%
  \BibitemOpen
  \bibfield  {author} {\bibinfo {author} {\bibfnamefont {M.}~\bibnamefont
  {Kjaergaard}}, \bibinfo {author} {\bibfnamefont {M.~E.}\ \bibnamefont
  {Schwartz}}, \bibinfo {author} {\bibfnamefont {J.}~\bibnamefont
  {Braum{\"{u}}ller}}, \bibinfo {author} {\bibfnamefont {P.}~\bibnamefont
  {Krantz}}, \bibinfo {author} {\bibfnamefont {J.~I.-J.}\ \bibnamefont {Wang}},
  \bibinfo {author} {\bibfnamefont {S.}~\bibnamefont {Gustavsson}}, \ and\
  \bibinfo {author} {\bibfnamefont {W.~D.}\ \bibnamefont {Oliver}},\ }\href
  {\doibase 10.1146/annurev-conmatphys-031119-050605} {\bibfield  {journal}
  {\bibinfo  {journal} {Annual Review of Condensed Matter Physics}\ }\textbf
  {\bibinfo {volume} {11}},\ \bibinfo {pages} {031119} (\bibinfo {year}
  {2019})},\ \Eprint {http://arxiv.org/abs/1905.13641} {arXiv:1905.13641}
  \BibitemShut {NoStop}%
\bibitem [{\citenamefont {Bruzewicz}\ \emph {et~al.}(2019)\citenamefont
  {Bruzewicz}, \citenamefont {Chiaverini}, \citenamefont {McConnell},\ and\
  \citenamefont {Sage}}]{Bruzewicz2019}%
  \BibitemOpen
  \bibfield  {author} {\bibinfo {author} {\bibfnamefont {C.~D.}\ \bibnamefont
  {Bruzewicz}}, \bibinfo {author} {\bibfnamefont {J.}~\bibnamefont
  {Chiaverini}}, \bibinfo {author} {\bibfnamefont {R.}~\bibnamefont
  {McConnell}}, \ and\ \bibinfo {author} {\bibfnamefont {J.~M.}\ \bibnamefont
  {Sage}},\ }\href {\doibase 10.1063/1.5088164} {\bibfield  {journal} {\bibinfo
   {journal} {Applied Physics Reviews}\ }\textbf {\bibinfo {volume} {6}},\
  \bibinfo {pages} {021314} (\bibinfo {year} {2019})},\ \Eprint
  {http://arxiv.org/abs/1904.04178} {arXiv:1904.04178} \BibitemShut {NoStop}%
\bibitem [{\citenamefont {Barredo}\ \emph {et~al.}(2018)\citenamefont
  {Barredo}, \citenamefont {Lienhard}, \citenamefont {de~L{\'{e}}s{\'{e}}leuc},
  \citenamefont {Lahaye},\ and\ \citenamefont
  {Browaeys}}]{barredo2018synthetic}%
  \BibitemOpen
  \bibfield  {author} {\bibinfo {author} {\bibfnamefont {D.}~\bibnamefont
  {Barredo}}, \bibinfo {author} {\bibfnamefont {V.}~\bibnamefont {Lienhard}},
  \bibinfo {author} {\bibfnamefont {S.}~\bibnamefont
  {de~L{\'{e}}s{\'{e}}leuc}}, \bibinfo {author} {\bibfnamefont
  {T.}~\bibnamefont {Lahaye}}, \ and\ \bibinfo {author} {\bibfnamefont
  {A.}~\bibnamefont {Browaeys}},\ }\href {\doibase 10.1038/s41586-018-0450-2}
  {\bibfield  {journal} {\bibinfo  {journal} {Nature}\ }\textbf {\bibinfo
  {volume} {561}},\ \bibinfo {pages} {79} (\bibinfo {year} {2018})},\ \Eprint
  {http://arxiv.org/abs/1712.02727} {arXiv:1712.02727} \BibitemShut {NoStop}%
\bibitem [{\citenamefont {{Ohl de Mello}}\ \emph {et~al.}(2019)\citenamefont
  {{Ohl de Mello}}, \citenamefont {Sch{\"{a}}ffner}, \citenamefont {Werkmann},
  \citenamefont {Preuschoff}, \citenamefont {Kohfahl}, \citenamefont
  {Schlosser},\ and\ \citenamefont {Birkl}}]{OhlDeMello2019}%
  \BibitemOpen
  \bibfield  {author} {\bibinfo {author} {\bibfnamefont {D.}~\bibnamefont {{Ohl
  de Mello}}}, \bibinfo {author} {\bibfnamefont {D.}~\bibnamefont
  {Sch{\"{a}}ffner}}, \bibinfo {author} {\bibfnamefont {J.}~\bibnamefont
  {Werkmann}}, \bibinfo {author} {\bibfnamefont {T.}~\bibnamefont
  {Preuschoff}}, \bibinfo {author} {\bibfnamefont {L.}~\bibnamefont {Kohfahl}},
  \bibinfo {author} {\bibfnamefont {M.}~\bibnamefont {Schlosser}}, \ and\
  \bibinfo {author} {\bibfnamefont {G.}~\bibnamefont {Birkl}},\ }\href
  {\doibase 10.1103/PhysRevLett.122.203601} {\bibfield  {journal} {\bibinfo
  {journal} {Physical Review Letters}\ }\textbf {\bibinfo {volume} {122}},\
  \bibinfo {pages} {203601} (\bibinfo {year} {2019})},\ \Eprint
  {http://arxiv.org/abs/arXiv:1902.00284v4} {arXiv:arXiv:1902.00284v4}
  \BibitemShut {NoStop}%
\bibitem [{\citenamefont {Lienhard}\ \emph {et~al.}(2018)\citenamefont
  {Lienhard}, \citenamefont {de~L{\'{e}}s{\'{e}}leuc}, \citenamefont {Barredo},
  \citenamefont {Lahaye}, \citenamefont {Browaeys}, \citenamefont {Schuler},
  \citenamefont {Henry},\ and\ \citenamefont
  {L{\"{a}}uchli}}]{lienhard2018observing}%
  \BibitemOpen
  \bibfield  {author} {\bibinfo {author} {\bibfnamefont {V.}~\bibnamefont
  {Lienhard}}, \bibinfo {author} {\bibfnamefont {S.}~\bibnamefont
  {de~L{\'{e}}s{\'{e}}leuc}}, \bibinfo {author} {\bibfnamefont
  {D.}~\bibnamefont {Barredo}}, \bibinfo {author} {\bibfnamefont
  {T.}~\bibnamefont {Lahaye}}, \bibinfo {author} {\bibfnamefont
  {A.}~\bibnamefont {Browaeys}}, \bibinfo {author} {\bibfnamefont
  {M.}~\bibnamefont {Schuler}}, \bibinfo {author} {\bibfnamefont {L.-p.}\
  \bibnamefont {Henry}}, \ and\ \bibinfo {author} {\bibfnamefont {A.~M.}\
  \bibnamefont {L{\"{a}}uchli}},\ }\href {\doibase 10.1103/PhysRevX.8.021070}
  {\bibfield  {journal} {\bibinfo  {journal} {Physical Review X}\ }\textbf
  {\bibinfo {volume} {8}},\ \bibinfo {pages} {021070} (\bibinfo {year}
  {2018})},\ \Eprint {http://arxiv.org/abs/1711.01185} {arXiv:1711.01185}
  \BibitemShut {NoStop}%
\bibitem [{\citenamefont {Barredo}\ \emph {et~al.}(2016)\citenamefont
  {Barredo}, \citenamefont {De~L{\'e}s{\'e}leuc}, \citenamefont {Lienhard},
  \citenamefont {Lahaye},\ and\ \citenamefont {Browaeys}}]{barredo2016atom}%
  \BibitemOpen
  \bibfield  {author} {\bibinfo {author} {\bibfnamefont {D.}~\bibnamefont
  {Barredo}}, \bibinfo {author} {\bibfnamefont {S.}~\bibnamefont
  {De~L{\'e}s{\'e}leuc}}, \bibinfo {author} {\bibfnamefont {V.}~\bibnamefont
  {Lienhard}}, \bibinfo {author} {\bibfnamefont {T.}~\bibnamefont {Lahaye}}, \
  and\ \bibinfo {author} {\bibfnamefont {A.}~\bibnamefont {Browaeys}},\
  }\href@noop {} {\bibfield  {journal} {\bibinfo  {journal} {Science}\ }\textbf
  {\bibinfo {volume} {354}},\ \bibinfo {pages} {1021} (\bibinfo {year}
  {2016})}\BibitemShut {NoStop}%
\bibitem [{\citenamefont {Labuhn}\ \emph {et~al.}(2016)\citenamefont {Labuhn},
  \citenamefont {Barredo}, \citenamefont {Ravets}, \citenamefont
  {de~L{\'{e}}s{\'{e}}leuc}, \citenamefont {Macr{\`{i}}}, \citenamefont
  {Lahaye},\ and\ \citenamefont {Browaeys}}]{labuhn2016tunable}%
  \BibitemOpen
  \bibfield  {author} {\bibinfo {author} {\bibfnamefont {H.}~\bibnamefont
  {Labuhn}}, \bibinfo {author} {\bibfnamefont {D.}~\bibnamefont {Barredo}},
  \bibinfo {author} {\bibfnamefont {S.}~\bibnamefont {Ravets}}, \bibinfo
  {author} {\bibfnamefont {S.}~\bibnamefont {de~L{\'{e}}s{\'{e}}leuc}},
  \bibinfo {author} {\bibfnamefont {T.}~\bibnamefont {Macr{\`{i}}}}, \bibinfo
  {author} {\bibfnamefont {T.}~\bibnamefont {Lahaye}}, \ and\ \bibinfo {author}
  {\bibfnamefont {A.}~\bibnamefont {Browaeys}},\ }\href {\doibase
  10.1038/nature18274} {\bibfield  {journal} {\bibinfo  {journal} {Nature}\
  }\textbf {\bibinfo {volume} {534}},\ \bibinfo {pages} {667} (\bibinfo {year}
  {2016})},\ \Eprint {http://arxiv.org/abs/1509.04543} {arXiv:1509.04543}
  \BibitemShut {NoStop}%
\bibitem [{\citenamefont {Matsui}(1998)}]{matsui1998approximation}%
  \BibitemOpen
  \bibfield  {author} {\bibinfo {author} {\bibfnamefont {T.}~\bibnamefont
  {Matsui}},\ }in\ \href@noop {} {\emph {\bibinfo {booktitle} {Japanese
  Conference on Discrete and Computational Geometry}}}\ (\bibinfo
  {organization} {Springer},\ \bibinfo {year} {1998})\ pp.\ \bibinfo {pages}
  {194--200}\BibitemShut {NoStop}%
\bibitem [{\citenamefont {Nieberg}\ \emph {et~al.}(2004)\citenamefont
  {Nieberg}, \citenamefont {Hurink},\ and\ \citenamefont
  {Kern}}]{nieberg2004robust}%
  \BibitemOpen
  \bibfield  {author} {\bibinfo {author} {\bibfnamefont {T.}~\bibnamefont
  {Nieberg}}, \bibinfo {author} {\bibfnamefont {J.}~\bibnamefont {Hurink}}, \
  and\ \bibinfo {author} {\bibfnamefont {W.}~\bibnamefont {Kern}},\ }in\
  \href@noop {} {\emph {\bibinfo {booktitle} {International Workshop on
  Graph-Theoretic Concepts in Computer Science}}}\ (\bibinfo {organization}
  {Springer},\ \bibinfo {year} {2004})\ pp.\ \bibinfo {pages}
  {214--221}\BibitemShut {NoStop}%
\bibitem [{\citenamefont {Das}\ \emph {et~al.}(2018)\citenamefont {Das},
  \citenamefont {da~Fonseca},\ and\ \citenamefont {Jallu}}]{das2018efficient}%
  \BibitemOpen
  \bibfield  {author} {\bibinfo {author} {\bibfnamefont {G.~K.}\ \bibnamefont
  {Das}}, \bibinfo {author} {\bibfnamefont {G.~D.}\ \bibnamefont {da~Fonseca}},
  \ and\ \bibinfo {author} {\bibfnamefont {R.~K.}\ \bibnamefont {Jallu}},\
  }\href@noop {} {\bibfield  {journal} {\bibinfo  {journal} {Discrete Applied
  Mathematics}\ } (\bibinfo {year} {2018})}\BibitemShut {NoStop}%
\bibitem [{\citenamefont {Hromkovi{\v{c}}}(2013)}]{hromkovivc2013algorithmics}%
  \BibitemOpen
  \bibfield  {author} {\bibinfo {author} {\bibfnamefont {J.}~\bibnamefont
  {Hromkovi{\v{c}}}},\ }\href@noop {} {\emph {\bibinfo {title} {Algorithmics
  for hard problems: introduction to combinatorial optimization, randomization,
  approximation, and heuristics}}}\ (\bibinfo  {publisher} {Springer Science \&
  Business Media},\ \bibinfo {year} {2013})\BibitemShut {NoStop}%
\bibitem [{\citenamefont {Cormen}\ \emph {et~al.}(2009)\citenamefont {Cormen},
  \citenamefont {Leiserson}, \citenamefont {Rivest},\ and\ \citenamefont
  {Stein}}]{cormen2009introduction}%
  \BibitemOpen
  \bibfield  {author} {\bibinfo {author} {\bibfnamefont {T.~H.}\ \bibnamefont
  {Cormen}}, \bibinfo {author} {\bibfnamefont {C.~E.}\ \bibnamefont
  {Leiserson}}, \bibinfo {author} {\bibfnamefont {R.~L.}\ \bibnamefont
  {Rivest}}, \ and\ \bibinfo {author} {\bibfnamefont {C.}~\bibnamefont
  {Stein}},\ }\href@noop {} {\emph {\bibinfo {title} {Introduction to
  algorithms}}}\ (\bibinfo  {publisher} {MIT press},\ \bibinfo {year}
  {2009})\BibitemShut {NoStop}%
\bibitem [{\citenamefont {van Leeuwen}(2005)}]{van2005approximation}%
  \BibitemOpen
  \bibfield  {author} {\bibinfo {author} {\bibfnamefont {E.~J.}\ \bibnamefont
  {van Leeuwen}},\ }in\ \href@noop {} {\emph {\bibinfo {booktitle}
  {International Workshop on Graph-Theoretic Concepts in Computer Science}}}\
  (\bibinfo {organization} {Springer},\ \bibinfo {year} {2005})\ pp.\ \bibinfo
  {pages} {351--361}\BibitemShut {NoStop}%
\bibitem [{\citenamefont {da~Fonseca}\ \emph {et~al.}(2017)\citenamefont
  {da~Fonseca}, \citenamefont {Pereira~de S{\'a}},\ and\ \citenamefont
  {de~Figueiredo}}]{da2017shifting}%
  \BibitemOpen
  \bibfield  {author} {\bibinfo {author} {\bibfnamefont {G.~D.}\ \bibnamefont
  {da~Fonseca}}, \bibinfo {author} {\bibfnamefont {V.~G.}\ \bibnamefont
  {Pereira~de S{\'a}}}, \ and\ \bibinfo {author} {\bibfnamefont {C.~M.~H.}\
  \bibnamefont {de~Figueiredo}},\ }\href@noop {} {\bibfield  {journal}
  {\bibinfo  {journal} {International Journal of Computational Geometry \&
  Applications}\ }\textbf {\bibinfo {volume} {27}},\ \bibinfo {pages} {255}
  (\bibinfo {year} {2017})}\BibitemShut {NoStop}%
\bibitem [{\citenamefont {Peruzzo}\ \emph {et~al.}(2014)\citenamefont
  {Peruzzo}, \citenamefont {McClean}, \citenamefont {Shadbolt}, \citenamefont
  {Yung}, \citenamefont {Zhou}, \citenamefont {Love}, \citenamefont
  {Aspuru-Guzik},\ and\ \citenamefont {O'Brien}}]{Peruzzo2014}%
  \BibitemOpen
  \bibfield  {author} {\bibinfo {author} {\bibfnamefont {A.}~\bibnamefont
  {Peruzzo}}, \bibinfo {author} {\bibfnamefont {J.}~\bibnamefont {McClean}},
  \bibinfo {author} {\bibfnamefont {P.}~\bibnamefont {Shadbolt}}, \bibinfo
  {author} {\bibfnamefont {M.-H.}\ \bibnamefont {Yung}}, \bibinfo {author}
  {\bibfnamefont {X.-Q.}\ \bibnamefont {Zhou}}, \bibinfo {author}
  {\bibfnamefont {P.~J.}\ \bibnamefont {Love}}, \bibinfo {author}
  {\bibfnamefont {A.}~\bibnamefont {Aspuru-Guzik}}, \ and\ \bibinfo {author}
  {\bibfnamefont {J.~L.}\ \bibnamefont {O'Brien}},\ }\href {\doibase
  10.1038/ncomms5213} {\bibfield  {journal} {\bibinfo  {journal} {Nature
  Communications}\ }\textbf {\bibinfo {volume} {5}},\ \bibinfo {pages} {4213}
  (\bibinfo {year} {2014})},\ \Eprint {http://arxiv.org/abs/1304.3061}
  {arXiv:1304.3061} \BibitemShut {NoStop}%
\bibitem [{\citenamefont {Pichler}\ \emph
  {et~al.}(2018{\natexlab{b}})\citenamefont {Pichler}, \citenamefont {Wang},
  \citenamefont {Zhou}, \citenamefont {Choi},\ and\ \citenamefont
  {Lukin}}]{pichler2018computational}%
  \BibitemOpen
  \bibfield  {author} {\bibinfo {author} {\bibfnamefont {H.}~\bibnamefont
  {Pichler}}, \bibinfo {author} {\bibfnamefont {S.-T.}\ \bibnamefont {Wang}},
  \bibinfo {author} {\bibfnamefont {L.}~\bibnamefont {Zhou}}, \bibinfo {author}
  {\bibfnamefont {S.}~\bibnamefont {Choi}}, \ and\ \bibinfo {author}
  {\bibfnamefont {M.~D.}\ \bibnamefont {Lukin}},\ }\href@noop {} {\bibfield
  {journal} {\bibinfo  {journal} {arXiv preprint arXiv:1809.04954}\ } (\bibinfo
  {year} {2018}{\natexlab{b}})}\BibitemShut {NoStop}%
\bibitem [{\citenamefont {Omran}\ \emph {et~al.}(2019)\citenamefont {Omran},
  \citenamefont {Levine}, \citenamefont {Keesling}, \citenamefont {Semeghini},
  \citenamefont {Wang}, \citenamefont {Ebadi}, \citenamefont {Bernien},
  \citenamefont {Zibrov}, \citenamefont {Pichler}, \citenamefont {Choi},
  \citenamefont {Cui}, \citenamefont {Rossignolo}, \citenamefont {Rembold},
  \citenamefont {Montangero}, \citenamefont {Calarco}, \citenamefont {Endres},
  \citenamefont {Greiner}, \citenamefont {Vuleti{\'{c}}},\ and\ \citenamefont
  {Lukin}}]{Omran2019}%
  \BibitemOpen
  \bibfield  {author} {\bibinfo {author} {\bibfnamefont {A.}~\bibnamefont
  {Omran}}, \bibinfo {author} {\bibfnamefont {H.}~\bibnamefont {Levine}},
  \bibinfo {author} {\bibfnamefont {A.}~\bibnamefont {Keesling}}, \bibinfo
  {author} {\bibfnamefont {G.}~\bibnamefont {Semeghini}}, \bibinfo {author}
  {\bibfnamefont {T.~T.}\ \bibnamefont {Wang}}, \bibinfo {author}
  {\bibfnamefont {S.}~\bibnamefont {Ebadi}}, \bibinfo {author} {\bibfnamefont
  {H.}~\bibnamefont {Bernien}}, \bibinfo {author} {\bibfnamefont {A.~S.}\
  \bibnamefont {Zibrov}}, \bibinfo {author} {\bibfnamefont {H.}~\bibnamefont
  {Pichler}}, \bibinfo {author} {\bibfnamefont {S.}~\bibnamefont {Choi}},
  \bibinfo {author} {\bibfnamefont {J.}~\bibnamefont {Cui}}, \bibinfo {author}
  {\bibfnamefont {M.}~\bibnamefont {Rossignolo}}, \bibinfo {author}
  {\bibfnamefont {P.}~\bibnamefont {Rembold}}, \bibinfo {author} {\bibfnamefont
  {S.}~\bibnamefont {Montangero}}, \bibinfo {author} {\bibfnamefont
  {T.}~\bibnamefont {Calarco}}, \bibinfo {author} {\bibfnamefont
  {M.}~\bibnamefont {Endres}}, \bibinfo {author} {\bibfnamefont
  {M.}~\bibnamefont {Greiner}}, \bibinfo {author} {\bibfnamefont
  {V.}~\bibnamefont {Vuleti{\'{c}}}}, \ and\ \bibinfo {author} {\bibfnamefont
  {M.~D.}\ \bibnamefont {Lukin}},\ }\href {\doibase 10.1126/science.aax9743}
  {\bibfield  {journal} {\bibinfo  {journal} {Science}\ }\textbf {\bibinfo
  {volume} {365}},\ \bibinfo {pages} {570} (\bibinfo {year} {2019})},\ \Eprint
  {http://arxiv.org/abs/1905.05721} {arXiv:1905.05721} \BibitemShut {NoStop}%
\bibitem [{\citenamefont {de~L{\'{e}}s{\'{e}}leuc}\ \emph
  {et~al.}(2018)\citenamefont {de~L{\'{e}}s{\'{e}}leuc}, \citenamefont
  {Barredo}, \citenamefont {Lienhard}, \citenamefont {Browaeys},\ and\
  \citenamefont {Lahaye}}]{DeLeseleuc2018a}%
  \BibitemOpen
  \bibfield  {author} {\bibinfo {author} {\bibfnamefont {S.}~\bibnamefont
  {de~L{\'{e}}s{\'{e}}leuc}}, \bibinfo {author} {\bibfnamefont
  {D.}~\bibnamefont {Barredo}}, \bibinfo {author} {\bibfnamefont
  {V.}~\bibnamefont {Lienhard}}, \bibinfo {author} {\bibfnamefont
  {A.}~\bibnamefont {Browaeys}}, \ and\ \bibinfo {author} {\bibfnamefont
  {T.}~\bibnamefont {Lahaye}},\ }\href {\doibase 10.1103/PhysRevA.97.053803}
  {\bibfield  {journal} {\bibinfo  {journal} {Physical Review A}\ }\textbf
  {\bibinfo {volume} {97}},\ \bibinfo {pages} {053803} (\bibinfo {year}
  {2018})},\ \Eprint {http://arxiv.org/abs/1802.10424} {arXiv:1802.10424}
  \BibitemShut {NoStop}%
\bibitem [{\citenamefont {Lahaye}\ and\ \citenamefont
  {Browaeys}()}]{ThierryPrivate}%
  \BibitemOpen
  \bibfield  {author} {\bibinfo {author} {\bibfnamefont {T.}~\bibnamefont
  {Lahaye}}\ and\ \bibinfo {author} {\bibfnamefont {A.}~\bibnamefont
  {Browaeys}},\ }\href@noop {} {}\bibinfo {howpublished} {Private
  communication}\BibitemShut {NoStop}%
\bibitem [{\citenamefont {Lienhard}(2019)}]{LienhardThesis}%
  \BibitemOpen
  \bibfield  {author} {\bibinfo {author} {\bibfnamefont {V.}~\bibnamefont
  {Lienhard}},\ }\emph {\bibinfo {title} {{Physique quantique exp\'erimentale
  \`a N corps dans des matrices d'atomes de Rydberg. Des mod\`eles de spins \`a
  la mati\`ere topologique.}}},\ \href {http://www.theses.fr/s164481} {Ph.D.
  thesis},\ \bibinfo  {school} {Institut d'Optique Graduate School} (\bibinfo
  {year} {2019})\BibitemShut {NoStop}%
\bibitem [{\citenamefont {Nandy}\ \emph {et~al.}(2017)\citenamefont {Nandy},
  \citenamefont {Pandit},\ and\ \citenamefont {Roy}}]{nandy2017faster}%
  \BibitemOpen
  \bibfield  {author} {\bibinfo {author} {\bibfnamefont {S.~C.}\ \bibnamefont
  {Nandy}}, \bibinfo {author} {\bibfnamefont {S.}~\bibnamefont {Pandit}}, \
  and\ \bibinfo {author} {\bibfnamefont {S.}~\bibnamefont {Roy}},\ }\href@noop
  {} {\bibfield  {journal} {\bibinfo  {journal} {Information Processing
  Letters}\ }\textbf {\bibinfo {volume} {127}},\ \bibinfo {pages} {58}
  (\bibinfo {year} {2017})}\BibitemShut {NoStop}%
\bibitem [{\citenamefont {Das}\ \emph {et~al.}(2015)\citenamefont {Das},
  \citenamefont {De}, \citenamefont {Kolay}, \citenamefont {Nandy},\ and\
  \citenamefont {Sur-Kolay}}]{das2015approximation}%
  \BibitemOpen
  \bibfield  {author} {\bibinfo {author} {\bibfnamefont {G.~K.}\ \bibnamefont
  {Das}}, \bibinfo {author} {\bibfnamefont {M.}~\bibnamefont {De}}, \bibinfo
  {author} {\bibfnamefont {S.}~\bibnamefont {Kolay}}, \bibinfo {author}
  {\bibfnamefont {S.~C.}\ \bibnamefont {Nandy}}, \ and\ \bibinfo {author}
  {\bibfnamefont {S.}~\bibnamefont {Sur-Kolay}},\ }\href@noop {} {\bibfield
  {journal} {\bibinfo  {journal} {Information Processing Letters}\ }\textbf
  {\bibinfo {volume} {115}},\ \bibinfo {pages} {439} (\bibinfo {year}
  {2015})}\BibitemShut {NoStop}%
\bibitem [{\citenamefont {Li}\ \emph {et~al.}(2017)\citenamefont {Li},
  \citenamefont {Jiang},\ and\ \citenamefont {Many{\`a}}}]{li2017minimization}%
  \BibitemOpen
  \bibfield  {author} {\bibinfo {author} {\bibfnamefont {C.-M.}\ \bibnamefont
  {Li}}, \bibinfo {author} {\bibfnamefont {H.}~\bibnamefont {Jiang}}, \ and\
  \bibinfo {author} {\bibfnamefont {F.}~\bibnamefont {Many{\`a}}},\ }\href@noop
  {} {\bibfield  {journal} {\bibinfo  {journal} {Computers \& Operations
  Research}\ }\textbf {\bibinfo {volume} {84}},\ \bibinfo {pages} {1} (\bibinfo
  {year} {2017})}\BibitemShut {NoStop}%
\bibitem [{\citenamefont {Endres}\ \emph {et~al.}(2016)\citenamefont {Endres},
  \citenamefont {Bernien}, \citenamefont {Keesling}, \citenamefont {Levine},
  \citenamefont {Anschuetz}, \citenamefont {Krajenbrink}, \citenamefont
  {Senko}, \citenamefont {Vuletic}, \citenamefont {Greiner},\ and\
  \citenamefont {Lukin}}]{Endres2016}%
  \BibitemOpen
  \bibfield  {author} {\bibinfo {author} {\bibfnamefont {M.}~\bibnamefont
  {Endres}}, \bibinfo {author} {\bibfnamefont {H.}~\bibnamefont {Bernien}},
  \bibinfo {author} {\bibfnamefont {A.}~\bibnamefont {Keesling}}, \bibinfo
  {author} {\bibfnamefont {H.}~\bibnamefont {Levine}}, \bibinfo {author}
  {\bibfnamefont {E.~R.}\ \bibnamefont {Anschuetz}}, \bibinfo {author}
  {\bibfnamefont {A.}~\bibnamefont {Krajenbrink}}, \bibinfo {author}
  {\bibfnamefont {C.}~\bibnamefont {Senko}}, \bibinfo {author} {\bibfnamefont
  {V.}~\bibnamefont {Vuletic}}, \bibinfo {author} {\bibfnamefont
  {M.}~\bibnamefont {Greiner}}, \ and\ \bibinfo {author} {\bibfnamefont
  {M.~D.}\ \bibnamefont {Lukin}},\ }\href {http://arxiv.org/abs/1607.03044} {\
  (\bibinfo {year} {2016})},\ \Eprint {http://arxiv.org/abs/1607.03044}
  {arXiv:1607.03044} \BibitemShut {NoStop}%
\bibitem [{\citenamefont {Henriet}\ \emph {et~al.}(2020)\citenamefont
  {Henriet}, \citenamefont {Beguin}, \citenamefont {Signoles}, \citenamefont
  {Lahaye}, \citenamefont {Browaeys}, \citenamefont {Reymond},\ and\
  \citenamefont {Jurczak}}]{Henriet2020a}%
  \BibitemOpen
  \bibfield  {author} {\bibinfo {author} {\bibfnamefont {L.}~\bibnamefont
  {Henriet}}, \bibinfo {author} {\bibfnamefont {L.}~\bibnamefont {Beguin}},
  \bibinfo {author} {\bibfnamefont {A.}~\bibnamefont {Signoles}}, \bibinfo
  {author} {\bibfnamefont {T.}~\bibnamefont {Lahaye}}, \bibinfo {author}
  {\bibfnamefont {A.}~\bibnamefont {Browaeys}}, \bibinfo {author}
  {\bibfnamefont {G.-O.}\ \bibnamefont {Reymond}}, \ and\ \bibinfo {author}
  {\bibfnamefont {C.}~\bibnamefont {Jurczak}},\ }\href
  {http://arxiv.org/abs/2006.12326} {\  (\bibinfo {year} {2020})},\ \Eprint
  {http://arxiv.org/abs/2006.12326} {arXiv:2006.12326} \BibitemShut {NoStop}%
\bibitem [{\citenamefont {Barkoutsos}\ \emph {et~al.}(2017)\citenamefont
  {Barkoutsos}, \citenamefont {Moll}, \citenamefont {Staar}, \citenamefont
  {Mueller}, \citenamefont {Fuhrer}, \citenamefont {Filipp}, \citenamefont
  {Troyer},\ and\ \citenamefont {Tavernelli}}]{Barkoutsos2017}%
  \BibitemOpen
  \bibfield  {author} {\bibinfo {author} {\bibfnamefont {P.~K.}\ \bibnamefont
  {Barkoutsos}}, \bibinfo {author} {\bibfnamefont {N.}~\bibnamefont {Moll}},
  \bibinfo {author} {\bibfnamefont {P.~W.~J.}\ \bibnamefont {Staar}}, \bibinfo
  {author} {\bibfnamefont {P.}~\bibnamefont {Mueller}}, \bibinfo {author}
  {\bibfnamefont {A.}~\bibnamefont {Fuhrer}}, \bibinfo {author} {\bibfnamefont
  {S.}~\bibnamefont {Filipp}}, \bibinfo {author} {\bibfnamefont
  {M.}~\bibnamefont {Troyer}}, \ and\ \bibinfo {author} {\bibfnamefont
  {I.}~\bibnamefont {Tavernelli}},\ }\href {http://arxiv.org/abs/1706.03637} {\
   (\bibinfo {year} {2017})},\ \Eprint {http://arxiv.org/abs/1706.03637}
  {arXiv:1706.03637} \BibitemShut {NoStop}%
\bibitem [{\citenamefont {Yu}\ \emph {et~al.}(2020)\citenamefont {Yu},
  \citenamefont {Wilczek},\ and\ \citenamefont {Wu}}]{Yu2020}%
  \BibitemOpen
  \bibfield  {author} {\bibinfo {author} {\bibfnamefont {H.}~\bibnamefont
  {Yu}}, \bibinfo {author} {\bibfnamefont {F.}~\bibnamefont {Wilczek}}, \ and\
  \bibinfo {author} {\bibfnamefont {B.}~\bibnamefont {Wu}},\ }\href
  {http://arxiv.org/abs/2005.13089} {\  (\bibinfo {year} {2020})},\ \Eprint
  {http://arxiv.org/abs/2005.13089} {arXiv:2005.13089} \BibitemShut {NoStop}%
\bibitem [{\citenamefont {Downey}\ and\ \citenamefont
  {Fellows}(2013)}]{downey2013fundamentals}%
  \BibitemOpen
  \bibfield  {author} {\bibinfo {author} {\bibfnamefont {R.~G.}\ \bibnamefont
  {Downey}}\ and\ \bibinfo {author} {\bibfnamefont {M.~R.}\ \bibnamefont
  {Fellows}},\ }\href@noop {} {\emph {\bibinfo {title} {Fundamentals of
  parameterized complexity}}},\ Vol.~\bibinfo {volume} {4}\ (\bibinfo
  {publisher} {Springer},\ \bibinfo {year} {2013})\BibitemShut {NoStop}%
\bibitem [{\citenamefont {Lemar{\'{e}}chal}(2001)}]{Lemarechal2001}%
  \BibitemOpen
  \bibfield  {author} {\bibinfo {author} {\bibfnamefont {C.}~\bibnamefont
  {Lemar{\'{e}}chal}},\ }in\ \href {\doibase 10.1007/3-540-45586-8_4} {\emph
  {\bibinfo {booktitle} {Lecture Notes in Computer Science (including subseries
  Lecture Notes in Artificial Intelligence and Lecture Notes in
  Bioinformatics)}}},\ Vol.\ \bibinfo {volume} {2241}\ (\bibinfo {year}
  {2001})\ pp.\ \bibinfo {pages} {112--156}\BibitemShut {NoStop}%
\bibitem [{\citenamefont {Dalibard}\ \emph {et~al.}(1992)\citenamefont
  {Dalibard}, \citenamefont {Castin},\ and\ \citenamefont
  {M{\o}lmer}}]{Dalibard1992}%
  \BibitemOpen
  \bibfield  {author} {\bibinfo {author} {\bibfnamefont {J.}~\bibnamefont
  {Dalibard}}, \bibinfo {author} {\bibfnamefont {Y.}~\bibnamefont {Castin}}, \
  and\ \bibinfo {author} {\bibfnamefont {K.}~\bibnamefont {M{\o}lmer}},\ }\href
  {\doibase 10.1103/PhysRevLett.68.580} {\bibfield  {journal} {\bibinfo
  {journal} {Physical Review Letters}\ }\textbf {\bibinfo {volume} {68}},\
  \bibinfo {pages} {580} (\bibinfo {year} {1992})}\BibitemShut {NoStop}%
\bibitem [{\citenamefont {Johansson}\ \emph {et~al.}(2013)\citenamefont
  {Johansson}, \citenamefont {Nation},\ and\ \citenamefont
  {Nori}}]{Johansson2013}%
  \BibitemOpen
  \bibfield  {author} {\bibinfo {author} {\bibfnamefont {J.~R.}\ \bibnamefont
  {Johansson}}, \bibinfo {author} {\bibfnamefont {P.~D.}\ \bibnamefont
  {Nation}}, \ and\ \bibinfo {author} {\bibfnamefont {F.}~\bibnamefont
  {Nori}},\ }\href {\doibase 10.1016/j.cpc.2012.11.019} {\bibfield  {journal}
  {\bibinfo  {journal} {Computer Physics Communications}\ }\textbf {\bibinfo
  {volume} {184}},\ \bibinfo {pages} {1234} (\bibinfo {year} {2013})},\ \Eprint
  {http://arxiv.org/abs/1211.6518} {arXiv:1211.6518} \BibitemShut {NoStop}%
\bibitem [{\citenamefont {{Virtanen}}\ \emph {et~al.}(2020)\citenamefont
  {{Virtanen}}, \citenamefont {{Gommers}}, \citenamefont {{Oliphant}},
  \citenamefont {{Haberland}}, \citenamefont {{Reddy}}, \citenamefont
  {{Cournapeau}}, \citenamefont {{Burovski}}, \citenamefont {{Peterson}},
  \citenamefont {{Weckesser}}, \citenamefont {{Bright}}, \citenamefont {{van
  der Walt}}, \citenamefont {{Brett}}, \citenamefont {{Wilson}}, \citenamefont
  {{Jarrod Millman}}, \citenamefont {{Mayorov}}, \citenamefont {{Nelson}},
  \citenamefont {{Jones}}, \citenamefont {{Kern}}, \citenamefont {{Larson}},
  \citenamefont {{Carey}}, \citenamefont {{Polat}}, \citenamefont {{Feng}},
  \citenamefont {{Moore}}, \citenamefont {{Vand erPlas}}, \citenamefont
  {{Laxalde}}, \citenamefont {{Perktold}}, \citenamefont {{Cimrman}},
  \citenamefont {{Henriksen}}, \citenamefont {{Quintero}}, \citenamefont
  {{Harris}}, \citenamefont {{Archibald}}, \citenamefont {{Ribeiro}},
  \citenamefont {{Pedregosa}}, \citenamefont {{van Mulbregt}},\ and\
  \citenamefont {{Contributors}}}]{2020SciPy-NMeth}%
  \BibitemOpen
  \bibfield  {author} {\bibinfo {author} {\bibfnamefont {P.}~\bibnamefont
  {{Virtanen}}}, \bibinfo {author} {\bibfnamefont {R.}~\bibnamefont
  {{Gommers}}}, \bibinfo {author} {\bibfnamefont {T.~E.}\ \bibnamefont
  {{Oliphant}}}, \bibinfo {author} {\bibfnamefont {M.}~\bibnamefont
  {{Haberland}}}, \bibinfo {author} {\bibfnamefont {T.}~\bibnamefont
  {{Reddy}}}, \bibinfo {author} {\bibfnamefont {D.}~\bibnamefont
  {{Cournapeau}}}, \bibinfo {author} {\bibfnamefont {E.}~\bibnamefont
  {{Burovski}}}, \bibinfo {author} {\bibfnamefont {P.}~\bibnamefont
  {{Peterson}}}, \bibinfo {author} {\bibfnamefont {W.}~\bibnamefont
  {{Weckesser}}}, \bibinfo {author} {\bibfnamefont {J.}~\bibnamefont
  {{Bright}}}, \bibinfo {author} {\bibfnamefont {S.~J.}\ \bibnamefont {{van der
  Walt}}}, \bibinfo {author} {\bibfnamefont {M.}~\bibnamefont {{Brett}}},
  \bibinfo {author} {\bibfnamefont {J.}~\bibnamefont {{Wilson}}}, \bibinfo
  {author} {\bibfnamefont {K.}~\bibnamefont {{Jarrod Millman}}}, \bibinfo
  {author} {\bibfnamefont {N.}~\bibnamefont {{Mayorov}}}, \bibinfo {author}
  {\bibfnamefont {A.~R.~J.}\ \bibnamefont {{Nelson}}}, \bibinfo {author}
  {\bibfnamefont {E.}~\bibnamefont {{Jones}}}, \bibinfo {author} {\bibfnamefont
  {R.}~\bibnamefont {{Kern}}}, \bibinfo {author} {\bibfnamefont
  {E.}~\bibnamefont {{Larson}}}, \bibinfo {author} {\bibfnamefont
  {C.}~\bibnamefont {{Carey}}}, \bibinfo {author} {\bibfnamefont
  {{\.I}.}~\bibnamefont {{Polat}}}, \bibinfo {author} {\bibfnamefont
  {Y.}~\bibnamefont {{Feng}}}, \bibinfo {author} {\bibfnamefont {E.~W.}\
  \bibnamefont {{Moore}}}, \bibinfo {author} {\bibfnamefont {J.}~\bibnamefont
  {{Vand erPlas}}}, \bibinfo {author} {\bibfnamefont {D.}~\bibnamefont
  {{Laxalde}}}, \bibinfo {author} {\bibfnamefont {J.}~\bibnamefont
  {{Perktold}}}, \bibinfo {author} {\bibfnamefont {R.}~\bibnamefont
  {{Cimrman}}}, \bibinfo {author} {\bibfnamefont {I.}~\bibnamefont
  {{Henriksen}}}, \bibinfo {author} {\bibfnamefont {E.~A.}\ \bibnamefont
  {{Quintero}}}, \bibinfo {author} {\bibfnamefont {C.~R.}\ \bibnamefont
  {{Harris}}}, \bibinfo {author} {\bibfnamefont {A.~M.}\ \bibnamefont
  {{Archibald}}}, \bibinfo {author} {\bibfnamefont {A.~H.}\ \bibnamefont
  {{Ribeiro}}}, \bibinfo {author} {\bibfnamefont {F.}~\bibnamefont
  {{Pedregosa}}}, \bibinfo {author} {\bibfnamefont {P.}~\bibnamefont {{van
  Mulbregt}}}, \ and\ \bibinfo {author} {\bibfnamefont {S.~.~.}\ \bibnamefont
  {{Contributors}}},\ }\href {\doibase
  https://doi.org/10.1038/s41592-019-0686-2} {\bibfield  {journal} {\bibinfo
  {journal} {Nature Methods}\ }\textbf {\bibinfo {volume} {17}},\ \bibinfo
  {pages} {261} (\bibinfo {year} {2020})}\BibitemShut {NoStop}%
\end{thebibliography}%

\end{document}